\documentstyle[psfig]{article}
\newcommand{\tbsp}{\rule{0pt}{12pt}}
\def\hlines{\hline \tbsp}
\newcommand{\nin}{\in \hspace{-2.8mm} / \hspace{1.5mm}}
\begin{document}
\pagestyle{headings}
\pagenumbering{arabic}
\begin{center}
{\large {\bf Active Optics in Modern, Large Optical Telescopes}}\\
\vspace*{3mm}
Lothar Noethe \\
{\it European Southern Observatory, \\
Karl-Schwarzschild-Str.2,\\
85748 Garching,\\
Germany}
\end{center}
\tableofcontents
\section{Introduction}
\label{sec:introduction}
The goal of large astronomical telescopes is the concentration of
large amounts of light in small areas, that is with optimum image
quality. This requires that the optical configuration of the
telescope be always close to an optimum state. The optimum state is
defined with respect to the environment in which the telescope is
operated.
In space it is
the diffraction image of the telescope and on the ground the image which can
be obtained with a large optically perfect telescope in the presence of atmospheric
disturbances, the so-called seeing disc. Deviations from this optimum state,
due to wavefront aberrations generated by the optics of the telescope, are
unavoidable. But, the telescope is still defined as diffraction limited or seeing
limited if the degradation of the image is smaller than accepted limits. The
criterion for a diffraction limited performance is that the ratio of the intensity
of the real image at its center to the intensity of the diffraction image at its center,
the so-called Strehl ratio, be larger than 0.8. This is achieved if the root mean square
(r.m.s.) $\sigma_{{\rm w}}$ of the wavefront aberrations is less than $\lambda/14$,
where $\lambda$ is the wavelength of the observed radiation. For a ground based
operation, where the atmospheric effects are not corrected, the telescope can be
defined as seeing limited if the equivalent
ratio of the intensity at the center of the real image to the one at
the center of the optimum image, the so
called central intensity ratio (CIR) (Dierickx [1992]), is also
greater than 0.8.
Whereas, for small wavefront aberrations, the Strehl ratio
depends on the square of the r.m.s. of the wavefront error, the CIR depends on the
square of the r.m.s. $\sigma_{{\rm t}}$ of the slopes of the wavefront error, and also
on the current seeing, expressed as the full width at half maximum $\Theta$ of the
seeing disk :
\begin{equation}
  CIR = 1 - 2.89 \bigg( \frac{\sigma_{{\rm t}}}{\Theta}\bigg) ^{2}
          \label{eq:CIR}
\end{equation}
$\Theta$ depends on the wavelength $\lambda$ of the light and is proportional to
$\lambda^{-\frac{1}{5}}$.
The goal of the design of a telescope is therefore to
limit the wavefront aberrations to amounts which will guarantee a
diffraction or seeing limited performance. In old {\em passive} telescopes this was
attempted by using special constructional design features. With the increase in size
this proved to be no longer sufficient (indeed, significant
extrapolation beyond 5\,m was possible neither technically nor
costwise), but with the introduction of active
elements, which can correct the aberrations during operation in a systematic
way, the goals can nowadays be achieved also for very large
telescopes.
Such ground based telescopes with the goal
of a seeing limited performance will be called {\em active}, those with the goal of
diffraction limited performance adaptive. In space, the goal of active
optics would be a diffraction limited performance. This article will only deal
with active optics, which by definition does not include the
correction of pointing errors, that is guiding and tracking.\\
Chapter \ref{sec:principles} gives an overview of the principles of
active optics.
Chapter \ref{sec:system} introduces the relationships
between the various components and parameters of an active optics
system with special emphasis on telescopes with a monolithic primary
mirror.
Chapter \ref{sec:wavefrontSensing} describes the properties and design
of one type of wavefront analyser customised for an active optics system.
Chapter \ref{sec:minEnergyModes} summarizes the major characteristics of
the elastic modes of a meniscus mirror, which are of central
importance for the control of a thin monolithic mirror,
and chapter \ref{sec:support} deals with the theory of the support of
such mirrors.
Chapter \ref{sec:alignment} shows how the alignment can be
controlled by active optics
and chapter \ref{sec:modificationOpticalConfig} the possibilities of
changing the optical configuration and the plate scale of a
telescope.
Chapter \ref{sec:AODesignNTTVLT} describes the designs of the active
optics systems of the New Technology Telescope (NTT) and the Very
Large Telescope (VLT) of the European Southern Observatory
and chapter \ref{sec:practicalExperience} summarizes some practical
experience with these active optics systems.
Chapter \ref{sec:existingAOTelescopes} gives a short overview of
existing telescopes working with active optics and
chapter \ref{sec:outlook} presents an outlook for the implementation
of active optics into future telescopes with even larger mirror
diameters and more than two optical components.\\
Most of the review deals with two mirror telescopes with altazimuth
mountings and strong emphasis is put on the systems aspects.
Earlier reviews have been given by Hubin and Noethe [1993] and by
Wilson [1996], the latter also with a detailed presentation of the 
historical developments and an extensive list of references. More
details about active optics with thin meniscus mirrors are given by
Noethe [2000].
\section{Principles of active optics}
\label{sec:principles}
\subsection{Error sources}
\begin{figure}[b]
     \centerline{\hbox{
      \psfig{figure=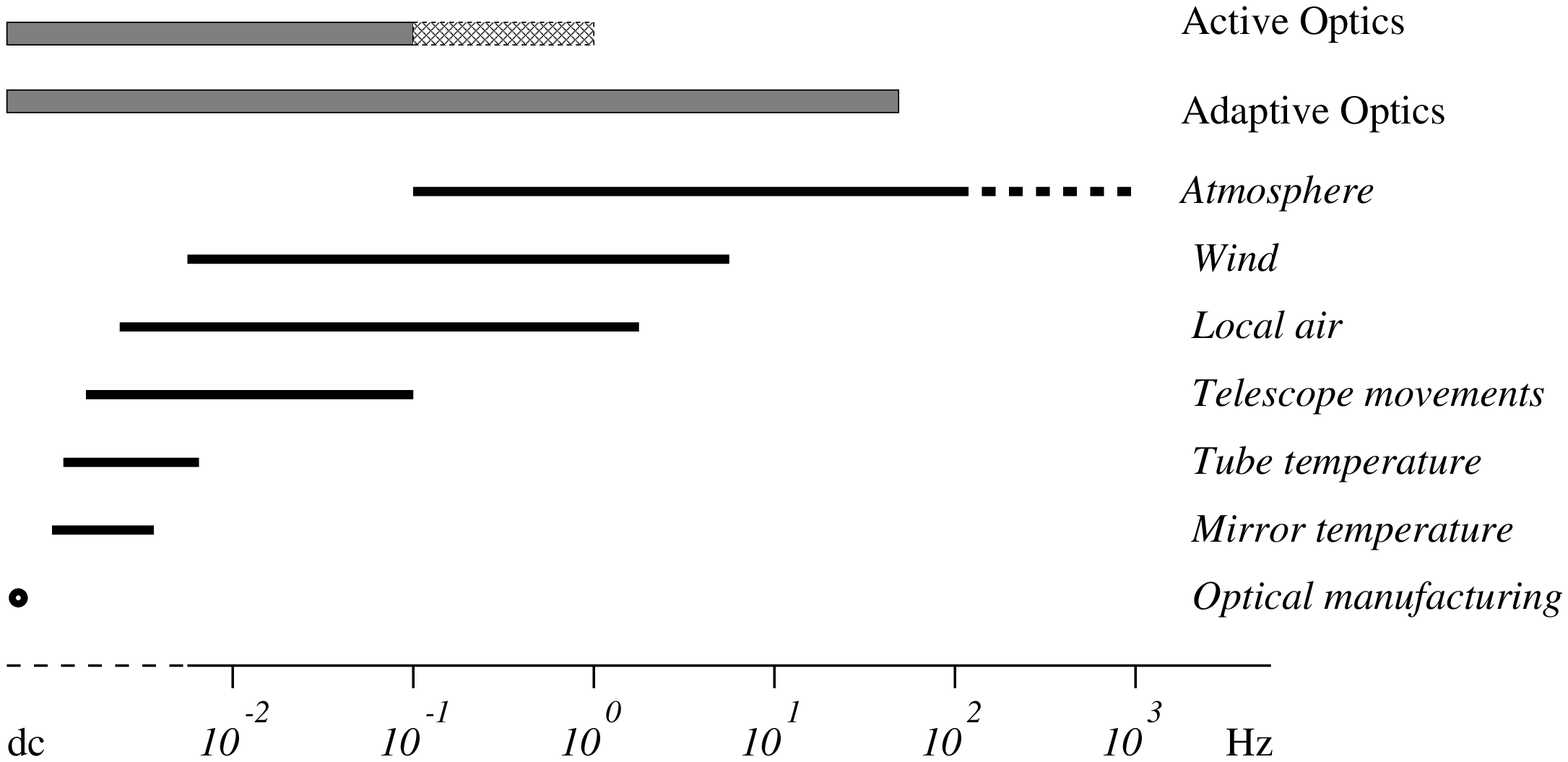,width=130mm}}}
      \caption{\label{fig:bandpass} {\small Bandpasses of sources of wavefront
            aberrations in optical telescopes}}
\end{figure}

Since the design of a telescope is strongly based on the avoidance of
wavefront aberrations, we discuss first the possible error sources,
shown and classified according to their frequency bandpasses, in figure
\ref{fig:bandpass}.\\
{\it 1. Optical manufacturing}.
These errors are constant in time. During the polishing phase
the mirrors can usually not be tested together as one system.
But, alone, neither of the
two mirrors produces a sharp image, in particular not with an incoming spherical
wavefront generated by a small pinhole. Therefore, interferometric testing is only
possible with so-called null lenses which generate wavefronts
which are identical to the required shapes of the mirrors.
Predominantly rotationally symmetric errors in the manufacturing of
these null lenses can then lead to severe errors in the shape of the
mirrors in the form of spherical aberration. However, testing of null lenses
is nowadays possible and, independently, the
spherical aberration of the combined system can be measured in the
manufacturing plant with the pentaprism test (Wetthauer and Brodhun [1920]).\\
{\it 2. Mirror temperature.}
Owing to their huge inertia and the ineffective heat exchange with the air,
large telescope mirrors follow temperature variations only slowly, that is the
mirrors filter out all but the lowest temporal frequencies of the air temperature
variations. Nevertheless, the day to night changes of the air temperature
result in temperature changes of the mirrors of possibly a few degrees.
This is, unless an extremely low expansion glass
is used, sufficient for a noticeable change of the focus position and
other aberrations.\\
{\it 3. Tube temperature.}
Owing to its much smaller mass and therefore lower inertia, and because
of a faster heat exchange due to radiative cooling, the changes of the tube
temperature are much faster and larger than the ones of the primary mirror.
Again, as in the case of the change of the mirror temperature,
the main and possibly only significant effect is a change of the focus position.\\
{\it 4. Telescope movements.}
Any movement of the telescope tube, for example a change of the zenith
angle in telescopes with an altazimuth mounting,
will change to some extent the alignment of the telescope and the forces
acting on the primary mirror, both effects generating wavefront aberrations.
While small telescopes can be intrinsically sufficiently rigid for these
effects not to play a role, large telescopes with diameters of the primary
mirrors of more than, say, two meters are always noticeably affected by
elastic deformations unless they are actively controlled.\\      
{\it 5. Local air.}
Local air is defined here as the air inside the telescope enclosure and
the air in the ground layer in the vicinity of the telescope enclosure.
The local air conditions in the enclosure can
be influenced by the design of the enclosure, avoidance of heat sources
and active devices to maintain small temperature differences between
various parts of the telescope and the ambient air (Racine, Salmon,
Cowley and Sovka [1991]).\\
{\it 6. Wind.}
Wind generates both movements and elastic deformations of the telescope
structure, especially of the telescope tube, as well as elastic deformations
of the primary mirrors if these are sufficiently thin.
Inside enclosures the peak of the energy spectrum is at approximately 2\,Hz.\\
{\it 7. Free atmosphere.}
The effects of the free atmosphere above the ground layer on the image
quality are predominantly generated by a layer at an altitude of
approximately 10 km. The frequency range is very large, ranging from
approximately 0.03\,Hz to 1000\,Hz.\\
The natural frequency for splitting the errors into two groups is the approximate
lower frequency limit of the errors generated by the free atmosphere.
Wavefront aberrations generated in the free atmosphere, especially at
high altitude, are strongly dependent
on the field angle, that is they are anisoplanatic. With integrations times
larger than 30 seconds the wavefront aberrations due to the free atmosphere
are effectively integrated out and the remaining aberrations are then
independent of the field angle, that is, are isoplanatic. 
This important condition allows that the information about the wavefront
aberrations obtained with a star anywhere in the field can be used to correct the
images over the whole field. 
The lower frequency range up to the limit of 0.03\,Hz includes completely the first
four sources and partially the sources five to seven. Systems which systematically
attempt to correct these telescope errors during operation leaving only the errors
generated by the free atmosphere, and therefore to achieve a seeing limited performance
will be called {\em active optics} systems, those which are predominantly designed
to correct the aberrations generated by the free atmosphere and to achieve diffraction
limited performance will be called {\em adaptive optics} systems. The latter work at much
higher frequencies and are not the subject of this paper.
\subsection{Classification of active telescopes}
Up to the 1980s all telescopes were passive in the sense that after
the initial setup the optical configuration was, apart from focusing,
never or very rarely, and then only manually, modified. Active telescopes,
on the other hand, are capable of modifying the optical configuration
systematically even during operation, based on data obtained with
measurements with the final, completely installed system.
They can be classified according to the type of control loops and the
type of correction strategies and capabilities.
\subsubsection{Control loops}
From a design point of view, the major differences between a passive
and an active telescope are the time periods for the stability
requirements of the system defining the optics on the one hand, and
the role of absolute versus differential requirements on the other hand.
To illustrate this point,
consider first the design of a {\em passive} telescope with two mirrors. The
optical configuration is fully defined by the shape of the primary mirror
and the relative positions of the two mirrors.
One therefore has to
find support systems for both mirrors which maintain
the shape and the relative positions independently of the
telescope attitude for time periods of hours. The positions are mainly
influenced by deformations of the telescope tube and the shape of the
primary mirror by deformations of its cell.
For large telescopes neither structural
component can be built with sufficient stiffness since this would
require deformations of the telescope tube of only a few micrometers and
deformations of the primary mirror cell of less than the wavelength of
light. But the variations of relative positions can be reduced by the
use of Serrurier struts, which, despite the deformation of the
telescope tube, make the support structures of both mirrors move in
parallel when the telescope attitude is changed.
The deformations
of the primary mirror can be minimised by decoupling the primary
mirror from the deformations of its cell by using astatic supports,
which can be either mechanical levers (Lassell [1842]) or hydraulic or
pneumatic devices interconnected in three groups (Yoder [1986]).
All these apply forces which are independent of the distance between the
mirror and its cell.\\
Clearly, both of these design features will only guarantee the
stability of the force setting, that is the application of the correct
forces for any zenith angle, and the stability of the relative
positions to a certain degree.
Any force errors will generate deformations of the primary
mirror which are inversely proportional to its stiffness.
The specifications for the
tolerable wavefront aberrations will therefore define the minimum
stiffness of the primary mirror and, up to diameters of approximately two
meters,with the help of the scaling laws (\ref{eq:scalingPressure}),
(\ref{eq:scalingSag}) and (\ref{eq:scalingSingleForce}) for thin
mirrors given in \S \ref{sec:scalingLaws}, also its minimum
thickness.
For diameters of more than two meters, the mirrors become
prohibitively thick.
In addition, because of the influence of shearing forces in thick
mirrors, they are more flexible than suggested by the scaling laws
mentioned above.
The required stiffness can therefore not be achieved by simply
increasing the thickness of the mirrors. The diameters of
monolithic primary mirrors of passive telescopes capable of a
seeing limited or even a diffraction limited performance are consequently
limited to the order of two meters.\\
In addition, the telescope should ideally be made of materials which do not
deform under temperature variations, and, for the mirrors, guarantee a
stable shape over long periods of time. The main effect of the
temperature variations would be defocus, due both
to a change of the length of the tube and a deformation of the
mirrors. For the mirrors, the material which
fulfills both requirements is low expansion glass. 
But, defocusing as a result of the contraction or expansion of the
generally metallic structure cannot be avoided.\\
{\em Active} telescopes, on the other hand, do not need the stability of
the forces or positions to be maintained over long periods of
time. Instead, forces and positions can be changed depending on the
knowledge of the passively
generated deformations. This is a much easier requirement than the
passive stability over time periods of hours and allows the use of
less rigid elements, in
particular a less rigid and therefore thinner primary mirror. The
additional important question is whether these modifications are
carried out in open or closed loop. Open loop changes require the
knowledge and predictability of the optimum absolute
forces and positions for all sky positions.
A condition for this predictability is that the system
be free of significant friction and therefore hysteresis effects.
It should also be
capable of setting these absolute forces and positions with the required
accuracy over time periods of hours.
On the other hand,
pure closed loop operations require the stability of
the forces and positions only for small time periods between two
measurements of the wavefront analyser. High accuracy is then
predominantly required for differential force and position settings,
which can be done much more accurately than absolute settings. As a
consequence, the requirements for the stability and predictability of
the deformations of the optomechanical elements can, compared with
open loop operations, be further reduced.\\
Since the number of free design parameters is much larger in active
telescopes and, at least for a closed loop operation, the system also
needs a wavefront analyser adapted to the mechanics of the telescopes,
the design of an active telescope is more complex than the one of a
passive telescope. Clearly, from the considerations above, {\em the
goal should be a closed loop active optics operation based on
information from the image forming wavefront in the exit
pupil}. Nevertheless, open loop
or mixed open and closed loop operations are also feasible.
In both cases the full active optics system
consists of mechanical parts performing the corrections, and wavefront
sensors, which either online or offline measure the wavefront errors.
\subsubsection{Correction strategy}
A complete and perfect correction would, in principle, require the
capability of moving all elements in all necessary
degrees of freedom and
correcting the shapes of all optical components. The free
positioning would also enable a perfect alignment with the axis of the
adapter. Such a complete correction would require measuring devices to
determine the
shapes and relative positions of all components. For the shapes this could be
individual devices for each component and for the alignment devices for the
relative orientation of two neighbouring components. In practice, a sufficient
set of such devices is not always available. The alternative is to measure the combined
wavefront aberrations generated by the deformations and misalignments of all components.
This can be done and may only be possible by using the light from a star.
The aberrations generated by the individual elements and
the misalignments then have to be deduced from the total wavefront error.
If this is not possible, the correction may be incomplete. On the
other hand, if the errors cannot be attributed to individual elements, a correction
by a subset of the elements may be sufficient, for example the
correction of the deformations in a two mirror telescope with two
monolithic mirrors by deformations of the primary mirror alone.\\
The two extreme types of active telescopes are therefore, on the one hand, those which
require the control and correction of the shapes of individual
components and, on the other hand, those, operating as a system,
where one component can also correct errors introduced by other components.\\
An example of the first kind is a telescope with a segmented primary mirror with
comparatively large individual rigid segments and a monolithic movable secondary
mirror. The errors introduced by the primary mirror, that is the
phasing and the alignment of the segments, are very different from the
errors introduced by elastic deformations or the figuring of the
secondary mirror and can therefore not
compensate each other. As a consequence, the optical surface of both
elements have to be controlled individually.
An example of the latter kind is a telescope with a flexible monolithic
primary mirror and also a movable monolithic secondary mirror. Here,
the nature of the errors is similar and
one element can correct errors introduced by the other one.
The elastic and figuring errors of both mirrors are usually corrected
by the primary mirror, since it is more flexible, often defined as the
pupil of the telescope,
and anyway equipped with a large number of supports.\\
The correction of errors mainly introduced by incorrect positioning of
the elements, that is defocus and third order coma, has, in both types
of active telescopes, to be done by appropriate movements of the
optical elements.\\
For the type and support of flexible monolithic mirrors there are
several options.
On the one hand,
the traditional type is a comparatively
thick mirror with a force based support, which is basically passive
and astatic, with an additional capability of changing the forces
differentially.
Such a system is ideally suited for a pure closed loop operation
with time periods between consecutive corrections of the order of
minutes, and, possibly with a reduced quality, also for a pure open loop
operation.
On the other hand, with active optics {\em position} supports also become
feasible. Since these are fundamentally non-astatic
they require more frequent correction and therefore, if the
times between corrections are smaller than the minimum
integration times for the wavefront sensing, usually a mixture of an
open and a closed loop operation.\\
An important advantage of active telescopes is the freedom to relax
the requirements
for the figuring of all optical elements, since some low spatial frequency
aberrations can be corrected by the active optics system. This gives
the manufacturer the opportunity to concentrate on minimising
the high spatial frequency aberrations. For very
thin mirrors the shape of the mirror is, in a sense, only defined by
the support forces. During the polishing process these cannot be
controlled to the accuracy required for a perfect shape. The mirror
therefore only functions together with
the active optics system and its shape is only defined by that system.
\subsection{Modal control concept and choice of set of modes}
\label{sec:modalConcept}
Most error sources generate wavefront aberrations which can be well described
by certain sets of mathematical functions. Since, in many cases, a small number of
these functions is sufficient to describe a wavefront aberration, a modal concept
for the analysis and the correction of the wavefront errors is essential for an efficient,
practical system.
Which set of functions is used, depends on the dominant error sources and on the
type of telescope. The choice is mainly between purely optical functions like the
Zernike polynomials and vibration modes (Creedon and Lindgren [1970])
based on elastic properties of a flexible element, usually the primary mirror.\\
A general requirement is that the set of functions should be complete with all functions
mutually orthogonal. Although only a very limited number of functions is used in
practice, the completeness guarantees that, in principle, any arbitrary wavefront
aberration can be well approximated. The orthogonality ensures that the values obtained
for the coefficients of certain functions do not depend on other functions used in the
analysis. Another important feature is the thinking in terms of
Fourier modes, which means that
different rotational symmetries are considered separately.\\
The wavefront errors generated by misalignments are defocus, third order coma
and some field dependent functions, all expressible as simple polynomials. The most
commonly used complete set of orthogonal polynomials over the full or annular pupil are
the Zernike or annular Zernike polynomials. 
The errors generated
by deformations of thin monolithic mirrors, on the other hand, are best described
by elastic modes. These are functions with the property that the ratio of the elastic
energy to the r.m.s. of the deflection is minimised. Both the Zernike
polynomials and the elastic modes are also complete and orthogonal
within each individual rotational symmetry.\\
\subsection{Examples of active telescopes}
\begin{figure}[h]
   \centerline{
    \psfig{figure=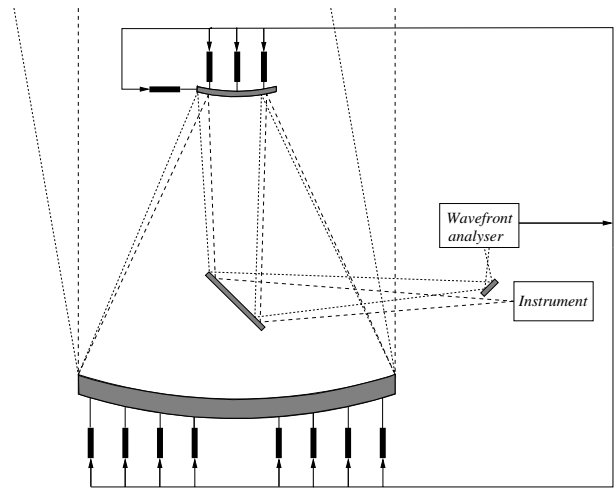,width=80mm}}
     \caption{\label{fig:principleAO} {\small Principle of active optics in
           telescopes with a thin meniscus primary mirror}}
    \end{figure}

Most modern large telescopes with diameters of the primary mirror of more than
two meters rely in some way on active optics.
The prototype of an active telescope with a system approach is the New Technology
Telescope of ESO. It is a Ritchey-Chretien telescope with a meniscus
primary mirror with a diameter of 3.5\,m and a thickness of 241\,mm.
It possesses Serrurier struts and
astatic mechanical levers for the support
of the primary mirror. The active elements are a motorized secondary mirror with
the capability to move in axial direction and to rotate around its center of
curvature, and movable counterweights in the supports of the primary mirror.
This allowed for
a correction of defocus, third order coma and a few of the lowest
order modes of the primary mirror. The principle of active optics as used in
the NTT is shown in Fig. \ref{fig:principleAO}.\\
Since the telescope has still the passive design features and, for
its diameter a fairly conservative thickness, corrections are
only necessary every few minutes. The telescope can therefore be operated in
closed loop. The additional features of its successor, the ESO Very Large Telescope (VLT),
a Ritchey-Chretien telescope with a meniscus primary mirror with a
diameter of 8.2\,m and a thickness of 175\,mm, are a motorised control of
the secondary mirror in six degrees and
also of the primary mirror in five degrees of freedom. Because of its
much lower rigidity due
to the larger diameter of eight meters and the reduced thickness of
175\,mm, corrections are necessary every minute, despite the use of the
usual passive design features.
This correction rate still allows a pure closed loop operation.\\
The 10\,m Keck Telescope is a Ritchey-Chretien design with a primary
mirror consisting of 36 hexagonal segments, each 1.8\,m across with a
thickness of 75\,mm and three position actuators.
The telescope optics including the segments of the primary mirror is
aligned approximately once per month based on data obtained from the
wavefront in the exit pupil generated by a star. Afterwards the shape
of the primary mirror, that is the relative positions of its segments,
is maintained by an internal closed loop based on piston measurements
at intersegment edges, whereas the position of the secondary mirror is
controlled in open loop
(Wizinowich, Mast, Nelson and DiVittorio [1994]).
\newpage
\section[Relationship between AO parameters]{Relationship between
active optics components and parameters}
\label{sec:system}
\begin{figure}[h]
 \centerline{
  \psfig{figure=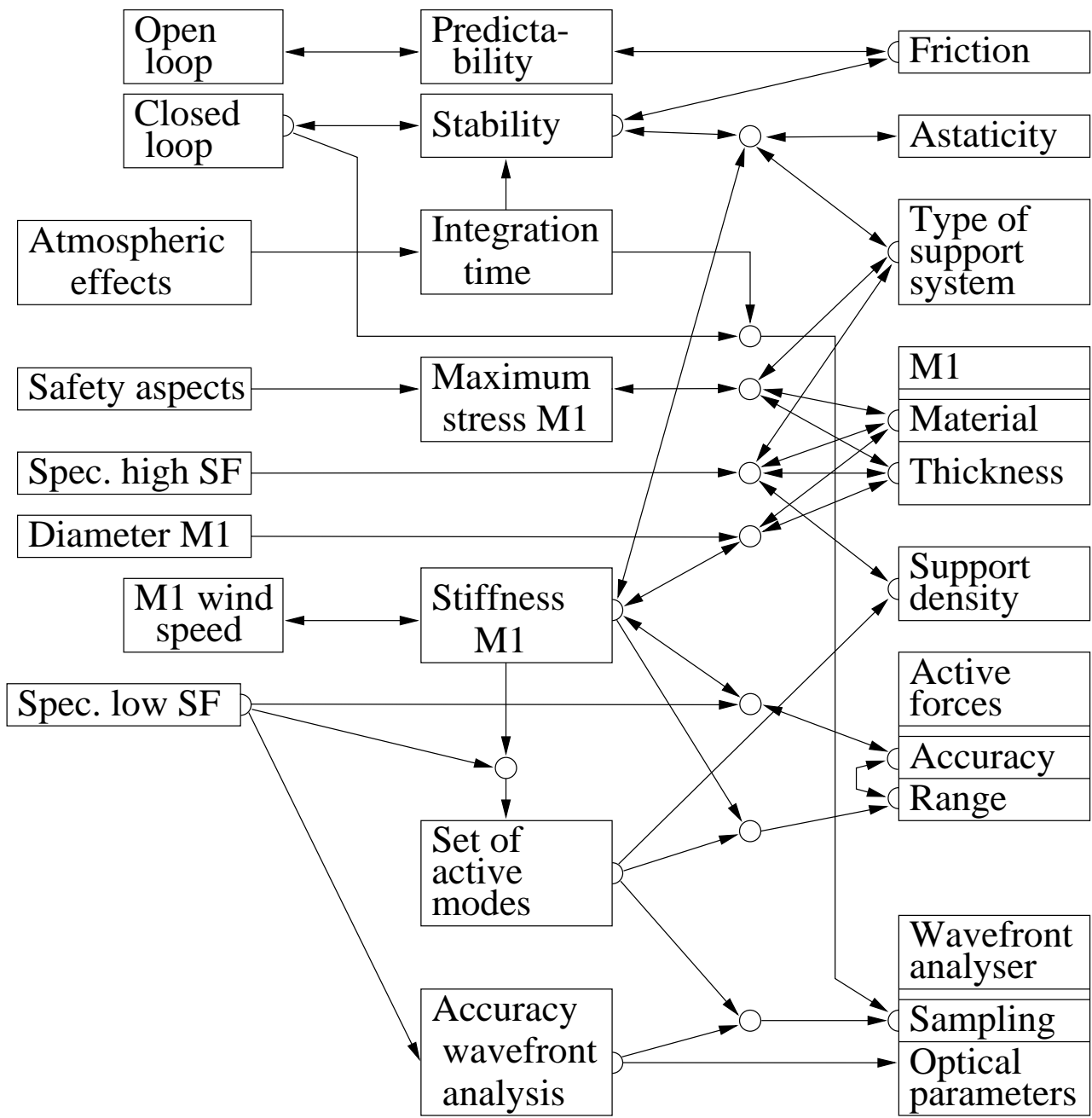,width=170mm}}
  \caption{\label{fig:AODependencies} {\small Dependencies between
  the specifications and the parameters used in an active optics design.}}
\end{figure}
If the active optics corrections are done on a system level, the
active optics system is not a feature added to the telescope system, but rather
an integral part of it, and for many design parameters the capability to do corrections
is even the driver. Fig. \ref{fig:AODependencies} shows for a
telescope with a thin meniscus mirror
the dependencies between various parameters and components of the telescope.\\
The column on the left contains fundamental parameters which are
independent of the particular
design like atmospheric effects, the safety of the mirrors under
exceptional conditions like earthquakes or failures of the support
systems. Also the light gathering power, defined by the diameter of
the primary mirror,
and the optical quality are fixed initial parameters.
The optical quality is, for active telescopes, conveniently defined by two
separate specifications for the high and low spatial frequency
wavefront aberrations (abbreviated 'Spec. low/high SF' in
fig. \ref{fig:AODependencies}).
The parameters in the second column, that is
the decision to operate in either closed or open loop and the wind
speed at the primary mirror, which is determined by the design of the
enclosure, can be either input parameters or the result of the system
analysis. The third column contains intermediate parameters which link
most of the input parameters with the parameters in the fourth column,
which define the properties of the mechanical and optical components
of the active optics system.\\
Arrows from a parameter A to another parameter B mean that B depends
either directly on A, as for example the stiffness of M1 on its
diameter, or that a requirement relating to B depends on A, as for example the
density of supports on the number of active modes, which are the modes
corrected by active optics.
Lines with arrows at both ends indicate that the connected parameters
can influence each other. It is then obvious from fig. \ref{fig:AODependencies}
that limitations on mechanical parameters like the achievable accuracy
of the force setting can have impacts on parameters like the allowable
wind speed at M1 or the decision to operate in open or closed loop.
The dependencies will be explained in detail in the following
chapters. The following example will show how the diagram should be read.
The required accuracy of the force setting under the primary mirror is
defined by the specification for the low spatial frequency aberration
and by the stiffness of the primary mirror, which determines how easily
these lowest modes can be generated. The minimum stiffness itself is
defined by the requirement to reduce the effects of wind pressure
variations to the level given by the specification for the low spatial
frequency errors of the wavefront.
\section{Wavefront sensing}
\label{sec:wavefrontSensing}
\subsection{General considerations}
\begin{figure}[h]
   \centerline{
    \psfig{figure=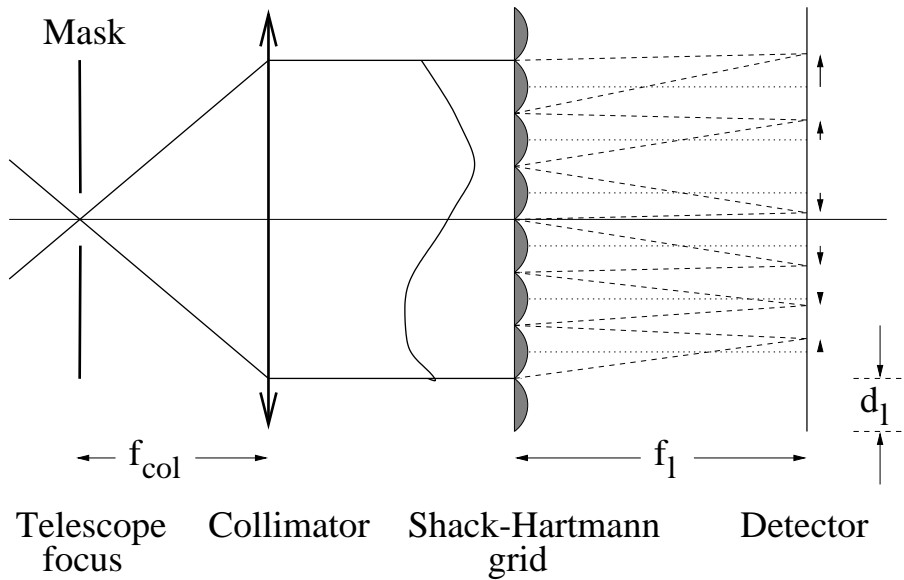,width=120mm}}
  \caption{\label{fig:ShackHartmannOptics} {\small Shack-Hartmann
          optics in a telescope.}} 
  \end{figure}

In particular for telescopes which operate in closed loop, the wavefront analyser
is an essential and critical part of the active optics system. In general, it is
much easier to obtain the
wavefront information from devices exploiting the pupil information than from
measurements of the characteristics of the image. The two most widely used methods
are the Shack-Hartmann method (Platt and Shack [1971]) and the curvature sensing
(Roddier and Roddier [1991])). A Shack-Hartmann device, which is shown
in fig. \ref{fig:ShackHartmannOptics}, measures the local tilts of the
wavefront of a star somewhere in the field.
A mask at the focus of the telescope prevents the light from other nearby
stars entering the sensor.
The telescope pupil is imaged on to an array of small
lenslets, each producing in its focal plane a spot on a detector. The
shift of the spot generated with light from a star compared with the
position of the spot generated with a point reference source placed
in the focus of the telescope is proportional to the average local
tilt of the wavefront over the subaperture sampled by a single
lenslet.
The curvature sensing method measures the intensity variations,
that is the Laplacian of the wavefront, and the shape of the edges, that is the first
derivatives of the wavefront, in defocussed intrafocal and extrafocal images.
Both methods work, in the end, with similar accuracy.\\
The wavefront sensor has to be adapted to the type of the telescope and the type of operation
of the active optics system, in particular the correction strategy. One important
criterion is that the measured coefficients of the modes are not dependent on the
particular number of modes. This requires that the modes fitted to the measured data
be orthogonal over the area of the pupil. The independence of the results for individual
modes gives, for example, the freedom to correct, depending on the results, only a
certain subset without the need to do another analysis with only the modes contained
in this subset. Another criterion is the question whether the r.m.s. of the wavefront
error or the slopes of the wavefront error should be minimised. The first choice would be
the optimum for a system aiming for diffraction limited performance,
the second for a system aiming for seeing limited performance.
For a system working with Zernike polynomials the first
choice requires a conversion of tilt data from the Shack-Hartmann device into wavefront
data and a subsequent fit of the orthogonal Zernike or, for annular pupils,
annular Zernike polynomials,
whereas the second choice requires a direct fit of Zernike type polynomials, whose
derivatives are orthogonal over the pupil, to the tilt data (Braat [1987]).
A system working with elastic modes of the primary mirror has to fit functions
to wavefront data, since the elastic modes, but not their derivatives, are orthogonal
over the area of the mirror.\\
Furthermore the wavefront sensor has to fulfill a number of requirements imposed by
the environment and the specification for the required accuracy, given usually in
terms of tolerable low spatial frequency wavefront errors. In the rest
of this section we will concentrate on the Shack-Hartmann method.
\subsection{Calculation of the wavefront coefficients}
\label{sec:algorithms}
The calculation of the coefficients is done in five steps.\\
{\it 1. Computation of local tilt values and indexing of the spots.}\\
The centroids of the Shack-Hartmann patterns obtained with the reference and
the star light are computed. A problem may be to find the reference spot
corresponding to a certain star spot. One possibility would be to mark certain
lenslets by reducing their transmission, another one to use the
irregularities of the lenslet array to find the relative shift between the two
patterns for which certain combinations of local distances give the best correlation.
The second method works well for well corrected systems and grids with
sufficient distortions.
In practice, with highly regular grids nowadays available, the errors introduced by
making a wrong correspondence are irrelevant for a first correction of strongly
aberrated wavefronts. After this initial correction the pattern is so regular, that
a well designed telescope with good pointing and tracking will almost always place the
star spots close to the corresponding reference spots.\\
{\it 2. Computation of the center of the pupil.}\\
The center of the pupil can be calculated as the simple weighted average of
the positions of the 
reference spots of all double spots. Other more complicated algorithms
may give a higher accuracy. The
major goal, apart from finding the proper center of the patterns is to
disregard distorted spots at the edges belonging to
subapertures which are not fully inside the pupil.\\
{\it 3. Interpolation of tilt data to regular positions in the pupil.}\\
In general, the Shack-Hartmann spot pattern is neither symmetric nor fixed
with respect to the pupil. Each fit of a set of modes to the data
involves the computation of the values of all modes at the relative
locations of all spots in the pupil. The alternative is to interpolate
the data to a fixed regular grid and calculate in advance the values of the
modes only once for the regular grid positions in the pupil. The
interpolation is done by
fitting a two dimensional polynomial to the data of the surrounding spots.
For a 23 by 23 pattern the optimum is the use of a second order polynomial
taking into account all spots within a distance from the regular spot position 
of 20\% of the radius of the full pattern.\\
{\it 4. Conversion of tilt data into wavefront data.}\\
The conversion of a shift of a centroid $\eta_{{\rm ccd}}$
into a slope of the wavefront is given by
  \begin{equation}
     \frac{{\rm d} w}{{\rm d} \rho} =
        \frac{1}{2N_{{\rm tel}}\; N_{{\rm l}}} 
               \frac{f_{{\rm col}}}{d_{{\rm l}}}
                        \eta_{{\rm ccd}}
   \label{eq:focalPlaneToWfNorm}
  \end{equation}
Here $w$ is the wavefront error, $\rho$ the normalised radius of the
entrance pupil, $f_{{\rm col}}$ the
focal length of the collimator, $d_{{\rm l}}$ the diameter,
$f_{{\rm l}}$ the focal length and
$N_{{\rm l}} = f_{{\rm l}}/d_{{\rm l}}$ the f-number of a lenslet,
and $N_{{\rm tel}}$ the f-number of the telescope.\\
The tilt data are then integrated to wavefront data. This is done by integrating,
for a square grid, along the $n_{{\rm rc}}$ rows and and $n_{{\rm rc}}$ columns,
stopping, if necessary, at the edge of the central hole with its
$n_{{\rm hole}}$ missing rows or columns, and starting a
new integration at the other side.
If the vector field was curl-free, for all spots the two values
obtained with the integration along the corresponding column
or the integration along the corresponding row, would, with a proper
choice of the integration constants, be identical. But with the noise added by
the measurement, this is not the case. Since the approximate number
of $0.75 n_{{\rm rc}}^{2}$ intersections is, for all practical grids, much larger
than the number of $2(n_{{\rm rc}}+n_{{\rm hole}})$ integration constants,
the optimum choice of the integration
constants can be obtained with a least squares fit.\\
{\it 5. Fit of chosen functions to the wavefront data.}\\
The next step is a straightforward least squares fit of the chosen
set of functions to the wavefront data on the regular grid. With a
fixed set of functions the fit is a multiplication with a
precalculated matrix. This yields the coefficients of the fitted modes
and, in addition, the residual
r.m.s. $\sigma_{{\rm resid}}$ of the wavefront aberration after
subtraction of the fitted modes.\\
{\it 6. Subtraction of field aberrations}.\\
Since the wavefront analysis is usually done in the field of the
telescope, but the active optics corrections require the coefficients
at the center of the field, the contributions from the field aberrations have
to be subtracted. In aligned systems these are rotationally symmetric,
but in misaligned systems the patterns are more complicated as
described in sec. \ref{sec:alignment}. An accurate subtraction of the
field effects therefore requires information on the actual
misalignment of the telescope.
\subsection{Definition of Shack-Hartmann parameters}
\label{sec:defSHParams}
The focal length $f_{{\rm col}}$ of the collimator is chosen such that
the image of the pupil on the Shack-Hartmann grid and therefore also
the spot pattern fits, with some margin, on the detector. This leaves then only
two adjustable parameters, namely the number of lenslets sampling the
pupil and the f-number of the lenslets.
\begin{itemize}
\item {\it Sampling of the wavefront.}\\
The sampling is determined by two requirements (see
fig. \ref{fig:AODependencies}).
First, it should
be sufficient to guarantee an accurate measurements of the coefficients of
all fitted modes.
For this, the major error sources are an inaccurate determination of the center
of the Shack-Hartmann pattern, the averaging of the tilts over
subapertures and the aliasing generated by the finite sampling. The
error due to the first source is of the order of 2.5\% for a 10 by 10
sampling, with the error being approximately inversely proportional to
the sampling $n_{{\rm rc}}$ in one direction. If the wavefront errors are
expanded in Zernike polynomials, the latter two sources lead only to crosstalk
into the next lower term in the same rotational
symmetry. This crosstalk is of the order of $\delta^{2}$, where
$\delta$ is the ratio of 
the diameter of the subaperture to the diameter of the pupil.\\
Second, to guarantee for a closed loop operation a full sky coverage with
field sizes of the order of 100 arcmin$^{2}$ available in most
telescopes,
the sampling should be sufficiently coarse, that is the
corresponding subapertures in the pupil should be large enough to gather,
with the chosen integration time, enough light from stars of magnitude 13.
Measurements with two wavefront analysers in different positions in the field have shown
that only with integration times of 30 seconds or more the differences
due to effects of the free atmosphere at high altitude are effectively
integrated out.
Measurements with these
integration times are therefore effectively isoplanatic and 30 seconds
is the minimum time
between active optics corrections in a closed loop operation.
With 30 seconds integration time sufficient maximum pixel values are,
at least for seeing values up to 1.5\,arcsec, guaranteed with
subapertures with diameters of approximately 400\,mm.
\item {\it f-number of the lenslets.}
This parameter is determined by the requirement that a wavefront
analysis can be done with high accuracy under all relevant external
conditions. 
The major external parameter is the atmospheric seeing. The image analysis should
function both under excellent seeing conditions with an expected minimum value
$\Theta \approx 0.2$\,arcsec and bad seeing conditions with seeing
values up to at least
$\Theta \approx 1.5$\,arcsec. Above these values the tolerable errors,
which would still guarantee a seeing limited performance, are so large
that a seeing limited performance can also be achieved with open loop
operations (see \S \ref{sec:openLoopPerformance}).
This leads to three conditions for the f-number of the 
Shack-Hartmann lenslets (Noethe [2000]).\\
{\it 1. Excellent seeing : Minimum spot size larger than 1.5 times the pixel size.}\\
For an accurate centroiding
the spot diameter has to be at least 1.5 times as large as the pixel
size $d_{{\rm p}}$. The minimum
spot size is generated by the reference light or possibly by star
light under optimum seeing conditions and is given by the diameter of
the Airy disk of the lenslets. This leads to the following condition
for the f-number $N_{{\rm l}}$ of the lenslets.
\begin{equation}
  N_{{\rm l}} \geq 1.25 \frac{d_{{\rm p}}}{\mu {\rm m}}
  \label{eq:condMinKF}
\end{equation}
{\it 2. Bad seeing : Avoidance of swamping.}\\ 
In bad seeing conditions swamping of the spots should be
avoided, i.e. the diameter of the spots due to
atmospheric effects should be smaller
than the diameter of the lenslets. For an assumed worst seeing of
$\Theta \approx 1.5$\,arcsec
and a diameter of the spot less than 0.7 times the lenslet diameter one gets
\begin{equation}
      N_{{\rm l}} \leq
            50000 \frac{n_{{\rm rc}}d_{{\rm l}}}{d_{{\rm M}}}
  \label{eq:conditionkFnumber}
\end{equation}
where $d_{{\rm ccd}}$ is the diameter of the CCD and $d_{{\rm M}}$ the outer
diameter of the primary mirror.\\
{\it 3. All seeing conditions : Maximisation of sensitivity to transverse aberration.}\\
The measuring accuracy of the Shack-Hartmann sensor is mainly limited by the
centroiding errors. The generated wavefront error
is proportional to the centroiding error with an r.m.s. $\sigma_{{\rm cen}}$, the f-number
$N_{{\rm l}}$ of the lenslets and, roughly, to the square root of the number of modes
used in the analysis. This leads to the condition for $N_{{\rm l}}$
 \begin{equation}
   N_{{\rm l}} \ge 0.2
                    \frac{\sigma_{{\rm cen}}}{\sigma_{{\rm wf},{\rm max}}}
                    \sqrt{n_{{\rm modes}}}
                          \label{eq:condKflNoise}
 \end{equation}
where $\sigma_{{\rm wf},{\rm max}}$ is the r.m.s. of the maximum
tolerable wavefront error
allocated to the wavefront analysis. Even with comparatively simple centroiding methods
the centroiding error is of the order of only 5\% of the pixel
size. If the maximum pixel value is constant, the centroiding error
does not depend on the spot size, but is only a function of the pixel
size. 
\end{itemize}
\subsection[Wavefront sensing in segmented mirror
telescopes]{Wavefront analysers for segmented mirror telescopes}
\label{sec:wfAnalysisSegments}
For telescopes with segmented mirrors, the wavefront analyser
should be capable of detecting the deformations of individual
segments, errors introduced by misalignments between mirrors, and
relative tilt and piston errors of individual segments.
These functions, most of them based
on the Shack-Hartmann principle, have been realised in the Phasing
Camera System (PCS) of the Keck telescope,
which can operate in four modes (Chanan, Nelson, Mast, Wizinowich and
Schaefer [1994]). The so-called {\em passive tilt mode}, where the
light from each segment is collected into one spot per segment,
can measure the tilt errors of the segments.\
The {\em fine screen mode}, where each of
the 36 segments is sampled in 13 places,
can measure the segment tilts,
but also the defocus and decentering coma aberrations of the telescope
optics, generated by a despace of the secondary mirror.
Global defocus and coma introduce, over each
subaperture corresponding to one segment, local defocus and
astigmatism, respectively. The axial error in the position of the
secondary mirror can then be calculated and corrected from the average
defocus, and the tilt or decenter from the distribution of
astigmatism over the subapertures. Both of these modes do not use
common Shack-Hartmann lenslets, but rather a combination of prisms and
a convex lens in the case of the passive tilt mode
(Chanan, Mast and Nelson [1988]), and a combination
of a mask, a defocusing lens and an objective with a focal length
about five times smaller than the one of the defocusing lens in the
case of the fine screen mode.
The {\em ultra fine screen mode}
samples just one segment with 217 closed packed hexagonal
Shack-Hartmann lenslets.\\
Finally, the {\em segment phase mode} (Chanan, Troy and Ohara [2000])
deduces the relative heights of adjacent segments from the
characteristic of either in-focus images or the difference between
intrafocal and extrafocal images, both with star light from apertures
with diameters of the order of 100 mm
centered at segment edges. The in-focus method uses two algorithms.
The narrowband algorithm is based on the diffraction pattern obtained
with quasi-monochromatic light. The pattern is a periodic function of
the relative displacement of the adjacent segments. The capture range,
which is the maximum difference between the heights for which the
algorithm can be applied, is of the order of 15\% of the wavelength
$\lambda$ of the light.
For $\lambda \approx 800\,$nm the accuracy is of the order
of 6\,nm. The broadband algorithm takes the effects of the finite
bandwidth into account. Both the capture range and the accuracy are
roughly inversely proportional to the bandwidth of the light. For 
$\lambda \approx 800\,$nm and a bandwidth of 200\,nm, the capture range
is 1\,$\mu$m and the accuracy 30\,nm.
Since both algorithms exploit interference effects, the coherence of
the light over the subaperture should not significantly be degraded by
atmospheric effects. This is guaranteed if the
diameter of the subaperture is smaller than atmospheric coherence
length $r_{0}$ for the wavelength used for the measurement.
Under this condition, the results of the relative height measurements are
largely independent of the current seeing.
The method using the differences between the intrafocal and extrafocal
images works at wavelengths of 3310\,nm with a bandwidth of 63\,nm.
The capture range is 400\,nm and the accuracy 40\,nm.\\
Piston errors start to limit the image quality if the atmospheric
coherence length $r_{0}$ for the observed wavelength $\lambda$ approaches the
dimensions of the individual segments (Chanan, Troy, Dekens, Michaels,
Nelson, Mast and Kirkman [1998]).
Since $r_{0}$ scales with the
wavelength as $\lambda^{6/5}$, phasing becomes increasingly important for
observations at longer wavelengths. For segments with diameters of
1.8\,m, as in the Keck telescope, phasing is effectively irrelevant for
observations with visible light, but at a wavelength of 5\,$\mu$m and an
r.m.s. piston error of 500\,nm, the central intensity is reduced by
approximately 60\%. At the Keck telescope the phasing
tolerances are set to $\le 100$\,nm for normal observing. However, for
observations also using adaptive optics to correct the atmospheric
disturbances or for telescopes in space, the tolerances should be much
tighter.
\section{Minimum elastic energy modes}
\label{sec:minEnergyModes}
The minimum-energy modes can be defined in the following way (Noethe [1991]).
Each rotational symmetry $m$ will be considered separately.
Let ${\cal F}_{m,0}$ be the set of all functions of rotational
symmetry $m$ defined over the area of the mirror.
The lowest mode $e_{m,1}$ is the one taken from the set ${\cal F}_{m,0}$
which minimises the ratio $\Gamma$ of the total elastic energy
${\cal J}$ of the mode to the r.m.s. ${\cal A}$ of its
deflection perpendicular to the surface.
Let ${\cal F}_{m,1}$ be the set of all functions of ${\cal F}_{m,0}$
which are orthogonal to $e_{m,1}$.
The second mode $e_{m,2}$ is the one taken from ${\cal F}_{m,1}$
which mimimizes the ratio $\Gamma$.
For an arbitrary $i$ let ${\cal F}_{m,i-1}$ be the set of all functions
of ${\cal F}_{m,0}$
which are orthogonal to all functions $e_{m,1},...,e_{m,i-1}$.
Then, the i-th mode $e_{m,i}$ is the one taken from ${\cal F}_{m,i-1}$
which mimimizes the ratio $\Gamma$.
The actual construction of the
minimum-energy modes requires the solution of the variational equation
 \begin{equation}
  \delta({\cal J} - \tilde{\xi} {\cal A}) = 0,
       \label{eq:var}
 \end{equation}
where $\tilde{\xi}$ is a free parameter which
can be interpreted as the energy per unit of the r.m.s. of the deflection. The use
of variational principles leads, together with the assumptions of a thin shallow
spherical shell, to a fourth order differential equation, which can be transformed
into two second order differential equations for each rotational
symmetry. Since the fixed points only define the position of the
mirror in space and have no impact on its shape,
the appropriate boundary conditions are
the ones for free inner and outer edges.
The solutions of
the differential equations form, within each rotational symmetry $m$, a
complete set $\{e\}$ of orthogonal functions, the elastic modes $e_{m,i}$.
The order of
a mode within each rotational symmetry is denoted by the index $i$.
If the eigenvalues $\xi_{m,i}$ are expressed as
 \begin{equation}
  \xi_{m,i} = \frac{1}{2} h \gamma \omega_{m,i}^{2},  \label{eq:dlam}
 \end{equation}
where $\gamma$ is the mass density of the mirror, $h$ its thickness,
and $\omega_{m,i}$ is interpreted as the circular frequency of a
vibration mode with the order $i$ within the rotational symmetry $m$,
the differential equations are identical to the 
equations describing vibrations of a thin shallow shell under
the assumption that in-plane inertial effects are neglected.
The eigenvalues $\xi_{m,i}$, which can be shown to be proportional
to the elastic energies of the modes, are therefore proportional to the square
of the eigenfrequencies of the corresponding vibration modes. For
geometrically similar mirrors of the same material, the
eigenfrequencies scale with $h/d_{\rm M}^{2}$.\\
\begin{figure}[h]
 \centerline{\hbox{
  \psfig{figure=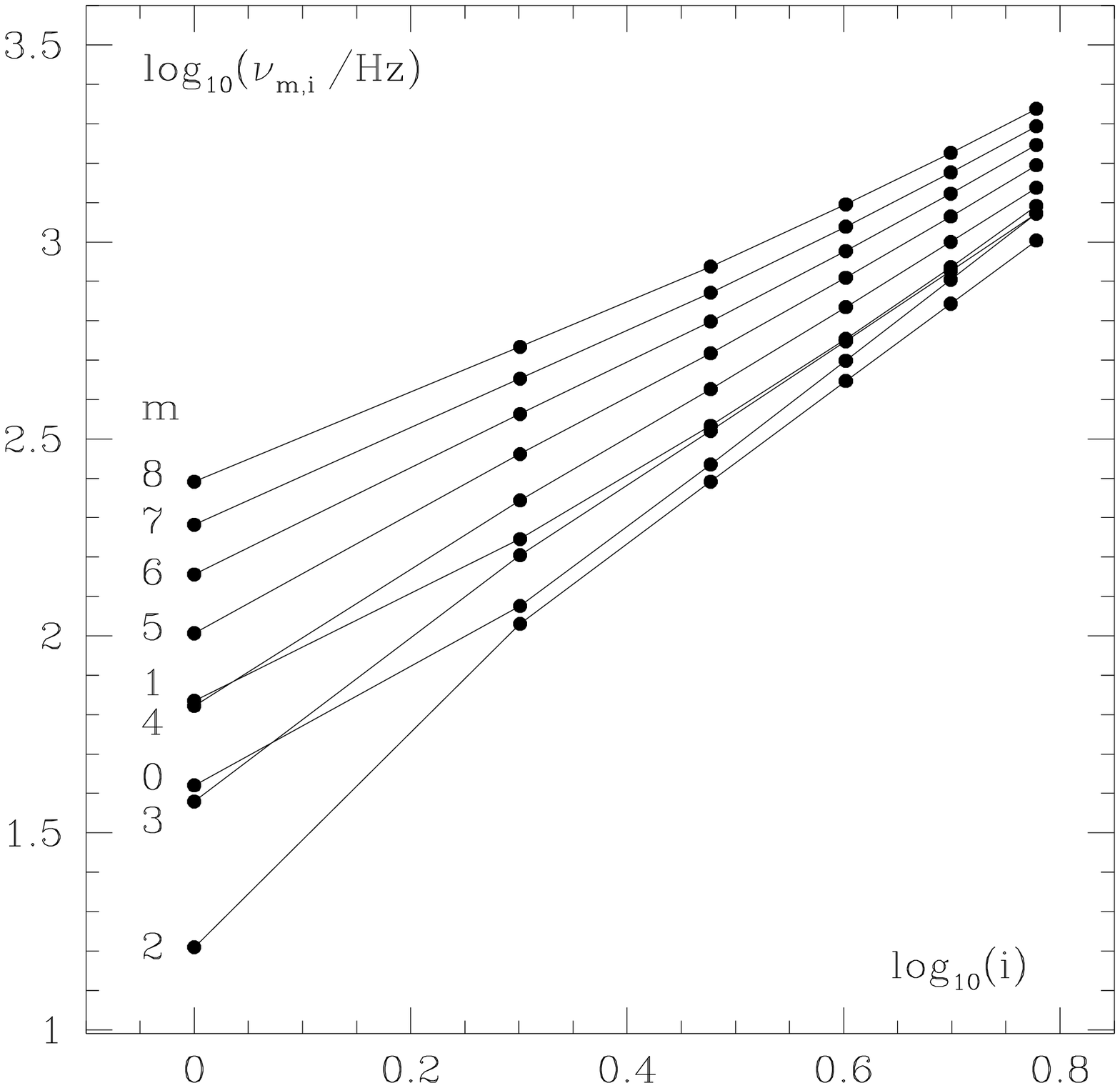,width=75mm}
  \psfig{figure=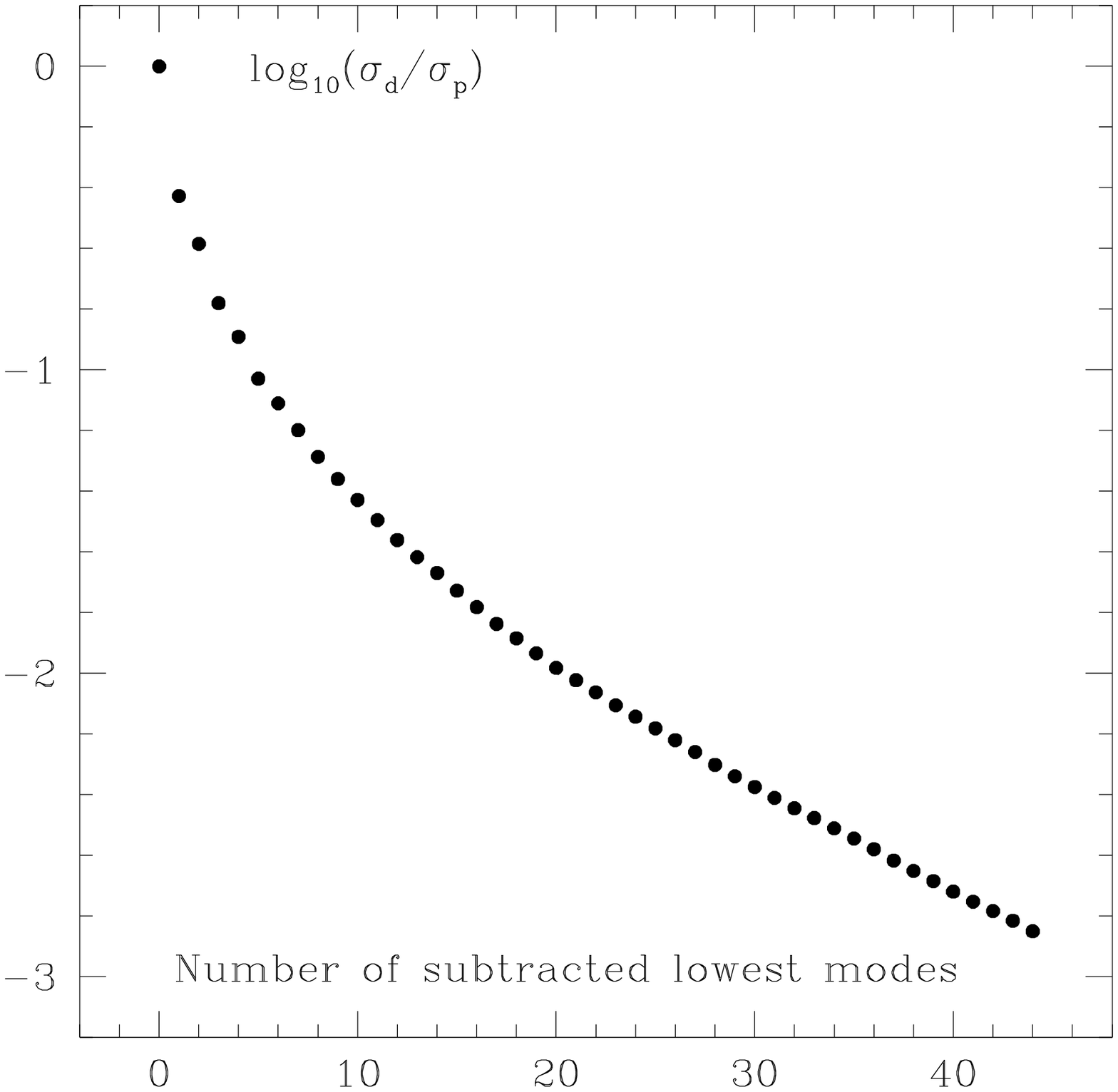,width=75mm}}}
  \caption{\label{fig:modeFrequencies} {\small {\em Left:} Eigenfrequencies of the
  elastic modes of the VLT primary mirror for the lowest nine
  rotational symmetries $m$ and lowest six orders $i$ within each
  rotational symmetry. {\em Right:} Logarithm of the fraction
  $\sigma_{{\rm d}}/\sigma_{{\rm p}}$ of the deflection
  generated by a random pressure field left after subtraction a
  certain number of lowest modes.}}
\end{figure}
Fig. \ref{fig:modeFrequencies} shows the eigenfrequencies of
the elastic modes of the VLT primary mirror with a diameter of 8.2\,m,
a thickness of 175\,mm and a radius of curvature of 28.8\,m as a
function of their order in a log-log plot.
Two features of elastic modes are very useful in the context of active optics.
First, the eigenfrequencies increase rapidly both with the symmetries $m$
and with the orders $i$. Within each rotational symmetry $m$ the increase in
the log-log plot is approximately linear, with the symmetry 2 having the
largest slope of approximately 2, that is the eigenfrequencies are
roughly proportional to $i^{2}$.
The lowest modes of the symmetries zero to three show, for the lowest
order, deviations from
the linear behaviour. For the rotational symmetries zero and one the
relative increase is due to membrane stresses induced by the thin
shell.
In a log-log plot of the eigenfrequencies against the rotational
symmetry $m$ the increases are, for $m \ge 2$, also linear, with a
largest slope of approximately 2 for the lowest order one. In this
order the eigenfrequencies are therefore proportional to $m^{2}$.
It is obvious from the plot in fig. \ref{fig:modeFrequencies} on the
left, the lowest elastic mode $e_{2,1}$ of rotational symmetry two 
is the by far softest and
therefore most easily excitable deformation.
Its control is therefore together with defocus and decentering coma,
which are generated by misalignments, the most important and demanding
task of active optics. 
Because of the fast increase of the stiffness of the modes with the
order and the rotational symmetry, any given
set of forces or any given pressure field
will generate significant deflections only in the lowest modes.
If $\sigma_{{\rm p}}$ is the r.m.s. of the deflections generated by random
white noise pressure fields,
figure \ref{fig:modeFrequencies} shows on the right the ratio
$\sigma_{{\rm d}}/\sigma_{{\rm p}}$, where $\sigma_{{\rm d}}$ is the
r.m.s. of the residual deflection after the subtraction of a given
number of elastic modes with the lowest eigenfrequencies.
A subtraction of the softest mode $e_{2,1}$
alone reduces the r.m.s. of the deflection to 40\%, and a subtraction
of the softest five modes to 10\%.\\
Second, a pressure
field, which is proportional to an elastic mode, will, since the mode
is an eigenfunction
of the underlying differential equation, generate a deflection with exactly the same
functional dependence. The coefficient of the deflection is then inversely
proportional to the eigenvalue, that is the elastic energy, of this mode. This
feature can be exploited to calculate the deflections generated by arbitrary pressure
fields or sets of forces. The pressure fields are directly expanded in terms of the
elastic modes, whereas the forces are described as delta functions and
then expanded. The
total deflection is obtained by summing up the deflections in the
individual modes, which are obtained by multiplying the expansion
coefficients of the pressure field by factors
inversely proportional to the elastic energies of the modes.\\
\begin{figure}[h]
 \centerline{
  \psfig{figure=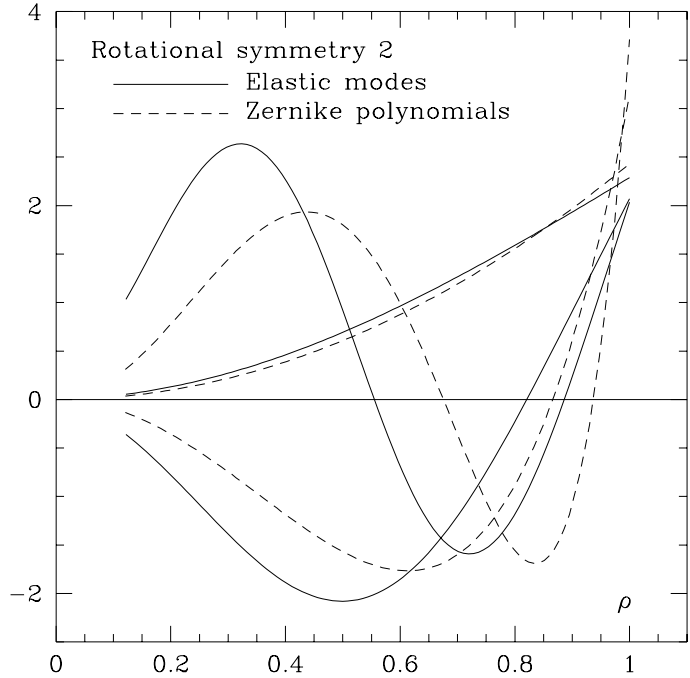,width=100mm}}
  \caption{\label{fig:zerElaSym2} {\small Lowest three eigenmodes of
  rotational symmetry two (dashed lines) and their corresponding
  annular Zernike polynomials (solid lines) as functions of the
  normalised radius $\rho$.}} 
\end{figure}
Zernike polynomials $z_{m,i}$ and elastic modes $e_{m,i}$ are very
similar in the respect that, in each
rotational symmetry $m$, the number of nodes of the radial function is defined by the
order $i$ of the mode. For rotational symmetries larger than one the
Zernike polynomials $z_{m,i}$ correspond to the elastic modes
$e_{m,i}$, but for rotational symmetries zero and one, where the
lowest Zernike polynomials piston and tilt represent full body
motions, the elastic modes $e_{m,i}$ correspond to the Zernike
polynomials $z_{m,i+1}$.
The major difference between the two sets of functions is that the
elastic modes are effectively
linear near the outer edge but show stronger variations near the inner
edge than the Zernike polynomials.
The consequence is that particularly higher order elastic modes
cannot be well approximated
by a small number of annular Zernike polynomials. Figure
\ref{fig:zerElaSym2} shows the first three annular
Zernike polynomials and elastic modes of rotational symmetry two. The
residual errors of fitting
the elastic mode of order $i$ with $i$ annular Zernike polynomials are, in fractions
of the r.m.s. of the elastic modes, 0.05 for $i=1$, 0.35 for $i=2$ and 0.62 for
$i=3$. To push the residual fraction below 0.05 for the modes $i=2$
and $i=3$ one needs to fit four and six annular Zernike polynomials,
respectively.
Nevertheless, at least the lowest elastic modes are, in vector
notation, effectively
parallel to their corresponding Zernike polynomials. Examples of such pairs are Zernike
defocus $z_{0,2}$ and the first elastic mode $e_{0,1}$ of rotational
symmetry zero,
Zernike third order coma $z_{1,2}$ and the first elastic mode
$e_{1,1}$ of rotational symmetry one, and Zernike third order
astigmatism $z_{2,1}$ and the first elastic mode $e_{2,1}$ of
rotational symmetry two. Clearly, members
of such pairs should not be fitted simultaneously to a wavefront.\\
The relative difference between the lowest mode $e_{2,1}$ of rotational symmetry two
and the equivalent Zernike third order astigmatism
($\sqrt{6}r^{2}\cos{2\varphi}$) is only of the order of 5\%. But the
forces to generate third order astigmatism
with an accuracy similar to the one achievable for the corresponding
elastic mode $e_{2,1}$ are significantly larger.
The elastic mode $e_{2,1}$ of the primary mirror of the VLT can
be generated, excluding print-through effects, with an accuracy of
0.00003 with maximum forces of $F_{{\rm max}} = 1.68$\,N for a coefficient of 1000\,nm.
The accuracy, with which Zernike
astigmatism can be generated, and the required forces depend on the number of elastic modes
used for the approximation. With two modes the accuracy is 0.012 with
$F_{{\rm max}} = 4.2$\,N and with six modes 0.0024 with $F_{{\rm max}} = 13.6$\,N.
This shows again the advantage working with elastic modes rather than
Zernike polynomials in the active optics corrections of elastically induced errors.\\
Since misalignment errors generate to first order nearly pure field
independent Zernike defocus and third
order coma, the Zernike polynomials $z_{0,2}$ and $z_{1,2}$ should be
included and, consequently, the corresponding elastic modes $e_{0,1}$
and $e_{1,1}$ excluded from the set of fitted modes.
The two Zernike modes will not exactly be orthogonal to the higher elastic modes within
their rotational symmetry, but this is in practice not a significant effect. The sets
of functions used in active optics then contain Zernike defocus and third order coma
and, if monolithic mirrors are used, some of the elastic modes with the lowest energies.
The number of elastic modes which will be considered depends mainly on the
forces which are required to correct these modes. The fact that these forces increase
much faster with the spatial frequencies of the modes than the coefficients of
these modes generated by noise effects in the wavefront analyser, puts a natural
limit on the number of modes which can be corrected. All modes which
are actually corrected during the active optics process will from now
on be called {\em active modes}.\\
The number of active modes can be defined in the following way.
One can assume that the forces applied by passive actuators are
accurate to approximately 5\% of the nominal load over the range of
usable zenith angles.
Further, the r.m.s. of the wavefront error introduced by any mode
should be well below the diffraction limit.
Therefore, one should correct all modes which are generated with
coefficients of more than, say, 15\,nm by random 
force errors evenly distributed in a range of $\pm 5$\% of the nominal
load of each support.
These coefficients, which will be inversely proportional to the square of the
eigenfrequencies of the corresponding modes, can be calculated by the
method mentioned above in this section from several runs with
independent sets of random forces.
\section{Support of large mirrors}
\label{sec:support}
\subsection{System dependencies}
The properties of the supports of large monolithic mirrors, in
particular of the primary 
mirrors, of large active telescopes are related to the basic
requirements in a complex way. In fig. \ref{fig:AODependencies} the mechanical
parameters, for which requirements are to be deduced from the input parameters,
are shown in the upper six boxes in the right column.
Friction is the only limitation for the predictability of the system. Clearly,
it also has an influence on the stability. The latter depends on the general type
of the support system, the astaticities of its components and the stiffness of the
primary mirror. The major safety requirement is the need to keep the stress
levels, in particular at the support points, well below the critical values.
These depend on the material of the mirror, and the values of the
generated stress
primarily on the thickness of the mirror and the type of the support system, in particular
the nature of the fixed points. The generated high spatial frequency aberrations,
the so-called print-through, depend clearly on the specific weight and the
elasticity module of the mirror material, on the thickness of M1 and the density of
supports. But it can also be influenced by the general type of the support system,
for example if part of the weight of the mirror is supported by a
continuous pressure field
at the back surface as realised in the support of the primary mirrors of the
Gemini telescopes (Stepp and Huang [1994]). The stiffness of M1, a central
parameter for the active optics design, is a function of the diameter
of M1, its thickness and its elasticity module. Another intermediate
parameter, the number of active modes, depends, as described at the
end of \S \ref{sec:minEnergyModes}, on the stiffness of M1 and the
tolerable low spatial frequency errors. The required accuracy of the force
setting depends on the stiffness of M1, that is predominantly on the stiffness of
the softest elastic mode $e_{2,1}$, and on the tolerated low spatial frequency
aberrations, again dominated by the mode $e_{2,1}$. The range of active forces
depends on the stiffnesses of the active modes, and since the accuracy of a load cell
is usually inversely proportional to its range, the active range is also directly
related to the accuracy of the force settings.
\subsection{Scaling laws for thin monolithic mirrors}
\label{sec:scalingLaws}
For a comparison of menisci with different diameters $d_{{\rm M}}$ and
thicknesses $h$
and of their support systems with $n_{{\rm s}}$ individual supports,
the following scaling laws can be used. They are
given for the wavefront error $w$ and the corresponding slope errors $t$
generated by deformations of the mirror.
\begin{itemize}
\item {\it Pressure field applied to the meniscus}\\
  If the pressure fields as functions of the normalized radii of the
  menisci are identical, the scaling law is given by
   \begin{equation}
     w \propto \frac{d_{{\rm M}}^{4}}{h^{3}}, \hspace{5mm}
     t \propto \frac{d_{{\rm M}}^{3}}{h^{3}}
       \label{eq:scalingPressure}
   \end{equation}
\item {\it Sag under the own weight}\\
  The sag between support points under the own weight of the meniscus
  obeys the scaling law
   \begin{equation}
     w \propto \frac{d_{{\rm M}}^{4}}{h^{2}n_{{\rm s}}^{2}},
  \hspace{5mm}
     t \propto \frac{d_{{\rm M}}^{3}}{h^{2}n_{{\rm s}}^{2}}
        \label{eq:scalingSag}
   \end{equation}
  These scaling laws can readily be derived from the ones for the
  pressure fields by noting that the forces applied by the own weight
  are proportional to the thickness $h$ of the meniscus. The factor
  $1/n_{{\rm s}}^{2}$ ensures that for a constant thickness the sag
  stays the same if the 
  number of the supports per area, which is proportional to
  $d_{{\rm M}}^{2}/n_{{\rm s}}$, remains constant.
\item {\it Single discrete force}\\
  For a single discrete differential force $\Delta F$ applied to the mirror the
  scaling law is given by
   \begin{equation}
     w \propto \frac{d_{{\rm M}}^{2}}{h^{3}} \Delta F, \hspace{5mm}
     t \propto \frac{d_{{\rm M}}}{h^{3}} \Delta F
        \label{eq:scalingSingleForce}
   \end{equation}
\item {\it Set of supports applying random force errors proportional to
the nominal loads}\\
If a force error is proportional to the nominal load of the support,
one has $\Delta F \propto d_{{\rm M}}^{2}h/n_{{\rm s}}$. Furthermore, if
$n_{{\rm s}}$ supports are applying forces with random errors
the expression for the effect of a single force has to be multiplied
by $\sqrt{n_{{\rm s}}}$. Together one then obtains from
eq. (\ref{eq:scalingSingleForce})
\begin{equation}
     w \propto \frac{d_{{\rm M}}^{4}}{h^{2}}\frac{1}{\sqrt{n_{{\rm s}}}},
 \hspace{5mm} 
     t \propto \frac{d_{{\rm M}}^{3}}{h^{2}}\frac{1}{\sqrt{n_{{\rm s}}}}
        \label{eq:scalingRandomForces}
\end{equation}
\end{itemize}
The design of the support system has to meet both the specifications
for the high and the low spatial frequency aberrations. The former are
dominated by the sag between the support points described by the
scaling law (\ref{eq:scalingSag}) and the latter by the
deformation in the shape of the mode $e_{2,1}$ generated by random
support forces. Since the achievable accuracy of the force setting
will be proportional to the total force range, one can apply the
scaling law (\ref{eq:scalingRandomForces}), if the nominal load is
understood as the force range. If, for different mirrors and their
supports, the high spatial frequency errors are taken to be identical, the
number of supports scales with $n_{{\rm s}} \propto d^{2}/h$ for
wavefront errors and $n_{{\rm s}} \propto d^{1.5}/h$ for slope
errors. The low spatial frequency errors then scale with
\begin{equation}
     w \propto \frac{d_{{\rm M}}^{3}}{h^{1.5}},
 \hspace{5mm} 
     t \propto \frac{d_{{\rm M}}^{2.25}}{h^{1.5}}
        \label{eq:scalingHSFconstant}
\end{equation}
This scaling law shows that the requirements for the accuracy of the
force setting increase strongly with the flexibility of the mirror.
For example, if the thickness of the mirror and the density of the
supports are kept constant, the required accuracy of the force setting
as a fraction of the total range is inversely proportional to $d^{3}$, if
the wavefront errors are to remain constant, and to $d_{2.25}$ for the
slope errors.
\subsection{Types of supports for thin monolithic mirrors}
\label{sec:typesMirrorSupports}
For thin monolithic mirrors there are three fundamental choices for the type of the
support system.\\\\
{\it 1. Force or position based systems.} While for passive telescopes
force based systems are the only option, for active telescopes both
types are feasible. The force
option is currently still preferred, since it allows a certain decoupling of the mirror
from the mirror cell and therefore a pure closed loop operation.
For large mirrors a position based support would, owing to the fast
deformations of the mirror cell, also require open loop corrections.\\
{\it 2. For force based systems : combination of passive and active
supports or purely active supports.} The combination is the best
and often the only solution for a pure closed loop system, since the
passive part, supporting the weight of the mirror, can be designed as
an astatic system, which guarantees the required stability over
sufficiently long time periods. Purely active supports usually have a
level of non-astaticity which requires also open loop corrections,
although possibly not as frequently as with a position based system.\\
{\it 3. Mechanical levers or, at least for the passive part, hydraulic or pneumatic
supports.} Mechanical levers add considerable additional weight and
require real fixed points, which, in certain emergency cases,
may have to support the full weight of the mirror.
Astatic hydraulic or pneumatic systems can work with supports connected in
sectors and therefore virtual fixed points.
Unwanted overloads are therefore distributed over several supports.
This generates, in case of failure, much smaller stresses than a support with real
fixed points and is, for very flexible mirrors, the safer and therefore preferred
solution.\\
Which of the above mentioned options is chosen depends on the maximum
tolerable stress levels
and the required stability of the optical
configuration of the telescope
system. With glass still the traditionally used, although not necessarily optimum
material, the maximum stress level plays an important role. The choice of a system with
real fixed points may then require a comparatively thick primary
mirror, whereas a system with virtual fixed points and therefore a
better distribution of the loads in exceptional circumstances may
allow the use of a much thinner mirror.\\
In the latter case a lower limit for the thickness of the mirror is
defined by the required stiffness
to limit deformations by wind buffeting to values defined by the specification for the
effects of wind buffeting expressed in terms of low spatial frequency aberrations.
This can be partially ameliorated by coupling the mirror for high temporal frequencies
to its, in general, stiffer mirror cell, for example by using a mirror
support with six fixed points as described in \S \ref{sec:fixedPoints}
(Stepp [1993]).\\
To be able to remove the mirror easily, for example for
realuminisation, from its cell,
it would be an advantage to have only push supports. While this is not
possible for the optimum solutions for the lateral supports presented
in \S \ref{sec:latSuppMeniscus}, it can be realised for the axial
supports. The only restriction will be a limitation for the maximum
zenith angle $\theta_{{\rm z,max}}$ for which the telescope optics can
be corrected with the active optics system. The reason is that the
largest required negative correction force $F_{{\rm corr}}$ has to be
smaller than the remaining gravity load which varies with the cosine
of the zenith angle. If $F_{{\rm G},0}$ is the nominal gravity load
at zenith angle zero, one gets 
$\theta_{{\rm z,max}} = \arccos(F_{{\rm corr}}/F_{{\rm G},0})$.
For larger zenith angles than $\theta_{{\rm z,max}}$ the mirror would,
at a given support point,
loose the contact with the support. One goal of the active optics
design should therefore also be to minimise the required range of the active
forces.\\
\subsection{Axial support of thin meniscus mirrors}
\subsubsection{Basic support geometry}
For the distribution of the axial supports one can choose between two
basic geometries. One would be a regular geometry with hexagonal
symmetry where neighbouring supports form equal lateral
triangles. This would be the most effective solution in terms of
the required number of supports, but the 
symmetry is not compatible with the circular shape of the mirror. The
other choice are discrete supports on circular rings. Over most of the
area the support geometry is then irregular, but near the edges the
deformations are more regular than those generated by the hexagonal
support. The
usual choice is the second option, also because analytical methods are
available at least for the optimisation of the ring radii.
\subsubsection{Minimisation of wavefront aberrations}
The theory for the analytical optimisation of the radii of the support rings for
thin plates has been developed by Couder [1931] and the one for thin
shallow shells by Schwesinger [1988].
Both calculate first the deflections for a support on a single continuous
concentric ring. The total deflection is then a superposition
of the deflections generated by $n$ rings, multiplied by the
appropriate load fractions.
Since the dependences of the deflections on the radii are not linear,
optimisations can only be done by trial and error methods. The final
result of the optimisation depends also
on two other parameters which are, in addition to the radii,
considered variable, namely
the load fractions and, introduced by
Schwesinger [1988], an overall
deformation in form of a paraboloid, which can
easily be corrected by an axial movement of the secondary mirror.
Compared with the results which are obtained under the condition that all
support forces are identical, that
is that the load fractions are fixed and no defocus is allowed, the
r.m.s. of the sag between the supports can be reduced by approximately
30\% if the additional degrees of freedom of the load fractions and,
more important, the defocus are used for the optimisation.
The reason for the strong effect of the defocus component is that the
deflections near the inner and outer edges are nearly linear and,
if the support forces generate an overall shape similar to a parabola,
a fitted parabola can intersect the deflection curve twice both
between the inner edge and the inner ring and the outer edge and the
outer ring.
\subsubsection{Effects of fixed points}
\label{sec:fixedPoints}
Any basically astatic axial system needs three fixed points for the
definition of the position
of the mirror in space. These can be either real, as in the case of
astatic mechanical lever supports,
or virtual, as in the case of hydraulic or pneumatic supports, where all supports in each
of three sectors are interconnected. Since the volume of the fluid or gas is constant in
each sector, the barycenter of the supports will stay constant. If the positions of the
virtual fixed points are defined as these barycenters, the two types of fixed points can
mathematically be treated in the same way. In the case of the real fixed points, they
usually replace, on one of the rings, three of the astatic or active supports
at angular separations of $120^{\circ}$.\\
The question now arises, whether modes of a given rotational symmetry
$m$ can be corrected with a
given number of supports $n_{{\rm s}}$ on one ring without exciting
appreciable deformations in other rotational symmetries. Let us assume
that the force changes at the actuators on one
ring follow the rotational symmetry $m$.
The reaction forces on the fixed points due to changes of
the actuator forces can easily be calculated from the conditions of
the equilibrium of the forces and the two moments around two
orthogonal axes perpendicular to the axis of the
mirror. It can then be shown (Noethe [2000]) that the sums of the
applied forces and the reaction forces on each support on the ring
do not follow the rotational symmetry $m$ any more,
if the rotational symmetries of the applied forces are zero, one,
$n_{{\rm s}}-1$, $n_{{\rm s}}$ or $n_{{\rm s}}+1$. For the rotational
symmetries zero and one the reaction forces can be made zero, if
more than one ring is used and, in addition, for $m=0$, the sum
of load fractions on the rings is zero or, for $m=1$, the sum of the
products of the load fractions and the corresponding ring radii, is
zero. The largest rotational symmetry
correctable with the axial support system is then $n_{{\rm s}}-2$,
where $n_{{\rm s}}-2$ is the smallest number of supports on any of the
rings.\\
Another effect of the fixed points is that the correction of modes
with all symmetries different from multiples of three lead to
additional tilt. The coefficient of the tilt is roughly equal to the
coefficient of the corrected mode. In practice, it is very small and
anyway quickly removed by the autoguider.\\
An interesting consideration, first suggested by the Gemini project
(Stepp [1993]), is the use of six
fixed points to couple the mirror to the, in general, stiffer mirror cell. This allows
the reduction of wind buffeting effects on the primary mirror. Of
course, the mirror should be
coupled to its cell only for high temporal frequencies. For low temporal frequencies it has to
be decoupled to facilitate active optics corrections and, if intended by the design, to
guarantee a basically astatic support. This can be achieved by splitting each of the three
sectors in a hydraulic support system with interconnected supports into two smaller sectors
and connecting the halves by a tunable valve. A straightforward
calculation (Noethe [2000])) shows that only modes with rotational symmetries
 \begin{equation}
  m=\;6i-1\;\; {\rm or} \;\; m=\; 6i \;\; {\rm or} \;\; m=\; 6i+1,\;\;\;\;\;\;\;
                      i=0,1,2,...          \label{eq:condn}
 \end{equation}
are compatible with a six sector support, that is, are decoupled from
the mirror cell. For all other rotational symmetries, in particular
the rotational symmetry two with the softest and therefore most easily
excitable first mode, the mirror is coupled to the cell and
deformations in form of these modes can therefore be reduced.
\subsubsection{Effect of support geometry on mode correction}
\label{sec:effectGeomOnCorr}
Not only the fixed point reactions, but also the number of supports alone on any of the
rings limits the correctability of certain modes. Let $m$ be the rotational symmetry
followed by the active forces on one support ring, $\theta$ the offset
angle, $n_{{\rm s}}$ the number of supports on the ring, and the set {\bf S} be
defined by ${\bf S} = \{p\, : \, p=j \cdot n_{{\rm s}}, j=0,1,2,...\}$.
The wavefront aberration generated in an arbitrary rotational symmetry
$\overline{m}$ and order $j$ is given by (Noethe [2000]) :
\begin{equation}
  w_{\overline{m}}(\varphi) \propto
   \left\{
    \begin{array}{ll}
         \cos{\overline{m}\varphi} \; \cos{m\theta}, &
             m+\overline{m} \in {\bf S} \;\; {\rm and} \;\;
             m-\overline{m} \in {\bf S}\\
         \cos{(\overline{m}\varphi - m\theta)},      &
             m+\overline{m} \in {\bf S} \;\; {\rm and} \;\;
             m-\overline{m} \nin {\bf S}\\
         \cos{(\overline{m}\varphi + m\theta)},      &
             m+\overline{m} \nin {\bf S} \;\; {\rm and} \;\;
             m-\overline{m} \in {\bf S}\\
       0,                                            &
             m+\overline{m} \nin {\bf S} \;\; {\rm and} \;\;
             m-\overline{m} \nin {\bf S}
    \end{array}
   \right\}
       \label{eq:totalDeformOneSymmB}
\end{equation}
The first three cases represent the combinations of the rotational
symmetry $m$ of the forces and the number $n_{\rm s}$ of the equidistant
supports on one ring which generate crosstalk into other rotational
symmetries $\overline{m}$. For example, a force pattern with a
rotational symmetry $m=4$ on a ring with $n_{{\rm s}}=9$ supports will generate
the required wavefront deformation $w_{4j}(r)\cos{4\varphi}$, but also
an unwanted crosstalk of the form $w_{5j}(r)\cos{5\varphi}$.
The same two wavefront
aberrations are generated by a force pattern with the same maximum
force but with a rotational symmetry $m=5$, since the forces with
rotational symmetries $m_{1}$ and $m_{2}$ on a ring with $m_{1} +
m_{2}$ supports are identical.
Most significant are couplings into the mode $e_{0,2}$. A support with, say,
nine supports on one of the rings will generate crosstalk into this
mode if a mode with the symmetry seven is corrected.
\subsection{Lateral support of thin meniscus mirrors}
\label{sec:latSuppMeniscus}
Lateral support systems are usually passive and should fulfill the
following two requirements. First, they should not,
for any inclination of the mirror, generate wavefront aberrations
which require significant active correction forces from the axial
support system, which would increase the range of active forces and
therefore reduce the maximum usable zenith angle as described at the
end of \S \ref{sec:typesMirrorSupports}.
Second, the mirror should be
supported at the outer edge only. Fortunately, a type of lateral
support with these characteristics exists.
The analytical theory has been developed by Schwesinger [1988, 1991]. Instead of
discrete forces it considers initially force densities at the edges
with the three components $f_{{\rm r}}$ in radial, 
$f_{{\rm t}}$ in tangential and $f_{{\rm a}}$ in axial direction. Thinking in Fourier terms
implies that any force densities which follow a given rotational symmetry
generate deformations
in only this symmetry. The only force densities which support the weight of the mirror are
those with the rotational symmetry one. A lateral support system should therefore only contain
force densities of rotational symmetry one.\\
The lateral support is greatly simplified for telescopes with altazimuth mountings. In this
case the directions of the forces with respect to a coordinate system which is fixed to the
mirror are constant. Only the moduli depend on the inclination of the
mirror cell.\\
If the mirror is neither too steep nor too thin, it can be laterally supported at the outer
rim under its center of gravity. But for steep and thin mirrors
this is not the case and axial forces at the outer edge have to be
used to balance the moment. The modulus of the axial force
density $f_{{\rm a}}$, which is proportional to $\sin{\varphi}$, where
$\varphi$ is the azimuth angle starting from the direction parallel to
the altitude axis, is then defined by the weight of the mirror, its
diameter and the distance between the plane of the supports and the
center of gravity of the mirror. The radial force densities
$f_{{\rm r}}$ have always to
be proportional to $\sin{\varphi}$ and the
tangential force densities $f_{{\rm r}}$ to $\cos{\varphi}$.
The only free parameter is then
the fraction $\beta$ of the weight supported by the tangential force
density, with the remaining weight $1-\beta$ supported
by the radial force density.\\
Schwesinger [1988, 1991] has derived analytical formulae for the dependence
of the radial function of the
deflection with the rotational symmetry one on the ratio $\beta$. The deflection may
contain third order coma, which can be corrected by a movement of the secondary mirror.
The residual wavefront error after fitting and subtracting third order coma should therefore
be the merit function for the optimisation with the ratio $\beta$. These wavefront errors
are, for an optimum choice of $\beta$, in practice so small that a possible further reduction with
additional supports at the inner edge is not necessary (Schwesinger
[1994]).\\
Schwesinger's theory for the rotational symmetry one can be extended
to all other rotational symmetries (Noethe [2000]). The deformation of
the mirror in
any rotational symmetry $m$ can be calculated for force densities
$f_{{\rm r}}$, $f_{{\rm t}}$ and $f_{{\rm a}}$ following the same
symmetry $m$.
Each of the three components in radial, tangential and axial
direction of any of the discrete forces at the edge can then be expanded in
an infinite series in all rotational symmetries.
Since, as in
the case of the elastic modes for axial deformations, the deflections
decrease rapidly with the rotational symmetry of the modes, the consideration of
the lowest symmetries will be sufficient to calculate the overall
deformations. This offers a fast and efficient alternative to
finite element calculations.\\
If the lateral supports are combined with the axial supports as in the
Subaru telescope (Iye [1991]), the actual locations of the
application of the forces have to be in the neutral surface to avoid
unwanted moments. For solid monolithic mirrors this requires the
drilling of additional holes. For mirrors with
a honeycomb structure it may be the natural and best solution.
\subsection{Segmented mirrors}
Although it is not a compulsary requirement, one goal of a segmented
mirror design is that the shapes of individual segments do not need active
corrections during the operation of the telescope.
With diameters as large as 2\,m they require passive, astatic supports
as, for example, multi-stage whiffle trees which apply both axial and
lateral forces. 
The deflections as functions of the number of supports per segment
area and thickness follow the scaling laws for monolithic
mirrors given in \S \ref{sec:scalingLaws}. An optimisation of the
distribution of supports is usually done with finite element calculations.
To correct figuring errors in a d.c. mode, static devices like warping
harnesses can be installed at the back surface of the segments.\\
If each segment is intrinsically stable, the major problem
is the alignment of the $n_{{\rm s}}$ segments both in piston and
tilt. Each segment therefore needs three actuators capable of changing
the axial positions of the three fixed points.
The support of a segmented mirror as a whole is therefore
position based and requires, owing to the normally strong flexure
of the cell with a change of the zenith angle, frequent corrections.
The alignment and control of a segmented mirror is dicussed in \S
\ref{sec:alignmentSegments}.
\section{Alignment}
\label{sec:alignment}
\subsection{Alignment of a two mirror telescope}
In a perfectly aligned two mirror telescope the axes of the primary
mirror, the secondary mirror and the rotator are congruent. This ideal
case can, in particular with large telescopes, only be achieved as an
approximation.
In particular, even if the alignment is sufficiently good for a certain
zenith angle, the mechanical deformations of the telescope structure
may generate misalignments at other zenith angles.
A complete alignment of the telecope can be done in three steps.
\begin{itemize}
 \item {\it Initial alignment with auxiliary equipment}\\
       Using autocollimation and finite focusing the axes of M2 and
       the rotator of a large telescope like the
       VLT can be aligned to an accuracy of approximately 3\,arcsec for
       the angles between the axes and less than 1\,mm for a
       shift of the vertex of M2 with respect to the axis of the
       rotator. But the position of M1 and therefore the angle between the
       axes of M1 and M2 and the shift between the vertex of M2 with
       respect to the axis of M1 are only defined within the
       mechanical tolerances of the M1 support, which are much larger
       than the accuracy
       of the alignment achieved for the relative alignment of the
       axes of the rotator and M2. The consequence will, in general,
       be a large amount of decentering coma.
 \item {\it Correction of decentering coma}\\
       The decentering coma generated by the
       misalignment between the axes of M1 and M2 can be measured by
       the wavefront analyser, and be corrected by a rotation of
       the secondary mirror around its center of curvature,
       by a full body movement of M1, or by a combination of both.
       After this
       operation the telescope may still be a schiefspiegler (WILSON
       [1996]) in which the axes of M1 and M2 are not aligned, but
       intersect at the so-called coma free point.
       This is a point around which the secondary
       mirror can be rotated without changing the
       value of field independent decentering coma.
 \item {\it Alignment of the axes of M1 and M2}\\
       The residual misalignment can be determined from 
       a mapping of the pattern of third order astigmatism, that is
       from measurements of this coefficient at a few field positions.
       A complete correction can be done by rotating either the
       secondary mirror or the primary mirror or both around
       the coma free point.
\end{itemize}
An overview of aberrations in misaligned telescopes can be found
in Wilson [1996] and of the alignment of telescopes in chapter~2
of Wilson [1999].
A general theory of low order field aberrations of decentered optical
systems has been given by Shack and Thompson [1980].
In particular it has been shown that the general field dependence of
third order astigmatism can be described by a binodal pattern,
known as ovals of Cassini.
Only for special cases such as a centered system do the two nodes
coincide and the field dependence reduces to the well known rotationally
symmetric pattern with a quadratic dependence on the distance to the
field center. These general geometrical properties have been used by
McLeod [1996], starting from equations by Schroeder [1987],
for the alignment of an aplanatic two mirror telescope.
McLeod showed that  the components $Z_{4}$
and $Z_{5}$ of third order astigmatism of a two mirror telescope
with the stop at the primary mirror for
a field angle $\theta$ with components $\theta_{x}$ and $\theta_{y}$
are given by
\begin{eqnarray}
 Z_{4} & = & B_{0}\, (\theta_{x}^{2} - \theta_{y}^{2})
               \, + \, B_{1}(\theta_{x}\alpha_{x}
                          \, - \, \theta_{y}\alpha_{y}) \nonumber\\
             && + \, B_{2}\, (\alpha_{x}^{2} - \alpha_{y}^{2})
                     \label{eq:z4sysGen}\\
 Z_{5} & = & 2B_{0}\, \theta_{x}\theta_{y}
               \, + \, B_{1}(\theta_{x}\alpha_{y}
                          \, + \, \theta_{y}\alpha_{x})
               \, + \, 2B_{2}\, \alpha_{x}\alpha_{y}
                     \label{eq:z5sysGen}
\end{eqnarray}
$B_{0}$ is the coefficient of field astigmatism for a centered
telescope, whereas $B_{1}$ and $B_{2}$ only appear in decentered systems.
Numerical values for $B_{0}$, $B_{1}$ and $B_{2}$ were obtained 
by using general formulae for field astigmatism of individual mirrors and
adding the effects of the two mirrors.
The values for $\alpha_{x}$ and $\alpha_{y}$ could then be obtained from
measurements of $Z_{4}$ and $Z_{5}$ in the
field of the telescope.\\
Explicit expressions for
the third order astigmatism parameters $B_{0}$, $B_{1}$ and
$B_{2}$ as functions of fundamental design parameters and optical
properties of the total telescope and of the position of the stop
along the optical axis give more insight into the characteristics
of field aberrations of two mirror telescopes. They have been derived
for centered two mirror telescopes by Wilson [1996] and for decentered
ones by Noethe and Guisard [2000].
In a decentered system one also has to take into account the
definition of the field center. The normal definition is the direction
parallel to the axis of M1, projected towards the sky. But in
decentered system the image of an object in
this field center is not in the center of the adapter, where the
instruments are located and which is therefore the practical field
center. If the field astigmatism is calculated with respect to this
practical field center, the structure of the Eqs. (\ref{eq:z4sysGen}) and
(\ref{eq:z5sysGen}) remains the same, but the parameters $B_{0}$,
$B_{1}$ and $B_{2}$ change and $\theta$ denotes the
field angle with respect to the center of the adapter (Noethe and Guisard
[2000]). To align the axes of the two mirrors and to put the intersection of this
axis with the focal plane to the center of the adapter, one has to reposition
both mirrors.\\
In principle, two wavefront analysers would be necessary and also
sufficient for a closed
loop alignment of a two mirror telescope. With only one wavefront
analyser available, mappings have to
be done at various zenith angles and the alignment can be controlled
only in open loop.
\subsection{Alignment of a segmented mirror}
\label{sec:alignmentSegments}
The alignment procedure described in this section is the one used
in the Keck telescope.
It is assumed that the shapes of the segments are not
affected by changes of the zenith angle.
The control of the position of a segment is restricted to three
degrees of freedom, a piston coordinate parallel to the optical
axis of the telescope, and two tilt components for rotations
around two orthogonal axes perpendicular to the optical axis.
The alignment of the segments is done in two steps.\\
First, the tilts of the $n_{{\rm s}}$ segments are
measured optically with the passive tilt or the fine screen mode 
of the phasing camera system (PCS) described in
\S \ref{sec:wfAnalysisSegments}.
The required corrections of the segment tilts are
done by appropriate differential movements of the $3n_{{\rm s}}$ piston
actuators.
Second, the differences in
height at midpoints of intersegment edges are measured optically by the
segment phase mode of the PCS.
The number of these sampling points is
larger than the number of degrees of freedom, which is equal to the number
of segments minus one. The optimum differential
piston movements, which are the ones that minimise the
r.m.s. of the differences in height of adjacent segment midpoints,
are obtained by a least squares fit.\\
The relative positions of the segments, and therefore
the overall shape of the mirror, are then maintained by the positions
actuators, but not in closed loop controlled by optical measurements
with the PCS, but by
measurements with position sensors located at intersegment boundaries.
These piston sensors are capable of measuring changes in the relative heights
of the adjacent segments perpendicular to the surfaces.
The actual readings after an alignment described above are defined as
target values for subsequent corrections. The number of the piston sensors
must be at least as large, but is
usually larger than the number of actuators, which is
$3n_{{\rm s}}$. 
The differences between the reference and the actual readings
are, via a least squares fit, converted into
actuator movements. Any noise in the sensor readings will lead to
errors in the relative tilt and piston values of the segments.
These aberrations can conveniently be expanded in so-called
{\em normal modes} (Troy, Chanan, Sirko and Leffert [1998]).
Similar to the elastic modes, which are
eigenvectors of a differential equation decribing the elastic
behaviour of the mirror, the normal modes are orthogonal
eigenvectors associated with a singular value decomposition of the
control matrix connecting actuator movements to the larger number of
sensor readings.
Apart from the discontinuities due to the segmentation, the
normal modes, in particular the lower order ones, approximate to
Zernike polynomials.\\
If random noise is assumed for the sensor readings, the average
of the coefficient of a normal mode contained in the wavefront error
decreases rapidly with the order of the mode. The normal mode which
can most easily be generated by the segmented primary mirror control
system of the Keck telescope is a defocus mode, followed by a 
mode similar to third order astigmatism. The defocus mode is
produced by a constant offset to all piston sensors, since the
corresponding changes of the actuator lengths will exactly follow a
parabola.
Compared with the equivalent case of
the average content of modes in a wavefront generated by random
pressure fields on a monolithic mirror, the 
decrease is, above all for the higher order modes, much weaker.
For example, the normal mode similar to the third order Zernike
polynomial of rotational symmetry two, that is seventh order
astigmatism, is only
ten times weaker than the strongest mode, whereas the corresponding
ratio for the elastic modes is of the order of 200 times weaker.
This slower convergence is of importance, since the higher order modes
generate stronger edge discontinuities. On the other hand, the
r.m.s. of the edge discontinuities related to tilt errors with a
given r.m.s. value are much smaller than expected from a random
distribution of the tilt errors over the segment, since most of the
tilt error is contained in the smooth modes with small edge
discontinuities.
\section[Modification of telescope optical
configurations]{Modification of the telescope optical configuration}
\label{sec:modificationOpticalConfig}
A defocus aberration can be introduced both by an axial movement of the secondary mirror
and a deformation of the primary mirror. This feature can be used to control the plate
scale of the telescope. The defocusing with the secondary mirror can also,
together with an elastic deformation of the primary mirror, be used to
maintain the optical quality of the telescope during a change of its
optical configuration.
\subsection{Control of the plate scale}
\label{sec:plateScale}
The focal length $f'_{2}$ of M2 is assumed to be constant. The plate scale is
therefore only affected by changes of the focal length $f'_{1}$ of M1
and the distance $d_{1}$ between M1 and M2.
A change of the shape of the primary mirror in the defocus mode,
described by the coefficient $c_{{\rm def}}$ of the equivalent
wavefront change, generates the following change $\delta f'_{1}$ of
the focal length of M1
\begin{equation}
  \delta f'_{1} = 8\, N_{1}^{2} \, c_{{\rm def}},
   \label{eq:ftocdef}
\end{equation}
where $N_{1}$ is the f-number of the primary
mirror.
The dependences of the variations of the back focal distance $b$ and
the focal length $f'$ on the variations of $f'_{1}$ and $d_{1}$
are given by Wilson [1996]
 \begin{eqnarray}
  \delta b       & = & (m_{2}^{2} + 1) \, \delta d_{1}
                        - m_{2}^{2} \, \delta f'_{1}
                              \label{eq:deltab} \\
  \delta f' & = &\frac{m_{2}^{2}}{f'_{2}} \;
                    (f'_{1} \,  \delta d_{1}
                      - (f'_{2} + d_{1}) \, \delta f'_{1}),
                            \label{eq:deltaftel}
 \end{eqnarray}
where $m_{2}$ is the magnification of the secondary mirror.\\
The two conditions for a control of the plate scale in the telescope
are the amount of the change $\delta f'$ of the focal length of
the telescope and the requirement that the distance $b$ between the
pole of M1 and the image remains unchanged, i.e. $\delta b = 0$.
These two conditions can be fulfilled
by variations of the two parameters $f'_{1}$ and $d_{1}$.\\
From (\ref{eq:deltab}) one then gets
 \begin{equation}
  \delta f'_{1} = \frac{m_{2}^{2} + 1}{m_{2}^{2}}  \delta d_{1}
 \end{equation}
Introducing this into (\ref{eq:deltaftel}) and solving for $\delta
d_{1}$ one gets
 \begin{equation}
  \delta d_{1} = \frac{1}{m_{2}^{2} - 1 - \frac{d_{1}}{f'_{2}} }
                   \delta f'
 \end{equation}
The accuracy of the control of the plate scale is limited by the accuracy of the force
setting under M1 and the axial positioning of M2, that is by the
individual contributions
from $\delta d_{1}$ and $\delta c_{{\rm def}}$ to $\delta f'$, and by the noise in the
wavefront measurements.
\subsection{Modification of the optical configuration}
\label{sec:NasmythToCassegrain}
If a two mirror telescope has both Nasmyth and Cassegrain foci, it may
not be possible to find a convenient design which places
both foci at the same distance from the secondary mirror. Switching from one focus to the
other therefore requires refocusing. In a classical Cassegrain
design this will generate 
field independent third order spherical aberration, which was
initially not present, and in a
Ritchey-Chretien design in addition field dependent third order coma.
Of those, the spherical aberration can be removed by a deformation of the primary mirror,
that is a change of its conic constant.\\
Changing the shape of M1 by a function proportional to $r^{4}$ requires comparatively
strong forces, since $r^{4}$ has strong curvature near the outer edge contrary to
elastic modes with effectively no curvature near the outer edge.
The curvature of $r^{4}$ near the outer edge can be greatly reduced by adding
an appropriate amount of defocus which will be compensated by an additional
axial movement of M2.
This new deformation can be better approximated by elastic modes and can therefore be
generated with much smaller forces.
\section[Active optics design for NTT, VLT and Keck]{Active optics design for the NTT, the VLT and the Keck telescope}
\label{sec:AODesignNTTVLT}
\subsection{General requirements and specifications}
\label{sec:genReqSpecsAO}
The NTT and VLT are examples of active two mirror telescopes with monolithic meniscus mirrors.
The NTT with a mechanical diameter of its primary mirror of 3.58\,m was the first telescope
with active optics as an integral part of its design. Nevertheless, since it was the first
attempt to build an active telescope, one conservative requirement was
that it could, with a reduced optical quality, also function in a
fully passive mode. The VLT with a diameter of its
primary mirror of 8.2\,m was envisaged to function only in the active
mode, since its primary mirror is about 40 times as flexible.
The designs of the active optics systems of these two telescopes
can serve as typical examples for two mirror telescopes of the four
and eight meter class with monolithic primary mirrors.\\
The specifications for the NTT were given in terms of the diameter
$d_{80}$ of the circle
containing 80\% of the geometrical energy. For a Gaussian point spread
function one has $d_{80} \approx 1.5 \theta \approx 2.54 \sigma_{t}$ and for an
atmospheric-seeing point spread function $d_{80} \approx 1.9 \theta$.
The specifications were then $d_{80}=0.15$\,arcsec for the active
and $d_{80}=0.40$\,arcsec for the passive mode. The figure for the active mode
can be split into $d_{80}=0.10$\,arcsec for the high and also
$d_{80}=0.10$\,arcsec for the low
spatial frequency aberrations. The specifications for the VLT were given in terms of the
central intensity ratio $CIR$, namely $CIR = 0.8$ for a seeing of 0.4\,arcsec. The relevant
contributions for the design of the active optics system were $CIR =
0.992$ or, according to eq. (\ref{eq:CIR}), an r.m.s. of the wavefront
slopes $\sigma_{{\rm t}} = 0.021$\,arcsec for the high
spatial frequencies of M1, $CIR = 0.979$ or $\sigma_{{\rm t}} =
0.034$\,arcsec for the active optics control
errors of the wavefront, including
both the wavefront analysis and the corrections, and
$CIR = 0.97$ or $\sigma_{{\rm t}} = 0.041$\,arcsec for the effects of wind
pressure variations on M1.
Since the major aberrations generated by wind and active optics control
errors are low spatial frequency aberrations, the $CIR$ figures for
these error sources can be
converted into approximate
r.m.s. values $\sigma_{{\rm w}}$ of wavefront aberrations dominated by
the mode $e_{2,1}$. $CIR = 0.99$ is then equivalent to
$\sigma_{{\rm w}} = 140$\,nm and $CIR = 0.98$ to $\sigma_{{\rm w}} = 200$\,nm.
Finally, all possibly occurring stresses in the mirrors had to be well
below the critical values for glass ceramics.\\ 
With these specifications all parameters in the first column of
fig. \ref{fig:AODependencies}, which form the basis
of the active optics design, were defined.
Two other parameters, which are in principle free in
fig. \ref{fig:AODependencies}, were also defined in advance.
First, the active optics systems of
both telescopes were required to work fully in closed loop with
integration times of the wavefront analyser of at least 30 seconds and
a full sky coverage, and, second,
the substrate of both primary mirrors was a glass ceramic.\\
The VLT had the additional requirement that it should work both with
the Nasmyth foci and a Cassegrain focus with the consequences
described in \S \ref{sec:NasmythToCassegrain}.
Furthermore, because of the strong impact of temperature differences
between the mirrors, the air in the enclosure and
the outside air on the image quality, the VLT was required to control
these differences within narrow limits instead of relying on natural
ventilation only as in the case of the NTT.\\
The specifications for the Keck telescope were given in terms of
$d_{80}$. The error budget for the total telescope was 0.41\,arcsec,
with 0.24\,arcsec for the segment figure being the largest
contribution. The total active optics error budget was split into a
contribution of 0.084\,arcsec from zenith distance independent and of
0.058 arsec from zenith distance dependent errors (Cohen, Mast and
Nelson [1994]).
\subsection{Active optics design of the NTT}
\subsubsection[M1 thickness, type of support system, active
modes]{Thickness of M1, type of support system and set of active modes}
The main driver for the thickness of M1 was the requirement, that the
telescope could, although with a reduced optical quality, be operated
also in a passive mode. 
Measurements at the equatorially mounted ESO 3.6\,m telescope showed that the $d_{80}$ values
due to low spatial frequency elastic aberrations were of the order of 0.5\,arcsec largely
independent of the sky position (Wilson [1999]).
Since the design of the M1 support of a telescope with an
altazimuth mounting like the NTT was significantly easier, it was estimated that the NTT
could, passively, achieve the same performance with a mirror of approximately half the
thickness, which was then finally defined as 241\,mm.
Another way of justifying this thickness of M1 is the following.
The average coefficient of the dominant low spatial frequency mode
$e_{2,1}$ generated with random forces in the range of
$\pm 1$\,N would be 5.5\,nm, with maximum values of the order of 15\,nm. Random
force errors in the range of $\pm5\%$ of the nominal forces of
approximately 760\,N would then generate on average coefficients of $e_{2,1}$ of
the order of 210\,nm, which is equivalent to an r.m.s. of the slope
errors of $\sigma_{{\rm t}} \approx 0.082$\,arcsec and $d_{80} \approx
0.20$\,arcsec. If equal
tolerances were also given to the defocus and decentering coma errors, one
would with a quadratic sum just fulfill the specification for
$d_{80} = 0.4$\,arcsec for the passive mode.\\
With such a thickness the stresses, which
arise if the mirror is unintentionally supported by three points only,
are well below the tolerable limit.
This then allowed the use of a conventional support with astatic levers and consequently
three real fixed points.
To define the set of active modes, one can apply the procedure
described at the end of \S \ref{sec:minEnergyModes}.
With the r.m.s. of 210\,nm for the average coefficient of $e_{2,1}$
generated with random forces in the range
$\pm 5\%$ of the nominal load, the frequency
limit for the modes to be considered is
$\nu_{2,1} \sqrt{210/15}\approx 430$\,Hz and the data in table
\ref{tab:eigenFrequencies}
show that for the NTT the modes up to $e_{4,1}$
should be corrected. But then, since the elastic mode $e_{0,1}$ is
replaced by defocus, there would be no possibility to correct rotationally
symmetric aberrations other than defocus. Because of the importance of
spherical aberration the elastic mode $e_{0,2}$ has to be added to the
set of active modes. The chosen force range of the active actuators of
$\pm30$\% of the nominal load was large enough
to allow also the use of the equivalent Zernike modes instead of the
more efficient elastic modes.
Spherical aberration can be generated with
much smaller forces
by combining it with a defocus deformation of M1. This defocus can
then easily be compensated by an appropriate axial movement of M2. For
the NTT the best combination is $r^{4} - 3.6r^{2}$.
\begin{center}
\begin{table}
\caption{{\small Eigenfrequencies of the lowest elastic modes of the
NTT and the VLT.}}
\begin{center}
\vspace{2mm}
\begin{tabular}{|l|r|r|r|r|r|r|r|r|r|}
\hline
Symmetry & 2 & 3 & 0 & 4 & 1 & 5 & 2 & 0 & 6\\
Order    & 1 & 1 & 1 & 1 & 1 & 1 & 2 & 2 & 1\\
\hline
NTT    & 115 & 273 & 192 & 479 & 434 & 732 & 737 & 852 & 1034\\
VLT    & 16 & 38 & 42 & 66 & 68 & 102 & 107 & 119 & 143\\
\hline
\hline
Symmetry & 3 & 1 & 7 & 4 & 8 & 2 & 0 & 5 & 3\\
Order    & 2 & 2 & 1 & 2 & 1 & 3 & 3 & 2 & 3\\
\hline
NTT    &  1131 & 1229 & 1383 & 1577 & 1779 & 1749 & 2050 & 2077 & 2366\\
VLT    &   160 & 176 & 192 & 221 & 246 & 246 & 272 & 289 & 331\\
\hline
\end{tabular}
\end{center}
\label{tab:eigenFrequencies}
\end{table}
\end{center}
\subsubsection{Axial support of M1}
\label{sec:actAxSuppM1}
The axial support of the primary mirror of the NTT consists of four
rings with 9, 15, 24 and 30 supports.
This distribution gives an r.m.s. $\sigma_{{\rm w}}$ of the high
spatial frequency wavefront
aberrations of approximately 7\,nm and an r.m.s. of
the slope error of the wavefront of $\sigma_{{\rm t}} \approx 0.02$
arcsec, well below the specification of $\sigma_{{\rm t}}  = 0.04$
arcsec, which is equivalent to $d_{80} = 0.1$\,arcsec.
The chosen density
of supports was also sufficient to generate all active modes
with high accuracy. The largest error in terms of the r.m.s. of
the relative difference between the requested and the actually
generated shapes is, not considering the effects of the print-through,
of the order of only 2\% for the second mode $e_{0,2}$
of rotational symmetry zero. For the other modes the relative errors
are of the order of 0.1\% or smaller.\\ 
The modification with respect to a passive support with astatic
mechanical levers were motorised
counterweights which could change the support force by approximately
$\pm 30$\% of the gravity load on the support. Since the gravity loads
are proportional to the cosine of the zenith angle, the
correction of errors which are independent of the zenith angle like
polishing errors would have required different positions of the
counterweights for
different zenith angles. For this reason additional springs were
introduced which could introduce correction forces independently of
the zenith angle. The springs had no motorised control and could only
be adjusted manually.\\
With the comparatively large thickness, wind buffeting on M1 was no
problem. In addition, the telescope optics was very stable over time
periods of one minute and could therefore be operated in closed loop.\\
The force setting accuracy to achieve wavefront errors of
$\sigma_{{\rm w}} < 50$\,nm is of the order of $\pm 10$\,N. To be sure
that the error is within the limit 95\% of the time and not only on
average, the force setting accuracy should be three times
better, that is $\pm 3$\,N.
\begin{figure}[h]
 \centerline{\hbox{
  \psfig{figure=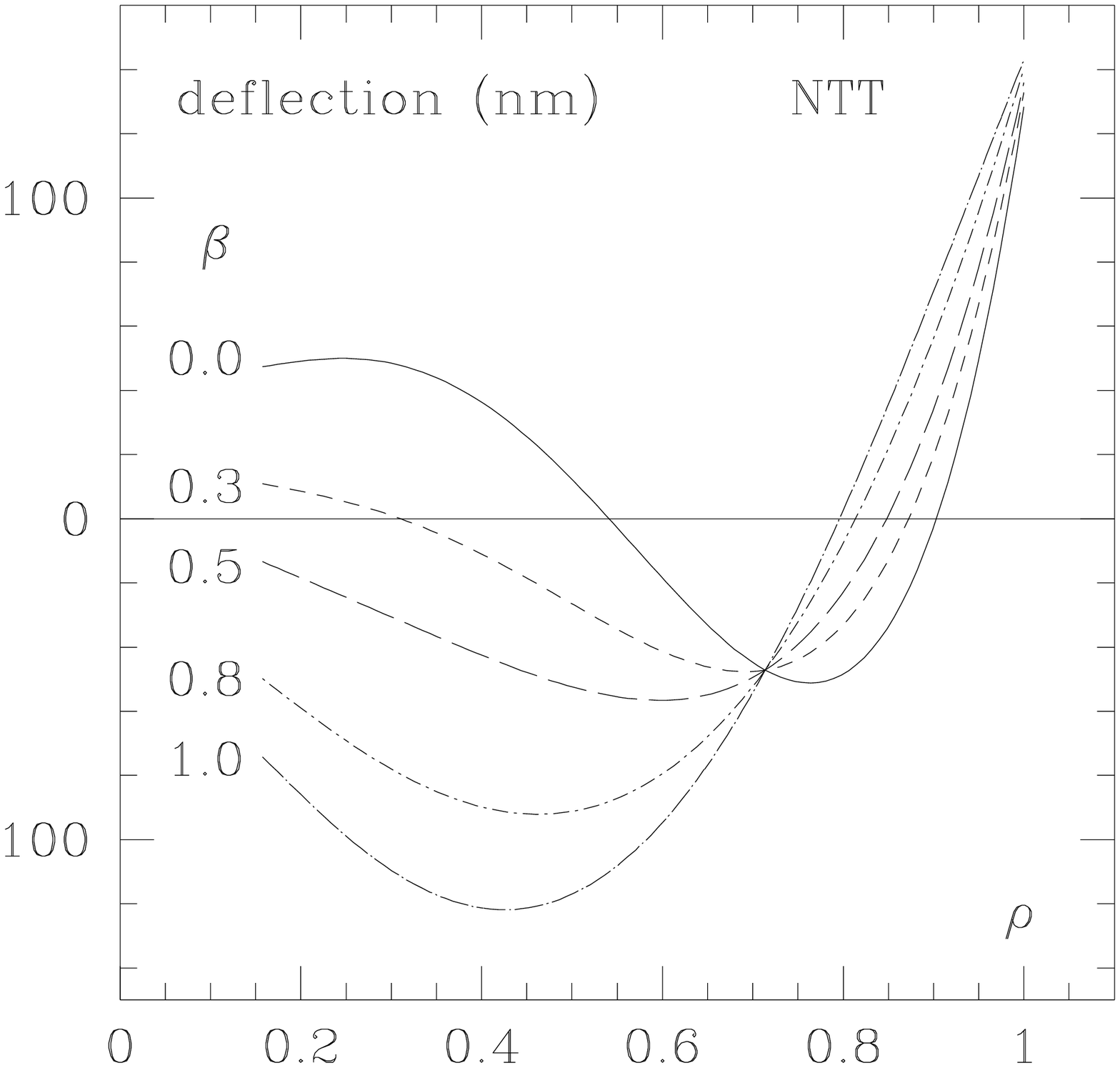,width=50mm}
  \psfig{figure=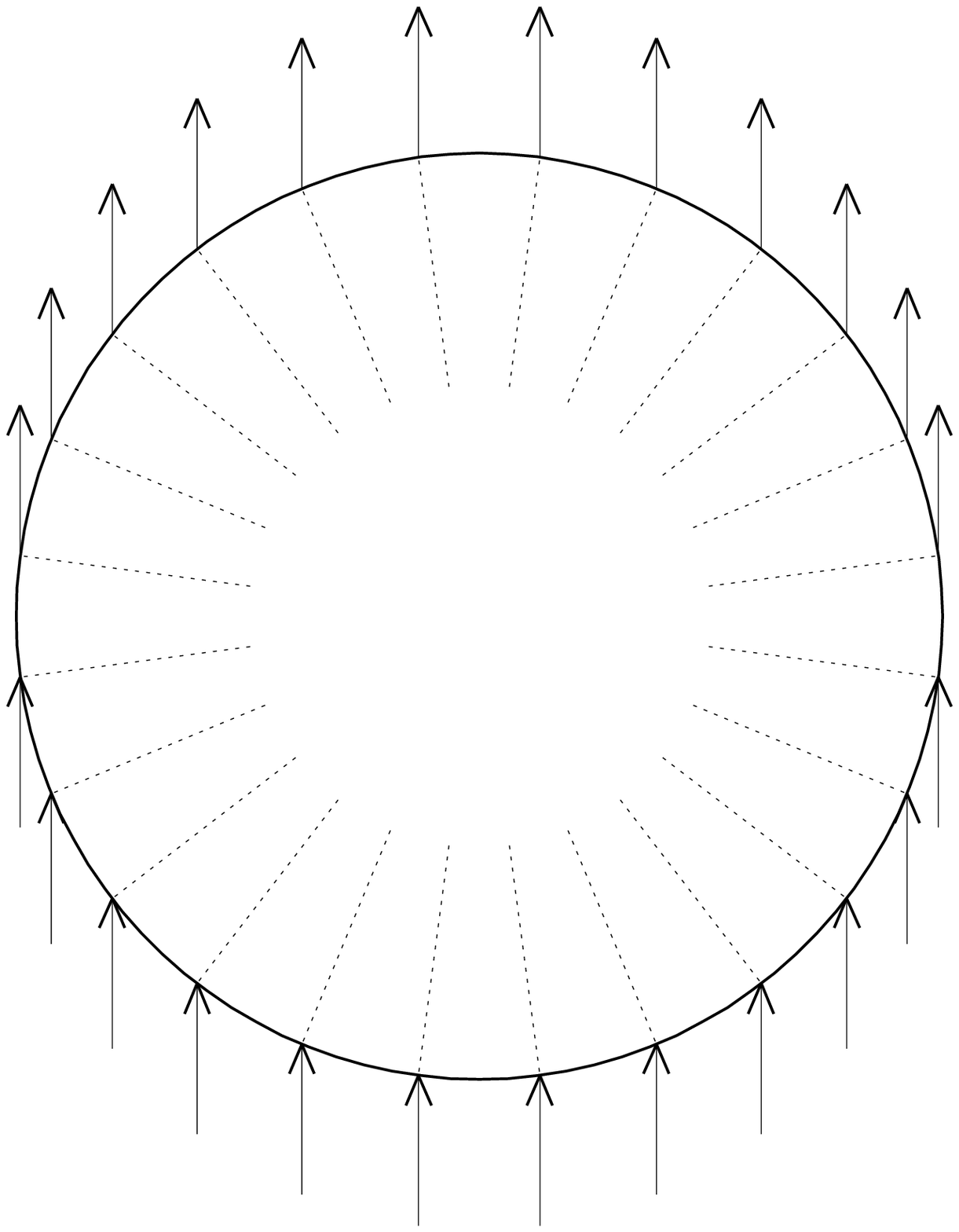,width=50mm}}}
  \caption{\label{fig:latNTT} {\small {\em Left} : Deflections of the
  NTT as functions of the normalised radius $\rho$ for
    various fractions $\beta$ of the weight supported by the
    tangential forces. {\em Right} Lateral forces for $\beta = 0.5$}}
\end{figure}
\subsubsection{Lateral support of M1}
Despite the relatively low f-number of 2.2 of the primary mirror the
plane perpendicular
to the axis of M1 through the center of gravity intersects the outer rim. M1 could therefore
be supported laterally under its center of gravity with all lateral forces in a plane
perpendicular to the axis of M1. Fig. \ref{fig:latNTT} shows on the
left the dependence of the surface deflection along a central vertical
line (Schwesinger [1988]) on the ratio $\beta$. Apparently, the
dependence on $\beta$ is not critical and for
$\beta = 0.5$ the deflection approximates to third order
coma, which can be corrected by a movement of the secondary mirror.
The r.m.s. of the residual
wavefront error is then approximately 20\,nm.
The choice of $\beta = 0.5$ is convenient, since for equidistant
positions of the lateral supports the
forces are all identical and parallel to the direction of the gravity
vector, as shown in figure \ref{fig:latNTT} on the right.
With 24 supports the forces are of the order of 2500\,N and the
stresses well below the critical values.
\subsubsection{Position control of M2}
The control of defocus and decentering coma requires an accurate
positioning of the secondary mirror. To reach an accuracy of
$\sigma_{{\rm t}} \approx 0.02$\,arcsec for both modes, one needs an
accuracy of the axial movement of M2 of approximately $2\,\mu$m for the
correction of defocus, and an accuracy of
the rotation around the center of curvature of approximately 3
arcsec for the correction of decentering coma.
The restricted number of motorised degrees of freedom
of the movements of M2 do
not allow a motorised correction of 
a misalignment, which would require a rotation around the coma free
point. This can, however, be done by a combination of a mechanical
adjustment of the M2 cell and a rotation of M2 around its center of
curvature.
\subsubsection{Wavefront analyser}
The wavefront analyser is a Shack-Hartmann device with a rectangular
25 by 25 lenslet array with lenslets of 1\,mm side length and a
f-number of 170. To fit the pattern on the CCD array with a side
length of 11\,mm, optics with a reduction factor of $m_{{\rm sh}} =
0.36$ had to be used. In the conditions (\ref{eq:condMinKF}),
(\ref{eq:conditionkFnumber}) and (\ref{eq:condKflNoise}) in
\S \ref{sec:defSHParams} the left hand sides then all have to be
replaced by the product $m_{{\rm sh}}N_{{\rm l}}$. With the chosen
parameters these conditions are all fulfilled.
The size of 150\,mm by 150\,mm of a subaperture on the primary mirror
corresponding to one lenslet may be too small
to find a sufficiently bright guide star in the
field for an arbitrary sky position. But the size of the subapertures
could be increased to 350\,mm by 350\,mm, since a sampling of 10
by 10 would easily be sufficient for an
accurate measurement of the small number of active modes.
\subsection{Active optics design of the VLT}
\subsubsection[M1 thickness, type of support system, active
modes]{Thickness of M1, type of support system and set of
active modes}
For an 8\,m mirror as thin as the one of the VLT the stresses generated
by an accidential support on three hard fixed points would have been
dangerous. Since a basic passive support was required for a pure
closed loop operation, a hydraulic support system with all supports
connected in each of the three sectors
was chosen as the passive part of the axial support system. To avoid
pressure differences due to gravity in inclined
positions, it was designed as a two chamber system (Schneermann, Cui,
Enard, Noethe and Postema [1990]). The active part
has electromechanical actuators which work in series with the
passive support and therefore add the correction forces to the passive
ones.\\
The lower limit of the thickness of the VLT was partially defined by
wind buffeting considerations. With expected wind pressure variations
of $1{\rm N}/{\rm m}^{2}$, the r.m.s. of the wavefront aberrations
could be limited to 150\,nm with a mirror thickness of
approximately 175\,mm. The wind pressure variations could have been
reduced further by reducing the wind flow in the enclosure, but this
could have generated local seeing effects due to insufficient flushing of
temperature inhomogeneities created inside the enclosure.
According to fig. \ref{fig:AODependencies} the
definition of the thickness defined the stiffness and therefore also
the rest of the active optics parameters.\\
The set of active modes is defined by the procedure described at the
end of \S \ref{sec:minEnergyModes}.
For the VLT the average coefficient of $e_{2,1}$ for random forces in
the range of $\pm 1$\,N is 85\,nm. With random force errors of 5\% of
the nominal load of 1500\,N the expected average coefficient
of $e_{2,1}$ is then 6375\,nm. The frequency limit for the active modes
to be considered is therefore $\nu_{2,1} \sqrt{6375/15} \approx 330$
Hz. The data in table \ref{tab:eigenFrequencies}
show that for the VLT the modes up to
$e_{5,2}$ should be corrected.
As discussed in \S \ref{sec:modalConcept} the modes $e_{0,1}$
and $e_{1,1}$ are replaced by the corresponding Zernike polynomials for defocus and third
order coma.
\subsubsection{Axial support of M1}
\label{sec:axialSupportVLT}
{\it Support density.} Since three is the highest order in the set
of active modes, six rings are sufficient to generate these modes with the
required
accuracy. A uniform distribution of supports on the rings together
with the requirement that the number of supports on each ring is a
multiple of three then leads to a total number of 150 supports with 9,
15, 21, 27, 36 and 42 supports on the six rings.
As a result of the scaling law (\ref{eq:scalingSag}), the
r.m.s. $\sigma_{{\rm w}}$ of sag of the mirror
between its support under its own weight would be approximately ten
times higher than the one at the NTT. Since the distances between the
supports are larger than in the NTT, the r.m.s. $\sigma_{{\rm t}}$ of
the slopes of the wavefront is only six times higher, that is
$\sigma_{{\rm t}} \approx 0.15$\,arcsec. For a seeing of $\Theta = 0.4$
arcsec this would give a central intensity ratio of $CIR \approx 0.6$,
far below the specification of $CIR = 0.992$ for the high spatial
frequency aberrations generated by the print-through. To reach the
$CIR$ specification, which is equivalent to $\sigma_{{\rm t}} \approx
0.02$\,arcsec,
the primary mirror would have required approximately 400 supports,
which would have added significant complexity and cost. Instead, the
specification could be reached by replacing each of the single point
supports by tripods (Schneermann, Cui, Enard, Noethe and Postema [1990]).\\
According to sects. \ref{sec:fixedPoints} and
\ref{sec:effectGeomOnCorr}, with nine supports on the inner ring a
correction of modes with rotational symmetries seven
and eight is not possible without generating crosstalk. Indeed, corrections of the modes
$e_{7,1}$ and $e_{8,1}$ with coefficients of 1000\,nm generate 470\,nm
of $e_{2,1}$ and 1053\,nm of $e_{1,1}$, respectively, since the
symmetries of the force distributions and the crosstalk mode add up to
the number nine of supports on the first ring. In addition, generating
1000\,nm of $e_{8,1}$ also produces 2217\,nm of $e_{2,1}$ and smaller
amounts of other aberrations because of reaction forces on the three
virtual fixed points.
These two modes should therefore not be corrected permanently in closed loop,
but only once after a preset to a new sky position. The crosstalk to lower
order modes is then removed by subsequent corrections.\\
{\it Accuracy of the force setting.}
To achieve on average an accuracy of 30\,nm r.m.s. for the softest mode
$e_{2,1}$ generated by
random force errors, the force setting accuracy has to be of the order of $\pm 0.4$\,N. To
obtain this accuracy 95\% of the times would require an accuracy of
$\pm 0.1$\,N.
Furthermore, to have some margin for this important and delicate part of the active
optics system, the value finally chosen was $\pm 0.05$\,N.\\
{\it Force range.}
The force range was primarily driven by three contributions. First,
the required switch 
from the Nasmyth to the Cassegrain configuration needs active
forces in the range from $-$180\,N to $+$470\,N.
\begin{figure}
     \centerline{\hbox{
      \psfig{figure=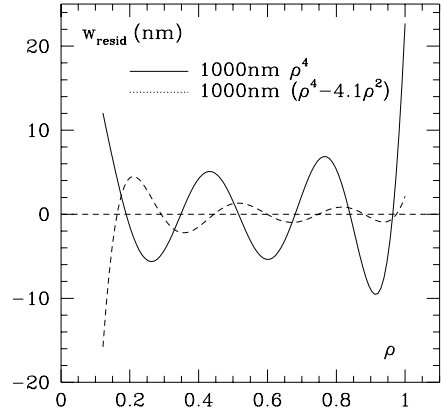,width=60mm}}}
      \caption{\label{fig:calSpher} {\small Differences between the required function
          and the one generated by using five elastic modes.
          {\em Solid line} : pure third order spherical aberration ($\rho^{4}$),
          {\em dashed line} : third order spherical aberration
           combined with defocus ($\rho^{4} - 4.1\rho^{2}$).}}
\end{figure}
Figure \ref{fig:calSpher} shows the residual wavefront errors
$w_{{\rm resid}}$ for attempts to generate
either pure third order spherical aberration ($\rho^{4}$) or third order spherical aberration
combined with an optimised defocus component($\rho^{4} - 4.1\rho^{2}$). The latter gives, as
discussed in \S \ref{sec:NasmythToCassegrain}, a residual r.m.s. of
the wavefront error 4.5 times smaller and also with approximately
45\% smaller maximum forces.
The forces can be further reduced by using less than the maximum
five elastic modes for the correction. For example, using only three
elastic modes reduces the maximum forces to 173\,N but increases the
residual r.m.s. from 46\,nm to 80\,nm.\\
The second major contribution of $\pm 120$\,N are the forces given to
the optical manufacturer for the
correction of low spatial frequency aberrations in form of the active
modes which were not removed during the figuring process of
M1 and M2. The third contribution are forces foreseen
for corrections of aberrations introduced by the support system and,
possibly, by local air effects. The total range of active forces was
then defined as $-500$N to $+800$N.\\
{\it Astaticity and friction.}
A closed loop operation requires a stability of the optical
configuration equivalent to an r.m.s. of the wavefront
errors of $\sigma_{{\rm w}} < 50$\,nm over time periods of
approximately one minute.
The major sources are the non-astaticity of the active
electromechnical actuators and friction effects both in
the lateral supports and the passive part of the axial supports. The
limits for friction were entered into the specifications for the
supports.
The astaticity of the active actuators is directly related
to the spring constant $D_{{\rm s}}$ of the springs in the
electromechanical actuators.
According to finite element calculations, for a change of the zenith angle
of $90^{\circ}$, the deformation of the mirror
cell with a rotational symmetry two due to its own 
weight and due to deformations of the centerpiece are of the order of
$d_{{\rm c}} \approx \pm 350 \, \mu$m at the outer edge of the cell.
The rate of change depends on the sky position.
During one minute the maximum deformation is, at the site of the VLT, 
$d_{{\rm c,minute}} = \pm 0.00343 \cdot d_{{\rm c}} = \pm 1.2\, \mu$m. Owing to the
non-astaticity of the active supports with a spring constants
$D_{{\rm s}}$ the deformations of the cell generate force changes of
$\pm d_{{\rm c,minute}}\; D_{{\rm s}}\, \mu{\rm m}$ over one
minute. These forces will predominantly generate a 
deformation of the mirror in form of the first elastic mode $e_{2,1}$ of
rotational symmetry two. If the forces have, over the area of
the mirror, roughly the functional dependence of this mode,
the coefficients of $e_{2,1}$
can be calculated by dividing the maximum forces at the outer edge
by the maximum calibration force $f_{{\rm max}}$ on the outer ring
needed to generate a specified amount of this mode.
If $\sigma_{2,1,{\rm max}}$ is the tolerable upper limit for the
change of the coefficient of $e_{2,1}$ over one minute,
the condition for the
spring constants of the active supports is given by
\begin{equation}
  D_{{\rm s}} \le \frac{\sigma_{2,1,{\rm max}} f_{{\rm max}}}
                         {0.00343 d_{{\rm c,minute}}}
\end{equation}
With $f_{{\rm max}} \approx 1.7 \,{\rm N}/\mu {\rm m}$ and
$\sigma_{2,1,{\rm max}} = 50$\,nm one obtains
$D_{{\rm s}} \le 0.07\, {\rm N} / \mu{\rm m}$.\\
{\it Coupling to the mirror cell.}
The condition (\ref{eq:condn}) shows that out of the six
elastic modes with the lowest eigenfrequencies the modes
$e_{2,1}$, $e_{3,1}$ and $e_{4,1}$ are non six sector modes, that is
they cannot be generated on a support with six fixed points.
This is, of course, only strictly true
if the support system is infinitely rigid. Otherwise the stiffnesses
of the mirror, of the passive hydraulic support system and of the
mirror cell
have to be properly combined (Noethe [2000]). One then gets for each
mode $e_{m,i}$ a
ratio $\eta_{m,i}$ of the deformations on a six sector support to the
ones on an astatic three sector support. For six sector modes like the
rotationally symmetric modes the ratio is one.\\
With respect to deformations in the form of the lowest elastic mode
$e_{2,1}$ the mirror cell of the VLT is approximately five times stiffer than
the primary mirror. Together with the stiffness of the passive support this
gives a ratio of $\eta_{2,1} = 0.33$. For the second softest mode
$e_{3,1}$ one gets $\eta_{3,1} = 0.70$, whereas the third mode
$e_{0,1}$ is a six sector mode with $\eta_{0,1} = 1.0$. Since the
first three modes account for a large fraction of the deformation
under wind pressure, a six sector support reduces the wavefront
aberration by approximately 50\%.\\
If the valves between the two halves of each sector are fully closed,
the filtering effect of the six sector support
on the non six sector modes applies to all temporal frequencies.
But the valves between the two subsectors
of each of the three sectors must be partially open to enable slow
active corrections of all modes.
In this context one can define a damping frequency $\nu_{{\rm d}}$
as the inverse of the 
relaxation time $t_{{\rm r}}$, that is the time after which an
instantaneously applied pressure difference between the two
subsectors drops to $1/{\rm e}$.
To assure that
90\% of an active optics correction is done after 10 sec, the
conditions for the relaxation times and the damping frequencies
are $t_{{\rm r}} \leq 4$ sec and $\nu_{{\rm d}} \geq 0.25$\,Hz,
respectively. Measurements of wind pressure variations on a 3.5 m
dummy mirror in the NTT enclosure have shown that the maximum of the power
spectrum inside the dome is at frequencies of approximately 2
Hz. Under the assumption
that the mirror can instantaneously follow these pressure variations,
a six sector support with a relaxation time of 4 sec will reduce the
deformations for most of the relevant frequencies. Calculations
with spectra obtained with pressure sensors on a dummy mirror inside
the NTT enclosure have shown that the reduction is at least of
the order of 40\%.
On the other hand, the coupling to the mirror cell over 4 sec will
generate, due to the flexure
of the mirror cell, wavefront errors in the mirror, but these are only of the order
of 17\,nm.
\subsubsection{Lateral support of M1}
\begin{figure}[h]
\centerline{\hbox{
 \psfig{figure=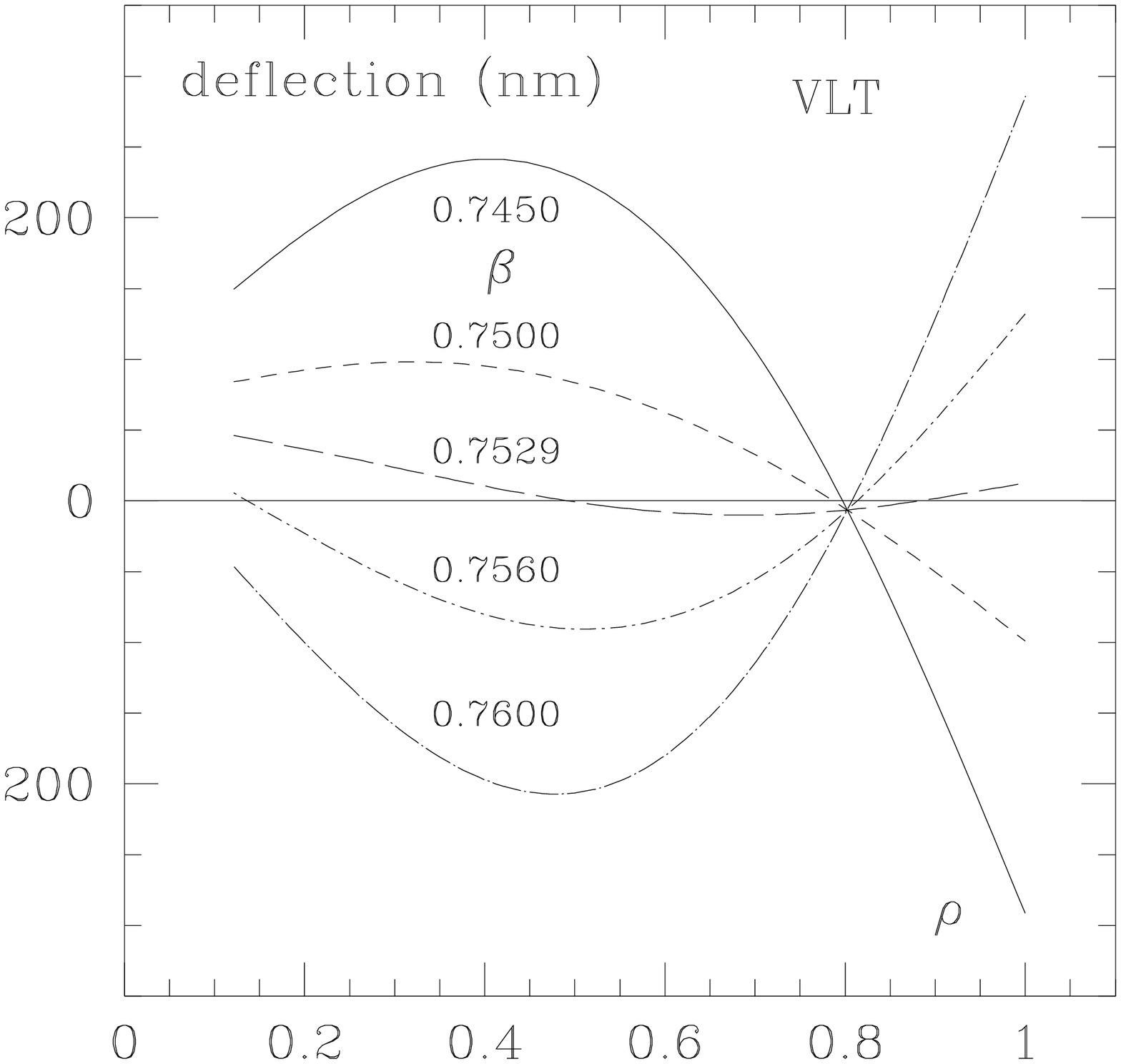,width=50mm}
 \psfig{figure=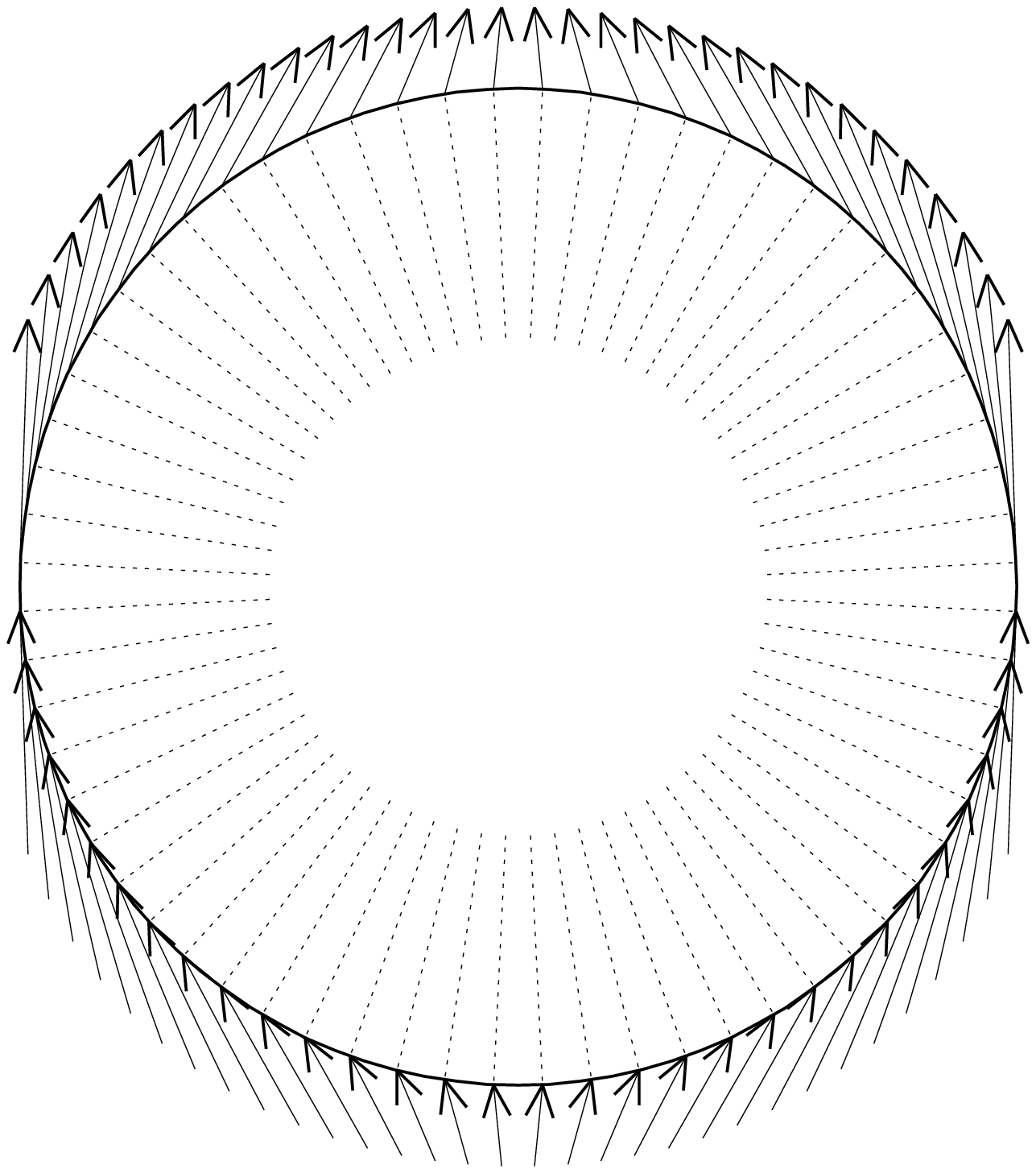,width=60mm}
 \psfig{figure=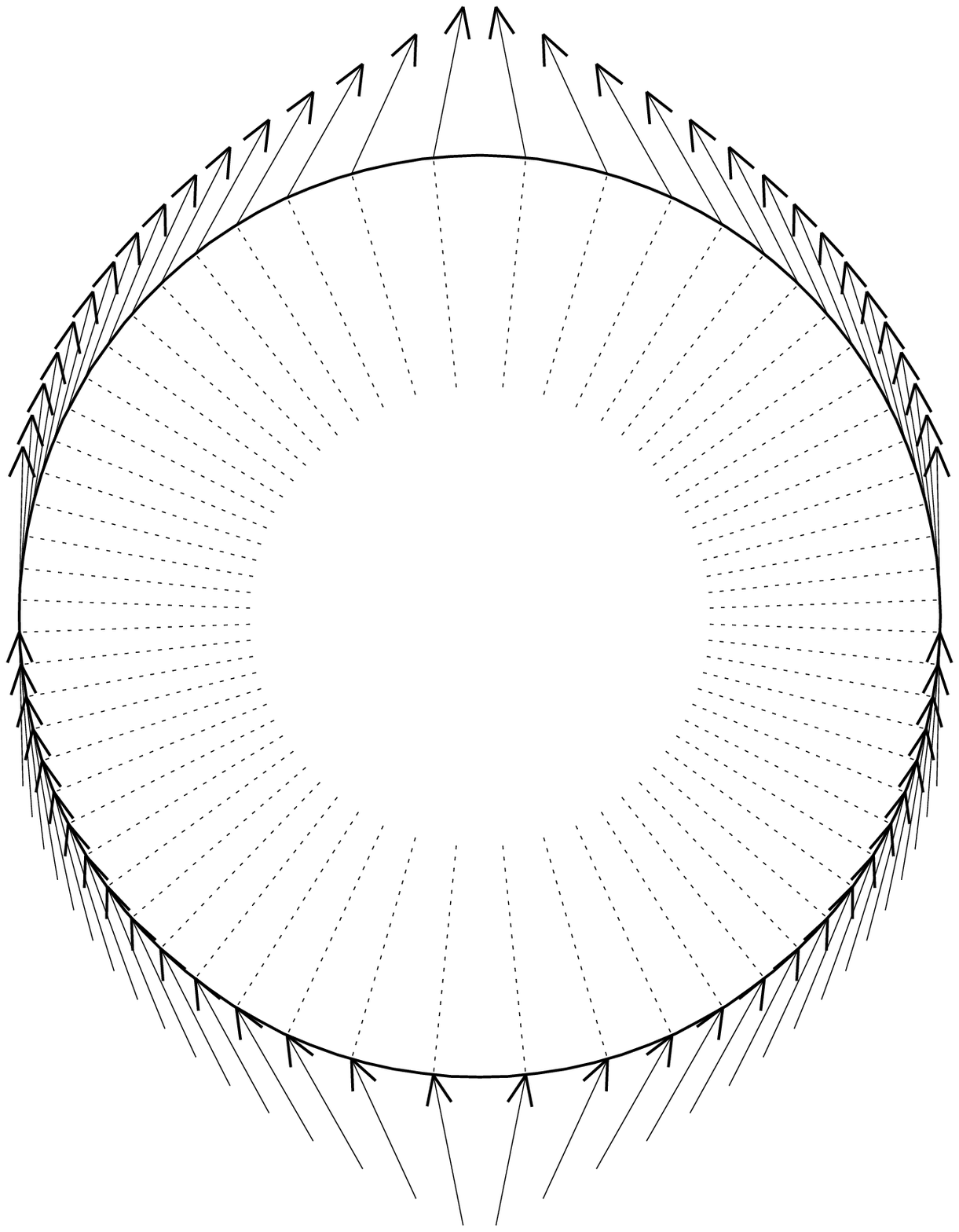,width=60mm}}}
 \caption{\label{fig:latVLT} {\small {\em Left} : Deflections of the
    VLT as functions of the normalised radius $\rho$ for
    various fractions $\beta$ of the weight supported by the
    tangential forces, {\em Middle} Lateral forces in the plane
    perpendicular to the axis of M1 with equidistant support points,
    {\em Right} Lateral forces in the plane
    perpendicular to the axis of M1 with identical moduli.}}
\end{figure}

The VLT primary mirror cannot be supported in the plane of the center
of gravity. Therefore, one needs axial forces around the edge to
balance the moment generated by supporting the mirror at the center of
the outer rim.
The deflections obtained with the standard VLT boundary conditions are
shown in figure \ref{fig:latVLT} (Schwesinger [1991]) on the left.
The r.m.s. values
$\sigma_{{\rm d}}$ of the deflections and $\sigma_{{\rm d,resid}}$
of the deflections after subtracting third order coma are shown in
table \ref{tab:deflLatSuppVLT}.
\begin{center}
\begin{table}
\caption{{\small r.m.s. values $\sigma_{{\rm d}}$ without and
$\sigma_{{\rm d,resid}}$ with subtraction of third order coma of the
deflections generated by the lateral support of the VLT with fractions
$\beta$ of the weight supported by tangential forces.}}
\begin{center}
\vspace{2mm}
\begin{tabular}{|l|r|r|r|r|r|}
  \hlines
   $\beta$ & 0.7450  &  0.7500  & 0.7529 & 0.7560 & 0.7600\\
  [1mm]\hline\tbsp
   $\sigma_{{\rm d}}$ (nm) & 124.4  &  46.5    &  8.7   &  49.2  & 111.5\\
  [1mm]\hline\tbsp
   $\sigma_{{\rm d,resid}}$ (nm) & 19.2 &  12.3 & 8.7 & 6.1  & 7.0\\
  [1mm]\hline
\end{tabular}
\end{center}
\label{tab:deflLatSuppVLT}
\end{table}
\end{center}
As in the case of the NTT, some of the deflections are very
similar to third order coma. 
But, contrary to the NTT, the deformations depend strongly
on the ratio $\beta$. It is therefore
necessary to choose a ratio
$\beta \approx 0.75$ to reduce the deformations to acceptable levels.
The lateral forces projected onto the plane perpendicular to the axis
of M1, that is the vectorial sum of the radial and tangential
components only, are shown in figure \ref{fig:latVLT} in the middle.
Unfortunately, the strong difference between the fractions of the
weight supported by radial and tangential
forces leads, with an equidistant distribution of the lateral
supports as shown in the drawing in the middle of fig. \ref{fig:latVLT},
to three times larger lateral forces and therefore
significantly larger stresses
near the altitude axis than at angles of $90^{\circ}$ from the altitude
axis. The requirements to have not more than 64 lateral supports and
to limit the lateral forces to 4000\,N required a redistribution of the
lateral supports. The new positions $\varphi_{i}$ of the supports $i$
were chosen such that the integrals of the force densities
between $\varphi_{i} - \delta_{i}$ and
$\varphi_{i} + \delta_{i} = \varphi_{i+1} - \delta_{i+1}$,
where $\delta_{i}$ is the identical distance between both the lower
and upper integration bounds and $\varphi_{i}$,
gave identical total lateral forces. The resulting components in a plane perpendicular to
the optical axis are shown in Fig. \ref{fig:latVLT} on the right.
\subsubsection{Position control of M1 and M2}
If the specification for the low spatial frequency errors of
$\sigma_{{\rm t}} = 0.034$\,arcsec is statistically split into three
contributions, namely from the elastic deformation of M1, from defocus
and from decentering coma, an r.m.s. slope error of
$\sigma_{{\rm t}} \approx 0.02$\,arcsec could be allocated to each. For
defocus and decentering coma this would require setting accuracies
with r.m.s. values of $\sigma_{{\rm z}} \approx 1.2 \, \mu$m for
movements in axial direction and $\sigma_{\phi,{\rm coc}} \approx 14$
arcsec for a rotation around the center of curvature, respectively.
The specifications for the mechanical units were
much tighter, namely $\sigma_{{\rm z}} \approx 0.5 \, \mu$m and
$\sigma_{\phi,{\rm coc}} \approx 0.3$\,arcsec. With these accuracies,
which are also achieved in practice, the r.m.s. of the wavefront
errors are $\sigma_{{\rm w,def}} \approx 35$\,nm and
$\sigma_{{\rm w,coma}} \approx 3$\,nm. The control of defocus is
therefore much harder than the one of coma and also of the shape of
the primary mirror. If the mechanical specifications are fulfilled and
the three contributions are added up quadratically, the r.m.s. of the
wavefront error from the low spatial frequency aberrations is of the
order of 50\,nm.\\
Contrary to the secondary mirror of the NTT, the M2 of the VLT can be
moved in all degrees of freedom, which also allows a motorised control
of the alignment of the axes of M1 and M2. Furthermore, with the
capability of a motorised control of the position also of M1 in five
degrees of freedom,
the telescope can be aligned such that the optical
axis goes through the center of the adapter.
\subsubsection{Wavefront analyser}
The Shack-Hartmann analyser of the VLT has a 20 by 20 lenslet array
with lenslets with a side length of 0.5\,mm and a f-number of 45. The
lenslets therefore sample subapertures on M1 with a side length of 400
mm. The pattern fits on a CCD with a side length of 11\,mm without the
use of a reduction optics. With a pixel size of 23\,$\mu$m
all requirements listed in \S \ref{sec:defSHParams} are then
fulfilled for a specified limit of 20\,nm for the r.m.s. of the
wavefront error generated by the noise of the wavefront analyser.
\subsection{Active optics design of the Keck telescope}
\label{sec:actOptDesignKeck}
Each of the 36 segments of the primary mirror is supported by three 12
point whiffletrees. Low spatial frequency aberrations
in the shape of an individual segment, mainly due to the manufacturing
process, can manually be corrected by a warping harness.
Each harness consists of 30 leaf springs, which apply
moments about pivots of the whiffletree (Mast and Nelson [1990]).
Through the use of the springs, the axial support forces can be adjusted
at each of the support points, subject to the equilibrium
conditions that the net forces and moments on the mirror be zero.
The applied forces are also independent of the inclination of the segment.
The optimum 30 pivot moments are calculated with a least squares fit
of the deformations introduced by individual springs to the overall
deformations of the segment, taking into account several hardware
constraints.\\
For the positioning of a segment in three degrees of
freedom, each whiffletree is attached to a displacement actuator.\\
The integration times in the phase camera system (PCS) described
in \S \ref{sec:wfAnalysisSegments} are all of the order of 30 seconds,
using stars of magnitude nine in the passive tilt mode and of
magnitude four to five in the other three modes.
The segment phase mode uses 78 of the 84 segment edge midpoints.
The six points closest to the
center are omitted since the corresponding intersegment edges are
partially obscured by the telescope tertiary tower.
The diameter of the subapertures, centered at
intersegment edges, of 120\,mm is always smaller
than the atmospheric coherence length for infrared wavelengths
$\lambda \ge 2\,\mu$m.\\
A complete alignment of the telescope optics is then done in three steps.
First, the fine screen mode is used to measure and correct the defocus
and decentering coma aberrations introduced by a despace of the
secondary mirror as described in \S \ref{sec:wfAnalysisSegments}.
Without this step, these aberrations would be corrected by a then
non-perfect alignment of the segments of the primary mirror.
Second, either the fine screen mode
or the passive tilt mode are used to stack the images of the 36
segments, that is to correct errors in the tilts of the
segments. Finally, 78 relative piston errors of the segments are
measured with the segment phasing mode. The appropriate piston
movements to correct these errors are obtained from a least squares
fit of the 36 axial movements to the 78 available data with the
constraint of a zero mean movement. A full alignment takes
approximately one hour.
The need for bright stars prevents a
full sky coverage, and the long time required for an alignment
limits active optics correction based on data
obtained with star light to open loop control.\\
For a change of the zenith angle of $90^{\circ}$ the primary mirror
cell deforms primarily in the defocus mode by 170\,$\mu$m r.m.s.,
which is equivalent, in the worst sky position, to a change of 30\,nm r.m.s.
over one second. Since the support of the mirror as a whole is position
based, the positions of the segments have to be adjusted at least
once per second. Active optics corrections therefore have to be done
in open or in a combination of open and closed loop. An open loop control
based on measurements after alignments at different zenith angles is
probably not feasible, since the predictability to an accuracy of the
order of, say 30\,nm, for an overall deformation of 600\,$\mu$m is
not achievable,
above all due to certainly existing hysteresis in the deformation
of the mirror cell.
The active optics system of the Keck telescope therefore
works in two hierarchical levels. A lower level controls the
shape of the primary mirror by an internal closed loop, based on
internal measurements of the relative positions of the mirror,
and an upper level controls the residual deformations of the primary
mirror and the alignment of the primary and secondary mirrors in
continuous open and periodic closed loop based on measurements with
star light.
For the lower level control, capacitive devices measure
the changes in the relative height of
adjacent segments in the direction normal to the surfaces at
intersegment boundaries. Two sensors are located at every
intersegment edge close to the end of the edges.
After an alignment the readings of the, in total, 168 sensors are
stored as reference values. During operation the actual readings of
the sensors have to kept as close as possible to the reference values.
The required movements of the 108 actuators, maintaining the average
tilt and piston of all segments, are calculated from the
168 differences of the sensor readings via a least squares fit.
The corrections are done twice per second.
The quality of the correction depends, apart from the noise in the actuators,
predominatly on the characteristics and noise of the sensors.
The dependence of the sensor readings on the inclination and on the temperature
generates mirror deformations in the defocus mode. But these dependences can
be accurately calibrated. The unavoidable random noise in the readings will
only introduce random errors in the shape of the mirror.\\
Without any other systematic error sources, the shape of the mirror would be
stable, and the mirror could be regarded as a passive element without the
need for correcting the shape in the upper level active optics loop.
But systematic error sources exist in the form of drifts of the sensor readings
and other unknown effects.
Whether the corrections of the ensuing wavefront aberrations
can be done in open or closed loop depends the predictability
and stability of the errors. On the one hand, the unknown effects may be predictable,
for example from measurements of the deformations
as functions of the zenith distance. They are then correctable in
open loop, which in practice would be equivalent to a change of the reference values
of the sensor readings as functions of the zenith angle.
On the other hand, the drift of the readings is usually not predictable and
requires closed loop corrections, that is a new alignment based on measurements
with the PCS. The upper active optics level therefore consists on the
one hand of continuous
open loop corrections of the primary mirror and also of the alignment
of the secondary mirror, and on the other hand of closed loop
realignments of the primary mirror at longer time intervals, typically
of the order of one month.\\
The active optics systems in the Keck telescopes with their segmented primary
mirrors and the NTT and VLT telescopes with their thin meniscus mirrors with
force based supports are in principle similar, if the role of the basic
astatic support of a thin meniscus mirror is seen as equivalent to the lower
level closed loop control of a segmented mirror.
Both attempt to provide, at least to first
approximation, a stable shape of the primary mirror independent of the
inclination of the telescope. Whereas in the Keck telescope
the residual errors, as well as the alignment of the two mirrors,
are corrected in continuous open and sporadic closed loop,
in the NTT and VLT this is done in closed loop.
\section[Practical experience with active optics]{Practical experience
with active optics at the NTT, the VLT and the Keck telescope}
\label{sec:practicalExperience}
\subsection{Intrinsic accuracy of the wavefront analysis}
The intrinsic quality of the wavefront analysis depends strongly on
the number of photons in the brightest pixel of any of the
Shack-Hartmann spots. It can be measured by comparing two
Shack-Hartmann patterns, both generated with the reference light, but
with different light levels.
With maximum pixel values of the order
of a third of the saturation of the CCD, the coefficient of the mode
$e_{2,1}$ due to intrinsic measuring noise is less than 10\,nm and
therefore negligible compared with the variations of this coefficient
introduced by the air, even for integration times of 30 sec as
described in \S \ref{sec:openLoop}.\\
In the Keck telescope the relative piston wavefront values of adjacent
segments can be measured with an accuracy of 50\,nm in the broadband
and 12\,nm in the narrowband mode. The accuracy of the tip-tilt
measurements of the segments is of the order of only
 $d_{80} \approx 0.03$\,arcsec.
The uncertainties in the measurements of error of the segment
deformations with the ultra fine mode are of the order of
$d_{80} \approx 0.065$\,arcsec, or 20 to 25 nanometers r.m.s. for the
lowest second order Zernike modes.
\subsection{Active optics operation at NTT and VLT}
\subsubsection{NTT}
The NTT suffered from spherical aberration which
was caused by incorrect polishing of the primary
mirror due to an error in the assembly of the null lens.
In terms of third order spherical aberration the wavefront error
was of the order of $3500 \, {\rm nm}\; r^{4}$. This error alone
generated a point spread function with $d_{80} \approx 0.7$\,arcsec
exceeding the specification of $d_{80} = 0.4$
arcsec for an operation
in the passive mode. Without the use of the active optics system the 
primary mirror would have had to be repolished.\\
A correction required forces of 420\,N with the calibration
forces calculated by Schwesinger [1988] and 240\,N with a calibration
using the two lowest elastic modes of symmetry zero.
The force range for corrections with the mechanical levers for a
zenith distance $\theta_{{\rm z}}$ is approximately
$\pm 0.3\cdot 800\, \cos{\theta_{{\rm z}}}$\,N.
A correction of spherical aberration with the force adjustments of the
levers would therefore have been possible near the zenith
only, with little reserves left for the correction of other
aberrations.\\
Instead of using the adjustable counterweights, the bulk of the error is
therefore corrected with the springs which supply correction forces
independently of the zenith angle.
But, another problem caused by the strong correction forces
remains. Since the axial support system of the NTT is a pure push
system, a negative correction force $F_{{\rm corr}}$ at a given
support cannot be higher than the gravitational load $F_{{\rm G}}$ at
this support. Since the maximum negative active forces for the
correction of spherical aberration alone are of the order of $-200$\,N,
the maximum usable zenith angle $\theta_{{\rm z,max}}$ defined in
\S \ref{sec:actAxSuppM1} is at most of the order of $75^{\circ}$.\\
With a thickness of 241\,mm of the primary mirror the NTT can, under
average seeing condition, be operated with a few corrections per
night. But, under good seeing conditions of, say, $\Theta = 0.5$
arcsec the active optics system should operate in closed loop.
It has been shown that the optical quality of the NTT can then reach
the specification of $d_{80} = 0.15$\,arcsec for an operation in the
active mode (Wilson, Franza, Noethe and Andreoni [1991]).
\subsubsection{VLT}
According to the scaling laws 
for the wavefront and slope errors in
eq. (\ref{eq:scalingRandomForces}),
the flexibility of the primary mirror of the VLT exceeds the one
of the NTT by factors of 37 and 16, respectively. Therefore, the VLT has to be
operated in the active mode all the time, even under bad seeing
conditions. In principle, the corrections could be done in open or in
closed loop. Since the closed loop corrections work well with an extremely
low failure rate, initial open loop corrections are done only after
presets to new sky positions.\\
Just after the installation of the telescope, a single manual
intervention may be necessary to reduce the wavefront aberrations to
levels which allow analyses with the wavefront analyser and therefore
automatic corrections. The reason is, that
without any correction forces, that is when the mirror is
supported by the passive hydraulic system alone,
the transverse aberrations in the focal
plane may be so strong that the Shack-Hartmann pattern is heavily
distorted. Consequently, a significant number
of the spots may be vignetted by the
mask in the Shack-Hartmann sensor. It is
then necessary to remove manually, for an arbitrary zenith angle in a
trial and error mode, the bulk of the two largest
aberrations, namely third order coma and the lowest elastic mode of
rotational symmetry two. The coefficients of these modes can be estimated
from defocused images.
This may take an hour, after which the
transverse aberrations are sufficiently small to be analysed
automatically. Such a manual intervention is therefore
only necessary once after the installation of the telescope.
Accurate coefficients of the active modes for the initial open loop
correction after presets will then, for all zenith angles, be obtained with
the wavefront analyser, and stored as a lookup
table in the database.\\
After a preset to a new position in the sky the images are, without a
correction, visibly deformed. For example, if the image is well
corrected near the zenith, a change of the zenith angle of
$45^{\circ}$ will generate images with $d_{80} \approx 2$\,arcsec.
Although the corresponding wavefront errors, which are dominated by
the mode $e_{2,1}$, would not cause
any problems for the automatic wavefront analysis,
a first correction is always done in open loop based on the look-up
table mentioned above.
Afterwards continuous closed loop corrections will be started.
Since the maximum pixel values depend on the
magnitude of the guide star, its colour and the current seeing, the
actual integration time is adapted to reach for the brightest pixels
a level of at least 50\%
of the saturation level. If the CCD saturates with integration times
of 30 seconds, exposures with shorter
integration times are averaged. If the maximum pixel counts are too low, the
integration time may be increased up to 60 sec. Stars with magnitudes of
the order of 12 to 13, which already guarantee a full sky coverage,
are ideal, although stars with magnitude as
faint as 15 may, depending on the colour, be usable.
In addition the primary
mirror is kept in a fixed position with respect to the M1 cell by
changing the oil volumes in the axial and lateral hydraulic sectors of the
M1 support.\\
Approximately 12000 wavefront analyses and corrections are done on each
telescope per month.
All relevant data, in particular
the coefficients of the modes and the residual r.m.s. $\sigma_{{\rm resid}}$,
are logged for further off-line processing.
Apart from the correction of the optics these measurements are also an
important maintenance tool, since they can detect errors in the
telescope optics which may, because of the
strong influence of the atmosphere, not be easily visible in the final
image.\\
An important feature of the VLT is the control of the temperature
of the primary mirror by a cold plate under its back surface and of the air
inside the enclosure also during the day by a ventilation
system (Cullum and Spyromilio [2000]). Both temperatures are set to
the outside temperature expected at the beginning of the night and,
during the night, the mirror temperature is equilibrated to the
normally falling temperature of the ambient air with the cold plate.
The temperature differences are most of the time within a
narrow band of $\pm 1^{\circ}$, for which the effects of dome and mirror
seeing on the image quality are insignificant (Guisard and Noethe [2000]).
\subsection{Closed and open loop performance of the VLT}
\subsubsection{Purity of modes generated during correction}
An important criterion for the functioning of the corrections is the
purity with which the active modes can be generated. This can
be checked by generating large wavefront errors in a single mode and
measuring the generated coefficients of all modes. If at all,
crosstalk will mostly
occur into lower modes of the same symmetry, and most important, into
the softest mode $e_{2,1}$. Several measurements have to be averaged to
distinguish real crosstalk from the normal variations of the
coefficients generated by the air as described in
\S \ref{sec:openLoop}. The strong crosstalk of 47\% from the mode
$e_{7,1}$ into the mode $e_{2,1}$ mentioned in
\S \ref{sec:axialSupportVLT} could be verified. Other crosstalk of
the order of 20\% exists from some modes of higher order into lower
order modes of the same rotational symmetry or into the softest mode
$e_{2,1}$, that
is from $e_{4,2}$ into $e_{4,1}$ and $e_{2,1}$, from $e_{2,3}$ into
$e_{2,2}$ and $e_{2,1}$, and from $e_{0,3}$ into $e_{0,2}$ and
$e_{0,1}$. Since the coefficients of these higher order modes are
always small, the crosstalk is not significant. For the rest of the
active modes the crosstalk into other modes is smaller than 10\% and
therefore also negligible.
\subsubsection{Wavefront variations without corrections}
\label{sec:openLoop}
\begin{figure}[h]
     \centerline{\hbox{
      \psfig{figure=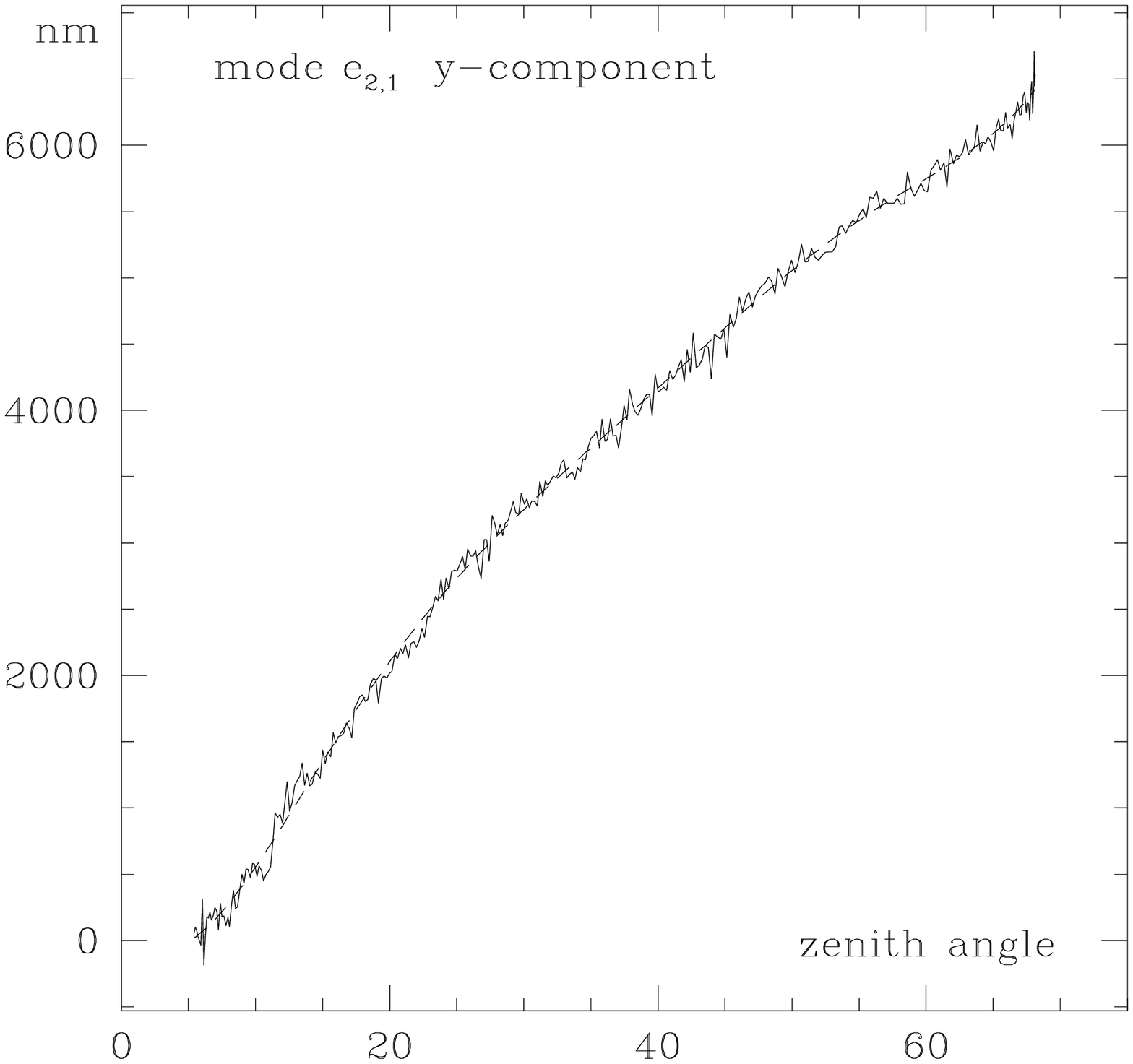,width=60mm}
      \psfig{figure=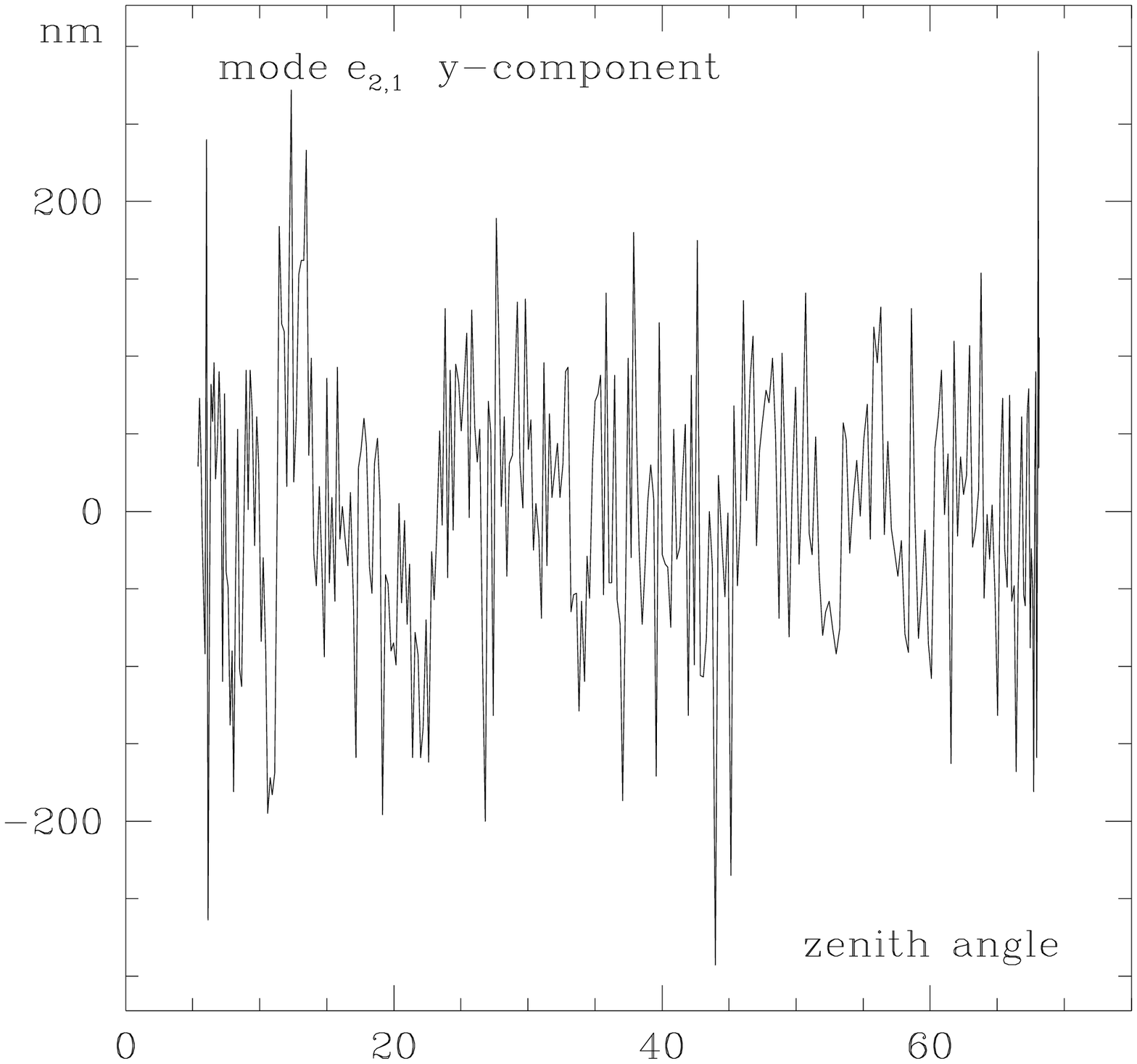,width=60mm}}}
   \caption{\label{fig:driftEla21y} {\small {\em Left :} y-component
      of the coefficient of the elastic mode $e_{2,1}$
     as a function of the zenith angle. The dashed
     line is a best fit of a sixth order polynomial. {\em Right :}
     Residual variations after the subtraction of the fitted polynomial.}}
\end{figure}

\begin{figure}
     \centerline{\hbox{
      \psfig{figure=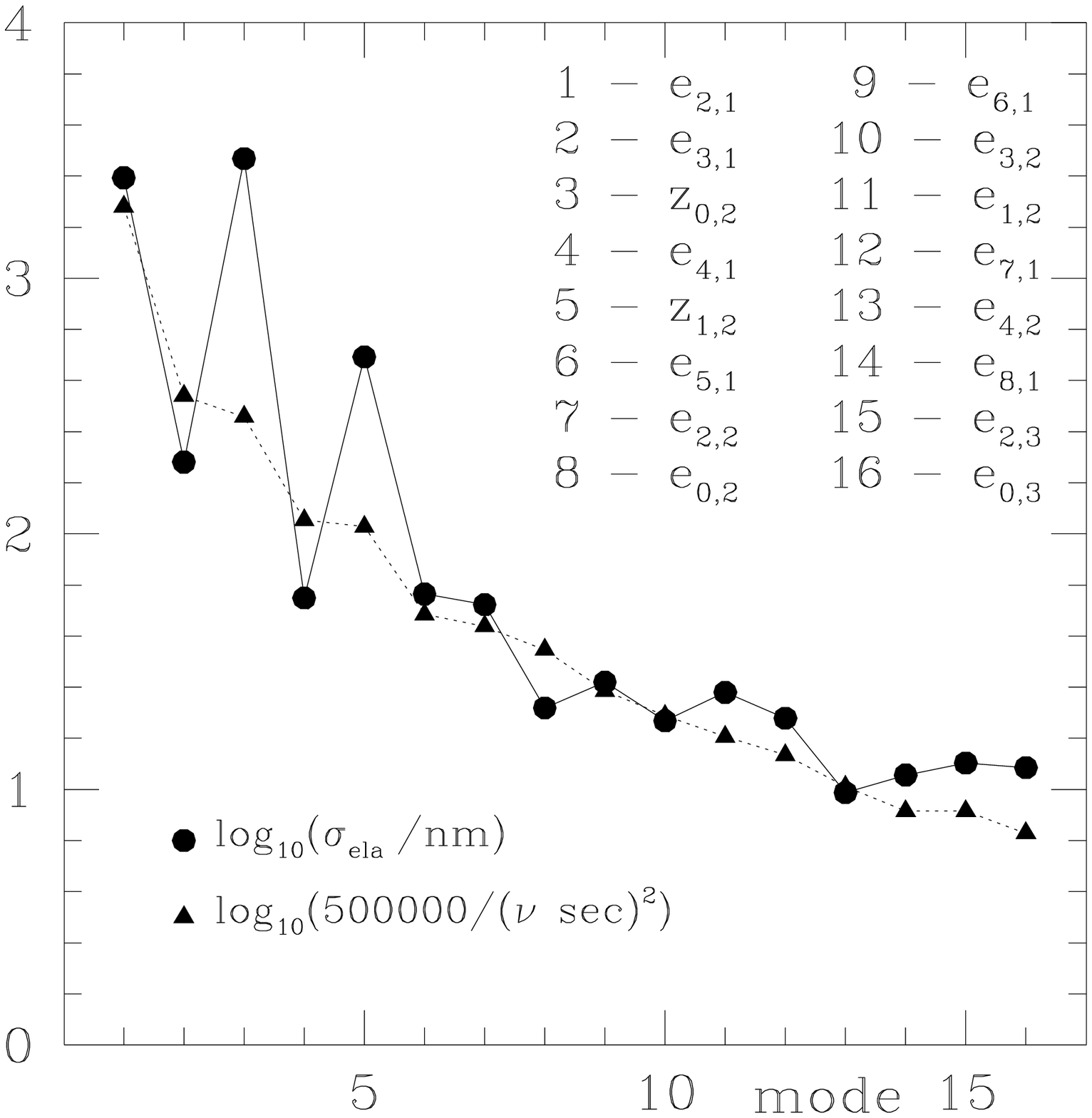,width=60mm}
      \psfig{figure=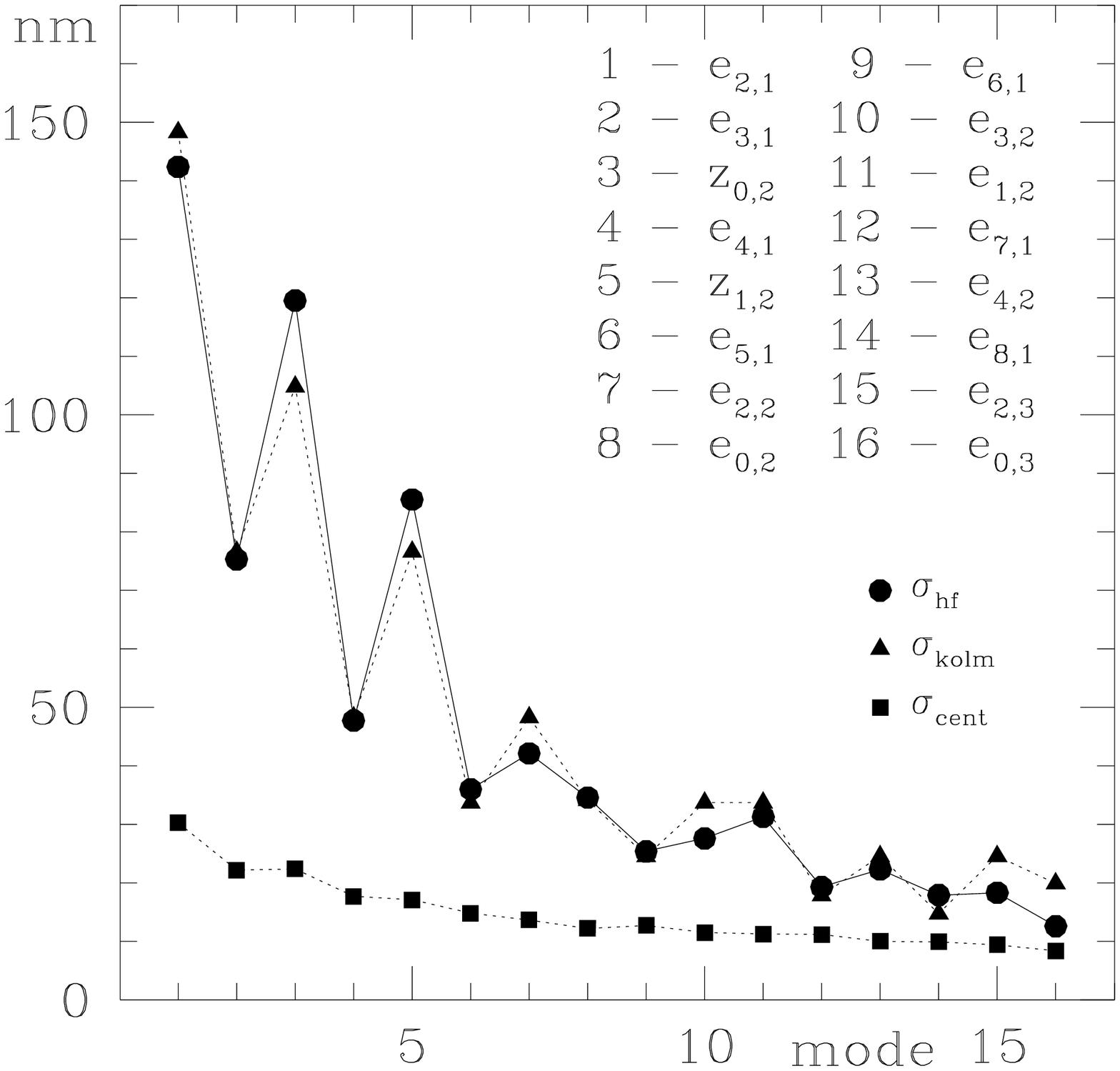,width=60mm}}}
   \caption{\label{fig:zenith} {\small {\it Left :} Correlations between
     inverse stiffness of modes and measured r.m.s. of the change of
     the coefficient. {\it Right} Correlations between expected
     r.m.s. values of the variations of the coefficients due to centroiding
     errors ($\sigma_{{\rm cent}}$) or Kolmogoroff 
     turbulence ($\sigma_{{\rm kolm}}$) and the r.m.s.
     $\sigma_{{\rm hf}}$ of the measured high frequency
     non-elastic variations.}}
\end{figure}

To measure the evolution of the wavefront errors primarily as a
function of the zenith angle, wavefront measurements have been done,
without performing any corrections, following a star going through
a position close to the zenith.
The y-component of the coefficient of the elastic mode $e_{2,1}$
obtained during such a drift measurement, which started at a position
near the zenith, is shown in figure \ref{fig:driftEla21y}
on the left.
Its evolution can clearly be
separated into a smooth low temporal frequency variation, representing
elastic effects, and high temporal frequency variations, representing
primarily, as will be shown later on, atmospheric effects.
The low temporal frequency behaviour is obtained by fitting a
sixth order polynomial, indicated by the broken line in
fig. \ref{fig:driftEla21y} on the left.
The difference between the measured data and the fitted curve is shown in
fig. \ref{fig:driftEla21y} on the right.
The average of the residual r.m.s. $\sigma_{{\rm resid}}$ during this
measurement was approximately 100\,nm.
From these data one can calculate r.m.s. values
$\sigma_{{\rm ela}}$ of the low
and $\sigma_{{\rm hf}}$ of the high temporal frequency variations
of the mode $e_{2,1}$, and similarly of all other measured modes.
These r.m.s. values $\sigma_{{\rm ela}}$ and $\sigma_{{\rm hf}}$
are displayed as circles in fig. \ref{fig:zenith} on the left
and on the right, respectively. Because of the large variation of the
figures for $\sigma_{{\rm ela}}$ for different modes, the data in
fig. \ref{fig:zenith} on the left have been plotted on a logarithmic
scale.
The figures for $\sigma_{{\rm ela}}$ should be
compared with parameters which describe the expected elastic overall
variation of the coefficients with the zenith angle, and those of
$\sigma_{{\rm hf}}$ with the expected measuring noise due to
centroiding and to high temporal frequency variations generated by the
air.
\\
{\it Elastic low temporal frequency variations}\\
According to eq. (\ref{eq:dlam}) the energy of a certain mode for a unit
r.m.s. deformation, and therefore also its stiffness are proportional
to the square of the eigenfrequency of the mode.
As a consequence, the coefficient of any mode contained in a deformation
generated by random forces will be inversely proportional to the
stiffness of this mode. In fig. \ref{fig:zenith} on the left the
inverse figures of the squares of the eigenfrequencies are represented by
triangles. The arbitrary scale factor of 50000 mentioned in the plot
has been chosen such that, with the exception of defocus and coma,
the triangles fall close to the corresponding circles, representing
the measured elastic r.m.s. variations $\sigma_{{\rm ela}}$.
The measured coefficients of $e_{3,1}$ and $e_{4,1}$ are smaller
than expected, whereas the coefficients of Zernike defocus and coma
are much larger than expected from the elastic properties of the
primary mirror. These comparatively large variations of defocus and
coma are due to the deformations of the structure
rather than elastic deformations of the primary mirror. Coma, in
addition, could also be generated by systematic effects in the
lateral support system.\\
The proportionality between the variations of the coefficients of the
modes and the inverse of their stiffnesses indicates that the
deformations are not generated by systematic effects, but rather by
random force errors. The coefficient of $e_{2,1}$ changes by
approximately 7000\,nm for a change of the zenith angle of
$45^{\circ}$. Assuming that the change is equally generated by random
force errors in the axial and the lateral support, the random force
errors in the axial support would be in the range of $\pm 60$\,N, which
is approximately 4\% of the nominal axial loads at zenith.\\
{\it Non-elastic high temporal frequency variations.}
The coefficients $\sigma_{{\rm cent}}$ of the considered modes
expected in a spot diagram generated by random
centroiding errors with an r.m.s. of 2\% of the pixel size are
represented by squares in fig. \ref{fig:zenith} on the right.
Not only the moduli, but also the relative
values of the coefficients expected from centroiding errors
are very different from the corresponding measured figures
$\sigma_{{\rm hf}}$, represented by the circles.
In particular the comparatively large
values for the low order coefficients
indicate that, at least for the lowest, say, eleven
modes, the measurements are not limited by the accuracy of the
wavefront analysis.
Instead, the relative ratios agree nearly perfectly
with the ones expected from fully developed Kolmogoroff
turbulence (Noll [1976]). This indicates that the high temporal
frequency variations of the wavefronts averaged over 30 seconds
are predominantly generated by the air.
For an assumed atmospheric coherence length or
Fried parameter of $r_{0} \approx 500$\,mm, the expected coefficients,
represented by triangles in the figure on the right,
are nearly identical to the measured coefficients.
\subsubsection{Open loop performance}
\label{sec:openLoopPerformance}
The performance of the telescope achievable with open loop active optics
corrections can be measured by presetting the telescope to zenith
distances in the range from $5^{\circ}$ to $65^{\circ}$ in steps of
$15^{\circ}$ several times down and up.
At each position the calibrated forces for the
specific zenith angle are applied and the wavefront aberration is
measured. The two important recorded figures are, at each zenith
angle, the differences $\Delta c_{{\rm hl}}$ between the
average values of the coefficients measured after a preset from a
previously higher or lower zenith angle, and the r.m.s. values of the
scatter of the coefficients around the average values.
Not considering the results for defocus, which are, in addition to
elastic effects, strongly influenced by temperature variations,
the only significant differences $\Delta c_{{\rm hl}}$ are of the
order of 600\,nm for $e_{2,1}$, 90\,nm for $e_{2,1}$ and 15\,nm for
$e_{4,1}$. The ratios follow quite well the ratios of the
corresponding stiffnesses. Similarly, the most important scatter comes
from $e_{2,1}$ with $\sigma \approx 300$\,nm, which has to be compared
with the scatter of $\sigma \approx 150$\,nm introduced by the
atmosphere. The additional errors arising from an open loop operation
therefore generate, given a good calibration of the active forces with
the zenith
angle and in the case of defocus with the relevant temperatures,
slope errors of the wavefront with $\sigma \approx 0.05$\,arcsec.
This is equivalent to a diameter $d_{50} \approx 0.1$\,arcsec for the
diameter of the circle containing 50\% of the geometrical energy,
which would have to be added
quadratically to the FWHM of the image obtained 
with an optimum closed loop correction. An open loop operation of the
VLT is therefore, with only a minor reduction of the image quality,
feasible.
It would still be mandatory to make frequent wavefront
analyses to detect any changes in the optomechanical system
invalidating the look-up table used for the open loop corrections.
\subsubsection{Closed loop performance}
\label{sec:closedLoop}
\begin{figure}
     \centerline{\hbox{
      \psfig{figure=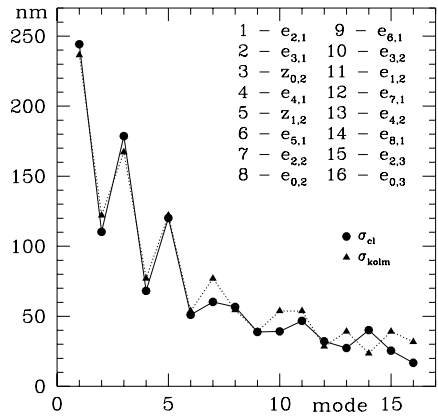,width=60mm}}}
   \caption{\label{fig:closedLoopRms} {\small Correlations between expected
     r.m.s. values of the variations of the coefficients due to
     Kolmogoroff turbulence ($\sigma_{{\rm kolm}}$) and the r.m.s.
     $\sigma_{{\rm cl}}$ of the coefficients in closed loop operation.}}
\end{figure}

Closed loop measurements showed that, on average, the coefficients of
the modes are proportional to the residual r.m.s. $\sigma_{{\rm resid}}$ 
of the wavefront error after the subtraction of the contributions
from the active modes (Guisard and Noethe [2000]). The
constants of proportionality are therefore a measure of the average
content of the modes in the measured
wavefront. Fig. \ref{fig:closedLoopRms} shows the r.m.s. values of the
coefficients measured in closed loop operation. To
be able to compare the data with the ones measured without
corrections, where during the drift measurement the residual
r.m.s. $\sigma_{{\rm resid}}$ was on
average of the order of 100\,nm (see \S \ref{sec:openLoop}),
the values $\sigma_{{\rm cl}}$ plotted in fig. \ref{fig:closedLoopRms}
also correspond to $\sigma_{{\rm resid}} \approx 100$\,nm.
Apparently, the correlation with the r.m.s. $\sigma_{{\rm kolm}}$ of
the coefficients expected from pure Kolmogoroff type turbulence,
represented with an appropriate scaling by the triangles, is as
good as in fig. \ref{fig:zenith} on the right. The only difference
between the two plots is that the coefficients from the closed loop
operation are, on average, 50\% larger than the ones measured without
corrections. Therefore not only the wavefront errors generated by the
air itself, but also the wavefront errors generated by the telescope
optics follow the Kolmogoroff statistics. This indicates, that the
errors in the telescope optics are mainly introduced through
the corrections, based on the measured wavefront errors
generated predominantly by the air.
Other mechanical error sources are effectively negligible,
since they would generate predominantly the lowest mode  $e_{2,1}$.
The coefficients of this mode would then, compared with the ones of
the other modes, be larger than expected from Kolmogoroff statistics.
As the correlation between successive data in fig. \ref{fig:driftEla21y}
on the right is only of the order of 25\%, the high temporal frequency
variations of the coefficients are effectively random.
A correction based on these data will therefore increase the r.m.s. of
the wavefront error by $\sqrt{2}$, but will not change the ratio of
the coefficients. This is exactly what can be seen from a comparison
of the plots in fig. \ref{fig:zenith} on the right and
fig. \ref{fig:closedLoopRms}.\\
Summing up quadratically the values of all the coefficients in
fig. \ref{fig:zenith} on the right, the r.m.s. of the total low
spatial frequency wavefront errors
generated by the telescope optics 
itself is on average of the order of 250\,nm.
Whereas the wavefront error due to
the air cannot be eliminated with active optics, the error in the
telescope optics can be avoided by using filtered data for the
correction. The filter should eliminate the high temporal
frequency variations due to the air,
and the active optics corrections would then follow the
dashed line in fig. \ref{fig:driftEla21y} on the left.
The quality of the telescope optics would then,
apart from the alignment, be limited by the accuracies of the force
and position settings, and
according to the specifications for the correction devices the
remaining wavefront aberrations would have r.m.s. values of the order of
50\,nm. The telescope optics itself can therefore approach
the diffraction limit even with the measuring limitations imposed by
the atmosphere.
As a consequence, the r.m.s. wavefront errors measured in closed loop
operations should be reduced by a factor of approximately $1/\sqrt{2}$,
and should then be identical to the ones of the high temporal
frequency aberrations shown in fig. \ref{fig:driftEla21y} on the
right.\\
The importance of a filter to remove the quickly varying local air
effects from the closed loop corrections is obvious also from three
other considerations. First, a comparison of the two plots of
fig. \ref{fig:zenith} shows that for all modes above mode 7,
i.e. $e_{2,2}$, the r.m.s. values of the high
temporal frequency variations are larger than the r.m.s. of the smooth
variations from small to large zenith angles. For these modes the
corrections are therefore much larger than the elastic variations
during the time between two corrections. It would even be sufficient
to correct these
modes only once after a preset to a new sky position. Second, the
maximum correction forces generated by the variations of the coefficients due
to local air effects are, if all modes except
$e_{7,1}$ and $e_{8,1}$ are corrected, of the order of 30\,N.
Although the support system
can apparently, despite the large force changes,
accurately correct even the softest mode $e_{2,1}$,
such correction with strong and frequently changing forces should
be avoided. If only the modes $e_{2,1}$, $e_{3,1}$ and $e_{4,1}$ were
corrected, the maximum forces would be of the order of 2.5\,N,
and if the corrections of all modes were based on filtered data,
the maximum forces can be further reduced to approximately 0.3\,N.
Third, the optical quality of the telescope optics itself due to low spatial
frequency errors is, if the corrections are based on unfiltered data,
approximately given by the quadratic sum of the r.m.s. values of the
slopes corresponding to the coefficients given in
fig. \ref{fig:zenith} on the right. The sum is equal to
$\sigma_{{\rm t}} \approx 0.06$\,arcsec, which is about twice as large
as the specified value for the active optics control errors of 0.034
arcsec given in \S \ref{sec:genReqSpecsAO}.\\
The best image obtained so far with the VLT with a long exposure time of
5 minutes had a FWHM of 0.25\,arcsec, during which time five closed loop
active optics corrections were done, and the best image with a short
exposure time of a few seconds had a FWHM of 0.18\,arcsec.
The latter figure is certainly close to the ultimate limit achievable
with ground based active telescopes. To approach the much smaller
diffraction limit of 0.015\,arcsec FWHM one has to use adaptive optics
in addition to correct the fast aberrations introduced by the atmosphere.
\subsection{Alignment of the VLT}
A misalignment introduces, in addition to the normal coma and
astigmatism effects, third order astigmatism with a linear field
dependence. This is, since the wavefront is measured with the guide
star in the field, interpreted as field independent astigmatism, and,
with an active optics correction, incorrectly introduced at the center.
The error at the center should be significantly smaller than the
error of $\sigma_{{\rm t}} = 0.034$\,arcsec allocated to the active
optics control, say $\sigma_{{\rm t}} = 0.01$\,arcsec,
which is roughly equivalent to a coefficient of 180\,nm of 
third order astigmatism $r^{2}\cos{2\varphi}$.
From eqs. (\ref{eq:z4sysGen}) and (\ref{eq:z5sysGen}) with the value
of $B_{1}$ for the VLT and a field angle of 13 arcmin the tolerable
misalignment angle $\alpha$ is then of the order of 30\,arcsec.\\
With just one wavefront analyser the misalignment angle can only be
detected by measuring the astigmatism at a minimum of two positions in the
field. Noise is introduced by two effects. First, inaccuracies in
the measurement of the coefficient of coma give rise to movements of
the secondary mirror which modify the true misalignment angle with
every correction.
Second, inaccuracies in the
measurement of the coefficient of astigmatism lead to errors in the
misalignment angle calculated from eqs. (\ref{eq:z4sysGen}) and
(\ref{eq:z5sysGen}). Together, they generate, under average conditions, errors
for $\alpha$ of the order of 50\,arcsec.
Simultaneous measurements with two wavefront analysers in the center
and in the field reduce the errors in the coefficients of astigmatism,
and consequently the error in $\alpha$ to about 30\,arcsec. On the
assumption that the misalignment does not show strong hysteresis,
the angle $\alpha$ can be determined as a function of the zenith angle
and be corrected in open loop. Otherwise, two permanently installed
wavefront analysers in the field are required for a closed loop
control of the alignment.
\subsection{Plate scale control}
If the plate scale were measured in a closed loop operation every
minute with integration times of one minute, its variations could be
similar to variations of the coefficients shown in
sec. \ref{sec:openLoop}, split into two contributions.
On the one hand, smooth systematic variations of the plate scale
as a function of zenith angle similar to the dashed line in
fig. \ref{fig:zenith} on the left will be generated by deformations of
M1 in form of the first mode of rotational symmetry zero, which is
similar to defocus, and a correction of this defocus by an axial
movement of M2. If this smooth variation can be measured as a function
of the zenith distance, it can be corrected in open loop.
On the other hand, fast defocus changes introduced by the air have
optically the same effect as a change of the curvature of M1 and will,
with a closed loop operation, also be corrected by an axial movement
of the secondary mirror, generating fast, random variations of the
plate scale.
Relative variations of the plate scale of approximately $10^{-5}$
at the VLT, measured with integration times of the 
order of one minute, can quantitavely be explained by defocus
variations due to the air.
By filtering the measured defocus data as described in
\S \ref{sec:closedLoop} these variations can be reduced by a factor
of $\sqrt{2}$. But for long exposures the random variations of the
plate scale are also, to some extent, averaged out.
\subsection[Active optics performance of the Keck
telescope]{Active optics operation and performance of the Keck
telescope}
A realignment of the telescope optics, which takes about one hour, is
done approximately once per month. The position of the secondary
mirror is controlled in open, whereas the shape of the primary mirror
is controlled with an internal closed loop running at 2\,Hz.\\
The combined effect of the wavefront errors attributed to the active
optics control system, which are independent of
the zenith distance, is $d_{80} \approx 0.28$\,arcsec. This is
dominated by the long term drift in the readings of the piston
sensors. 
Both the sensor noise and the actuator noise are of the order of 5\,nm
r.m.s., which, with random errors, would lead to an image blur with
$d_{80} = 0.027$\,arcsec.
The errors in the alignment of the individual segments are
$d_{80} \approx 0.10$\,arcsec in tilt and approximately 30\,nm r.m.s. in
piston. 
The distribution of the tilt errors over the segments is not entirely
random. The ratios of the coefficients of the normal modes in the
expansion of the wavefront error follow quite well the ratios of the
eigenvalues of the control matrix described in
\S \ref{sec:alignmentSegments}. The absolute values of the
coefficients immediately
after a mirror alignment are approximately three times larger than
expected from the random sensor noise, and increase by another factor
of two after 11 days without a realignment.
The r.m.s. of the edge discontinuities is 76\,nm, with
contributions of 24\,nm from tilt errors, 23\,nm from phasing errors
and, the largest figure, 59\,nm from segment aberrations.\\
The zenith distance dependent aberrations are, with the internal closed
loop control of the primary mirror and accurate calibrations of the
dependences of the piston sensor readings on inclination and
temperature, reduced to $d_{80} \approx 0.5$\,arcsec at a
zenith distance of $55^{\circ}$.
The deformations in non-focus-modes are, as functions of
the zenith angle, quite predictable. With a correction of these
aberrations in open loop, the image blur due to zenith-dependent active
optics control errors is anticipated to be of the order of $d_{80}
\approx 0.03$\,arcsec.
\section{Existing active telescopes}
\label{sec:existingAOTelescopes}
Apart from the telescopes NTT, VLT and Keck already mentioned,
other modern large telescopes also rely on active optics corrections.
The largest group of these telescopes are the ones with solid thin
meniscus primary mirrors. The active optics of the 2.5\,m Nordic
Optical Telescope (NOT) is operated in pure open loop. The 3.5 m
Galileo telescope (TNG) is similar in
design to the NTT. In the 8.2\,m Subaru telescope each of the 264
supports supplies active axial and passive lateral forces (Iye
[1991]).
To avoid
unwanted moments from the lateral forces, the application points of
the support forces have to be in the neutral surface of the
mirror, which required the drilling of 264 bores into the solid
meniscus blank. The two 8\,m Gemini
telescope are equipped with a three stage axial support system (Stepp
and Huang [1994]). The
passive astatic part consists of a continuous pneumatic support under
the back surface supporting 75\% of the weight, and a hydraulic
support like the one in the VLT supporting the remaining 25\%. The
third stage has active electromechanical actuators as in the VLT.
Telescopes with structured primary mirrors like the 3.5\,m WIYN, the
6.5\,m Magellan and the $2 \times 8$\,m LBT (Columbus) telescopes
with honeycomb mirrors have the advantage that the primary mirrors
are, for the same weight and diameter, stiffer than the thin meniscus
mirrors, making them more resistant to wind buffeting
deformations. But with diameters as large as 8\,m also these mirrors
require active optics control for the fine tuning of the optics. Each
actuator applies both axial and lateral forces in the neutral
surface of the mirror. The increased stiffness has the disadvantage of
reducing the dynamic range of active optics corrections and
corresponding reduction of the low spatial frequency tolerance
relaxation in fabrication.\\
The primary mirror of the Hubble Space Telescope was equipped with
a figure correction system, consisting of 24 actuators. These could
correct, for example, up to half a micron of astigmatism, but the
range was far too small to correct the spherical aberration due to
incorrect polishing (see the remark in Mast and Nelson [1990]).
Further details about existing active
telescopes are given by Wilson [1999].
\section{Outlook}
\label{sec:outlook}
The extension of the active optics principles for two mirror
telescopes to telescopes with three or more elements and also with
combinations of segmented and monolithic mirrors does not require new
principles and techniques.
For telescopes with more than two elements additional information
about the misalignment can be obtained from measurements of the field
dependence of Zernike polynomials with rotational symmetries larger
than two. In general,
the field dependence of aberrations with a rotational symmetry $m$ is
given by a generalisation of the ovals of Cassini for rotational symmetry
two, but now with $m$ nodes instead of two nodes. Usually, the
coefficients of the higher order polynomials in the field decrease
rapidly with the order. In current large Ritchey-Chretien two mirror
telescopes the only significant field aberration is third order
astigmatism, but in larger multi-element telescopes higher order
Zernike field aberrations may also be detectable.
If not, an alignment which only
corrects the field dependence of third order astigmatism, would, although
it is in principle not perfect, be sufficient. If they can be measured, an
alignment has to be done by moving more than only two elements.
The number of wavefront analysers has to equal at least the highest
rotational symmetry contained in the set of polynomials used for the
alignment analysis. Consequently, two mirror telescopes should also be
equipped with two wavefront analysers in the field to be able to do
closed loop alignment corrections.\\
Future extremely large telescopes may have more than one flexible
monolithic mirror (Dierickx, Delabre and Noethe [2000]) and therefore the
capability of correcting also field effects.
Segmented mirror technology will become increasingly important for the
future generation of very large telescopes. For these, the development
of active optics control systems,
which can operate in closed loop at time intervals of the order of one
minute, will be essential.\\
Telescopes with somewhat different applications of active optics include the
Chinese LAMOST project (Su, Cui, Wang and Yao [1998]) and the hexapod
telescope of the University of Bochum (see the overview given by Wilson
[1999]). LAMOST is
a 4\,m meridian type Schmidt telescope
with a thin reflecting aspheric corrector plate, where the optimum shape of
this plate depends on the zenith distance. The required deformations
as a function of the zenith distance could be introduced by active
forces at the back surface of the corrector plate.
In the 1.5\,m hexapod telescope the primary mirror is effectively the
thin front plate of the mirror cell. The support is position based and the
mirror behaves like mirrors used in adaptive optics, in the sense that
the actuation of one support generates a local deformation of the
mirror.\\
Active optics is particular suited to telescopes in space.
Since even with the disturbing effects of the atmosphere on the
wavefront analysis, diffraction limited performance of the telescope
optics itself can be achieved on the ground for telescopes with
mirror diameters of eight meters, it should be much easier to realise
this goal in space.\\
Low expansion glasses are still the favourite substrate for large
mirrors. In the case of monolithic mirrors the advantage of the
stability of the shape at different temperatures is, at least with a
closed loop active optics system, irrelevant.
Other substrates, especially
aluminium, may be better suited for large blanks. The advantages are a
higher thermal conductivity and therefore a faster equilibration with
the ambient temperature, a lower safety risk, and possibly a reduced
cost for the blank production. Long term instabilities of the shape
would most likely be in the low spatial frequency modes and could
easily be corrected by active optics.\\\\
\begin{center}
{\bf Acknowledgements}\\
\end{center}
The author would like to thank S. Guisard, J. Spyromilio and
R.N. Wilson for carefully reading the manuscript and for
suggestions.\\\\
\begin{center}
{\bf References}\\
\end{center}
Braat, J., 1987, JOSA A, {\bf 4}(4), 643.\\
Chanan, G., Mast, T., Nelson, J., 1988, Proc. ESO Conf. Very Large
Telescopes and their Instrumentation, Garching, Germany.\\
Chanan, G., Nelson, J., Mast, T., Wizinowich, P., Schaefer, B., 1994,
 Proc. Conf. Instrumentation in Astronomy VIII, SPIE Vol. 2198, 1139.\\
Chanan, G., Troy, M., Dekens, F., Michaels, S., Nelson, J., Mast, T.,
Kirkman, D., 1998, Applied Optics, Vol. 37(1), 140.\\
Chanan, G., Troy, M., Ohara, C., 2000,
 Proc. Conf. Optical Design, Materials, and Maintenance,
 SPIE Vol. 4003, 188.\\
Cohen, R., Mast, T., Nelson, J., 1994, Proc. Conf. Advanced Technology
 Optical Telescopes V, SPIE Vol. 2199, 105.\\
Couder, A., 1931, Bulletin Astron., 2me Serie,
  Tome VII, Fasc.VI, 266 and 275.\\
Creedon, J.F., Lindgren, A.G., 1970, Automatica, {\bf 6}, 643.\\
Cullum, M., Spyromilio, J., 2000,
 Proc. Conf. Telescope Structures, Enclosures, Controls,
 Assembly/Integration/Validation, and Commissioning,
 SPIE Vol. 4004, 194.\\
Dierickx, Ph., 1992, J. Mod. Optics, {\bf 39} (3), 569.\\
Dierickx, Ph., Delabre, B., Noethe, L., 2000,
 Proc. Conf. Optical Design, Materials, and Maintenance,
 SPIE Vol. 4003, 203.\\
Guisard, S., Noethe, L., Spyromilio, J., 2000,
 Proc. Conf. Optical Design, Materials, and Maintenance,
 SPIE Vol. 4003, 154.\\
Hubin, N., Noethe, L., 1993, Science, 262, 1390\\
Iye, M., 1991, JNLT Technical Report No. 2.\\
Mast, T. and Nelson, J., 1990, Proc. Conf. Advanced Technology Optical
 Telescopes IV, SPIE Vol. 1236, 670
McCleod B.A. 1996, PASP 108, 217.\\
Noethe. L., Franza, F., Giordano, G., Wilson, R.N.,
       Citterio, O., Conti, G., and Mattaini, E., 1988, 
       J. Mod. Optics, {\bf 35}(9), 1427.\\
Noethe. L., 1991, J. Mod. Optics, {\bf 38}(6), 1043.\\
Noethe, L., Guisard, S., 2000,
  Astron. Astrophys., Suppl. Ser. {\bf 144}, 157.\\
Noethe, L., 2000, Active optics in large telescopes with thin meniscus
 primary mirrors, Habilitationsschrift, Technische Universit\"{a}t Berlin\\
Noll, J.N., 1976, J. Opt. Soc. Am., {\bf 66}, 3, 1976.\\
Racine, R., Salmon, D., Cowley, D., Sovka, J., 1991,
  Proc. Astr. Soc. Pac., {\bf 103}, pp. 1020.\\
Roddier, F., Roddier, C., 1991,  Applied Optics, {\bf 30}(11), April 1991, 1325.\\
Shack, R.V., Platt, B.C., 1971, JOSA, 61, 656.\\
Shack, R.V., Thompson, K., 1980, Proc. Conf. Optical Alignment, SPIE Vol. 251, 146\\
Schneermann, M., Cui, X., Enard, D., Noethe, L., Postema, H., 1990,
 Proc. Conf. Advanced Technology Optical Telescopes IV,
 SPIE Vol. 1236, 920.\\
Schroeder, D.J., 1987, Astronomical Optics, Academic, San Diego.\\
Schwesinger, G., 1988, J. Mod. Optics, 35(7), 1988, 1117.\\
Schwesinger, G., 1991, J. Mod. Optics, 38(8), 1991, 1507.\\
Schwesinger, G., 1994, Applied Optics, 33(7), March 1994, 1198.\\
Stepp, L., 1993, Gemini report TN-O-G0002, 22.1.1993.\\
Stepp, L., Huang, E., 1994,
 Proc. Conf. Advanced Technology Optical Telescopes V, SPIE Vol. 2199, 223.\\
Su, D., Cui, X., Wang, Y., Yao, Z., 1998, Proc. Conf. Advanced Technology
 Optical/IR Telescopes, SPIE Vol. 3352, 76.\\
Troy, M., Chanan, G., Sirko, E., Leffert, E., 1998,
 Proc. Conf. Advanced Technology Optical/IR Telescopes VI, SPIE Vol. 3352, 307\\
Wetthauer, A., Brodhun, E., 1920, Zeitschr. f. Instrumentenkunde,
 {\bf 40}, 96.\\
Wilson, R.N., 1982, Optica Acta, {\bf 29}(7), 985.\\
Wilson, R.N., Franza, F., Noethe, L., 1987,
  J. Mod. Optics, {\bf 34}, 485.\\
Wilson, R.N., Franza, F., Giordano, P., Noethe, L., Tarenghi, M., 1989,
  J. Mod. Optics, {\bf 36}(11), 1415.\\
Wilson, R.N., Franza, F., Noethe, L., Andreoni, G, 1991,
  J. Mod. Optics, {\bf 38}(2), 219.\\
Wilson R.N., 1996,
  Reflecting Telescope Optics I, Springer-Verlag, Berlin.\\
Wilson R.N., 1999,
  Reflecting Telescope Optics II, Springer-Verlag, Berlin.\\
Yoder, P.R., 1986, Opto-Mechanical Systems Design,
 Marcel Dekker Inc.\\
Wizinowich, P., Mast, T., Nelson, J., DiVittorio, M., 1994,
 Proc. Conf. Advanced Technology Optical Telescopes V, SPIE Vol. 2199, 94.\\\\
\end{document}